\documentclass[twocolumn]{aastex63}
\usepackage{lineno}

\usepackage{graphicx}
\usepackage{wrapfig}
\usepackage{natbib}
\usepackage{color}
\graphicspath{{./figures/}} 
\usepackage{mathtools}
\usepackage{epstopdf}
\usepackage[autostyle]{csquotes}
\usepackage{hyperref} 
\def    \apjl  		{\rm {ApJL}}

\def    \apjl  		{\rm {ApJL}}

\def	\cm		{\,{\rm {cm}}}
\def	\K		{\,{\rm K}}
\def	\g		{\,{\rm {g}}}
\def	\mum	{\,{\mu \rm{m}}}
\def	\erg    		{\,{\rm {erg}}}
\def	\s		{\,{\rm {s}}}
\def	\H		{\,{\rm {H}}}

\def    \P     {\,\boldsymbol{P}\,}
\def    \B     {\,\boldsymbol{B}\,}
\def    \k     {\,\boldsymbol{k}\,}
\def    \J     {\,\boldsymbol{J}\,}

\def \bea {\begin{eqnarray}}
\def \ena {\end{eqnarray}}                 
     
\begin{document}
\shorttitle{ Grain dynamic during RATD period}
\title{Synthetic Modelling of Polarized Dust Emission in Intermediate-Mass YSOs: II: Effects of Radiative Torque Disruption on Dust Grains in Protostellar Jets/Outflows}
 
\author{Nguyen Chau Giang}
\affil{Korea Astronomy and Space Science Institute, Daejeon 34055, Republic of Korea}
\email{chaugiang@kasi.re.kr}
\affil{Department of Astronomy and Space Science, University of Science and Technology, 217 Gajeong-ro, Yuseong-gu, Daejeon, 34113, Republic of Korea}

\author{V. J. M. Le Gouellec}
\email{valentinlegouellec@gmail.com}
\affil{NASA Ames Research Center, Space Science and Astrobiology Division M.S. 245-6 Moffett Field, CA 94035, US}
\affil{NASA Postdoctoral Program Fellow}

\author{Thiem Hoang}
\affil{Korea Astronomy and Space Science Institute, Daejeon 34055, Republic of Korea}
\email{thiemhoang@kasi.re.kr}
\affil{Department of Astronomy and Space Science, University of Science and Technology, 217 Gajeong-ro, Yuseong-gu, Daejeon, 34113, Republic of Korea}

\author{A. J. Maury}
\affil{Institute of Space Sciences (ICE), CSIC, Campus UAB, Barcelona, Spain}
\affil{ICREA, Pg. Lluís Companys 23, Barcelona, Spain}

\author{P. Hennebelle}
\affil{AIM, CEA, CNRS, Université Paris-Saclay, Université Paris Diderot, Sorbonne Paris Cité, F-91191 Gif-sur-Yvette, France}

\begin{abstract}
One of the potential explanations for the existence of very large grains (VLGs) in the inner envelope of low/intermediate-mass Class 0/I Young Stellar Object is the migration of VLGs from the protostellar disk via a protostellar outflow. To understand whether the grain migration is prevented by RAdiative Torque Disruption (RATD), we perform the numerical modeling of RATD in parallel with the grain propagation, using the gas velocity and density structure inside the jet and outflow from an MHD simulation of an intermediate Class 0 protostar. We found that with the bolometric luminosity $\geq 20L_{\odot}$, RATD can destroy aggregate grains of size $1 \sim  500\mum$ having maximum tensile strength $S_{\rm max} \leq 10^{5}\erg\cm^{-3}$ inside the jet/outflow base after $< 2$ yrs. This effect lets sub-micron grains dominate the outflow and partially prevent the migration of large grains from the inner disk to inner envelope. In contrast, RATD cannot prevent the migration of composite VLGs and submillimeter grains having $S_{\rm max}\geq 10^{7}\erg\cm^{-3}$. Next, we incorporate RATD into POLARIS, assuming grains are not moving relative to the gas. We found that POLARIS works well in describing the disruption for aggregate grains, but overestimates the disruption efficiency for composite grains. The observed polarization degree can be reduced by twice when aggregate grains are removed from the outflow cavity wall and inner envelope by RATD. However, RATD is not an important factor controlling dust polarization properties as iron inclusions do.
 
\end{abstract}


\cite{Tsukamoto_Maury_review}
 \section{Introduction}\label{sec:intro}
Dust grains are the building blocks of planets, yet how dust grains grow from submicron size in dense molecular clouds (MCs) to kilometer size (planetesimals) in protostellar disks remains unclear (see a review by, i.e., (\citealt{Testi_2014}). Dust grains provide the surface for the formation of complex organic molecules (i.e., as CO, $\rm CO_{\rm 2}$, methanol, ethanol \citealt{Dishoeck_1998}, \citealt{Jorgensen_2020}), and influence the abundance of charged particles and the magnetic diffusivities of the ambipolar diffusion, Ohmic dissipation, and Hall effect inside protostellar cores/disks (\citealt{Zhao_2016}, \citealt{lebreuilly_2023}, \citealt{Tsukamoto_2020_dust_model}, \citealt{Tsukamoto_Satoshi_2022}, \citealt{Tsukamoto_2023_coevolution}). Dust grains also help to study the dynamic importance of magnetic fields ($\B$) in guiding star, disk, jet, and outflow formation (\citealt{Hull_2019}, \citealt{Maury_2022}, \citealt{Yusuke_2022}) thanks to polarized thermal dust emission from magnetically aligned dust grains (\citealt{Anderson_2015}, \citealt{Tram_Hoang_2022}). Consequently, accurately constraining the grain size distribution, grain physical properties, and grain dynamics is important to improve our understanding of the formation of stars, planets, and COMs. 

Related to the planet formation process, the early grain growth is usually studied via the dust opacity index $\beta$ inferred from observations of the spectral energy distribution of thermal dust emission. Larger grains produce lower $\beta$. Typically, one gets $\beta \sim 1.7-2$ inside the diffuse interstellar medium (ISM) and MCs (\citealt{Li_Draine_2001}, \citealt{Sadavoy_2016}), which corresponds to the typical maximum grain size of $\sim 0.1 - 0.25\mum$ in this region. In contrast, observations toward the protostellar cores (\citealt{Schnee_2014}, \citealt{Bracco_2017}) and Class 0/I Young Stellar Objects (YSOs) frequently report the reduction of $\beta$ with increasing gas column density, reaching $\beta \sim 1$ inside Class 0/I disks (\citealt{Kwon_2009}, \citealt{Chiang_2012}, \citealt{Miotello_2014}, \citealt{Galametz_2019}, \citealt{Cacciapuoti_2023a}). Grain growth inside protostellar disks is also indicated via the rising in thermal dust polarization spectrum (\citealt{Yen_2020}) and the detection of the uniform polarization pattern by self-dust scattering within these inner 100 au scale (\citealt{Kataoka_2015}, \citealt{Cox_2015}, \citealt{Kataoka_2016}, \citealt{Yang_2016a}, \citealt{Sadavoy_2019}, \citealt{Aso_2021}). \cite{Kataoka_2015} found that self-dust scattering is only effective if the maximum grain size inside the disk becomes comparable to the observed wavelength. As ALMA detects self-scattering features inside Class 0/I disks at Band 3 ($0.85-0.87\mu m$), it indicates that grains may already grow to $10\mum$ (very large grains, VLGs) and even submillimeter grain sizes (grains beyond $> 100\mum$) there.  

The low $\beta \sim 1$ is not only found inside the protostellar disk but also the inner envelope of $\sim 500$ au scale (\citealt{Kwon_2009}, \citealt{Chiang_2012}, \citealt{Miotello_2014}, \citealt{Cacciapuoti_2023a}). The presence of VLGs beyond the disk scale is also suggested (by, i.e., \citealt{Valentin_2019}, \citealt{Valdivia_2019}, \citealt{Hull_2020}, \citealt{Valentin_2023b}, \citealt{Giang_et_al_2024}) to explain $\geq 5\%$ of dust polarization and the efficient grain alignment inferred by ALMA in hundreds and thousands au of low/intermediate-mass Class 0/I YSOs (\citealt{Cox_2018}, \citealt{Maury_2018}, \citealt{Kwon_2019},  \citealt{Valentin_2023a}). Grain growth simulation inside the collapsing core showed that VLGs and submillimeter grains can form inside the protostellar disk after $\sim 1000$ yr (\citealt{Ormel_2009}, \citealt{Hirashita_2013}, \citealt{Li_2014}, \citealt{lebreuilly_2023}, \citealt{Tsukamoto_ashfall_2021}, \citealt{Tsukamoto_2023_coevolution}). Inside the inner envelope, sub-micron grains below $0.25\mum$ (typical maximum grain size inside ISM) can only grow to micron-sized grains of $\sim 1\mum$ due to the low gas density. This finding disfavors the fast grain growth scenario within 500 au scale around the protostar (\citealt{Wong_2016},  \citealt{Tsukamoto_2023_coevolution}, \citealt{Lebreuilly_2024}).  

A potential mechanism for the presence of VLGs inside the inner envelope is the migration of large grains formed inside the protostellar disk via the protostellar outflow. This scenario is first studied numerically by \cite{Wong_2016}, showing that VLGs and submillimeter grains up to $\sim 1\rm mm$ inside the disk can be uplifted to the outflow base by the gas drag force and only be destroyed $< 10\%$ by grain shattering when entering the envelope. Recently, non-ideal MHD simulations accounting for the grain dynamic by \cite{Tsukamoto_ashfall_2021} (see also \citealt{Marchand_2023}) confirmed that scenario. VLGs and submillimeter grains after entering the outflow can be quickly decoupled from outflowing gas and populate the inner envelope by centrifugal forces after $6\times 10^{3} - 10^{4}$ yr. Some of them can infall again to the outer disk region, forming the close dust cycle inside the protostellar core, named Ashfall dust model (\citealt{Tsukamoto_ashfall_2021}, see also \citealt{Koga_2023}). Recently, a negative correlation between $\beta$ determined in the inner envelope and the jet and outflow mass-loss rate is found from the CALYPSO samples of Class 0/I YSOs (\citealt{Maury_2019}) by \cite{Cacciapuoti_2023a}. As the outflow mass-loss rate is correlated with the envelope mass, the smaller $\beta$ in the inner envelope may relate to the outflow activities (\citealt{Galametz_2019}), which positively supports the migration of dust grains from disks to inner envelope supposed by \cite{Wong_2016} and \cite{Tsukamoto_ashfall_2021}.  

However, dust grains can face strong dust destruction by the mechanical and radiation feedback induced by the protostar when entering and traveling inside the jet and outflow. Firstly, they can be sublimated when entering the sublimation zone of the jet base where the dust temperature exceeds $\geq 1200$ K (sublimation threshold for astrosilicate grains) or $\geq 2100$ K (sublimation temperature for graphite grains). Note that for low/intermediate Class 0 protostars, the sublimation zone inside the jet is within $< 0.1$ au to the protostar (\citealt{Garcia_2001}). Therefore, dust grains entering the jet and outflow base beyond this zone can still propagate outward without being removed by thermal sublimation. Secondly, \cite{Gusdorf_2008a} and \cite{Gusdorf_2008b} pointed out that $\sim 5\%$ of outflowing grains can be destroyed by thermal sputtering when suffering from the mild shock with $v_{\rm shock} \sim 20-50$ km/s. \cite{Guillet_2007}, \cite{Guillet_2009}, \cite{Guillet_2011}, and \cite{Guillet_2013} found that inside the shock front where the gas density exceeds $n_{\rm H_{2}} \geq 10^{5}\cm^{-3}$, grain-grain shattering followed by vaporisation can also destroy dust grains whose temperature exceeds thousand Kelvin. The enhanced dust cross-section by shattering will strengthen the impact of sputtering on grains, which increases the grain destruction fraction inside the shock to $\sim 8\%$ (\citealt{Gusdorf_2008b}). Grain-grain shattering can also destroy $\sim 10\%$ of outflowing grains entering the inner envelope due to their relative motion with infalling grains (\citealt{Wong_2016}). Given $< 10\%$ of grains are destroyed while propagating inside the jet and outflow, and $< 10\%$ are removed when entering the inner envelope, $\sim 80\%$ of large grains uplifted from the disk surface can migrate toward the inner envelope, producing the large dust polarization as detected by ALMA there. 

Another mechanism that can destroy dust grains is the RAdiative Torque Disruption (RATD) proposed by \cite{Hoang_2019_nature}. Theoretically, irregular dust grains experiencing unbalanced interactions between their left- and right-hand sides with an anisotropic radiation field can be spun up to suprathermal rotation by radiative torques (RATs, \citealt{Dolginov_1976}, \citealt{Lazarian_Hoang_2007a}, \citealt{Hoang_Lazarian_2008}). The suprathermal rotation not only causes grain alignment (\citealt{Dolginov_1976}, \citealt{Lazarian_Hoang_2007a}, \citealt{Anderson_2015}, \citealt{Lazarian_2015}, \citealt{Hoang+2022}) but also the disruption of large grains into smaller fragments. The RATD effect occurs when the centrifugal force induced by grain suprathermal rotation exceeds the Van der Waals force that binds grain monomers together. In contrast to thermal sublimation that requires strong radiation field to break up chemical bond with the binding energy of $\sim 0.1$ eV inside grains, RATD only needs to break up the Van der Waals force with the lower binding energy of $\sim 0.01$ eV (\citealt{Hoang_2019}, \citealt{Hoang_2020}). Such a lower required disruption energy makes RATD more effective than thermal sublimation in processing grains beyond the sub-au scale. RATD is also independent of the drift velocity between gas-grain and grain-grain as thermal sputtering and grain-grain shattering. Therefore, it can affect dust grains beyond the shock regions inside the jet and outflow and infalling grains inside the inner envelope where the gas-grain coupling is still held on. However, the most important point is that RATD can work with low-energy optical-FIR photons and large grains formed via grain coagulation tend to have very low maximum tensile strength due to their highly porous structures (\citealt{Li_Greenberg_1997}, \citealt{Seizinger_2013}, \citealt{Tatsuuma_2019}, \citealt{Kimura_2020}, \citealt{Garcia_2020}). As the peak of RAT efficiency occurs when grains interact with radiation of the wavelength comparable to their size (\citealt{Lazarian_Hoang_2007a}, \citealt{Hoang_Lazarian_2008}), RATD may strongly affect the propagation of large grains up to hundreds and thousands au scale inside the jet and outflow.
 
The first study of RATD in the early stage of protostellar cores is performed by \cite{Hoang_2021_polarization_hole}. By considering the simple Boner-Ebert gas distribution with the maximum $n_{\rm H_{2}} \sim 10^{7}\cm^{-3}$, they found that VLGs up to $50\mum$ can be destroyed by RATD within 40 au around the protostar with $L_{\rm star} = 50L_{\odot}$. Inside the intermediate-mass Class 0 protostar, \cite{Valentin_2023b} found that RATD can destroy VLGs beyond $10\mum$ having aggregate-type structure up to $\sim 4000$ au inside the jet and outflow cavity if the bolometric luminosity exceeds $5L_{\odot}$. However, VLGs beyond the disk scale are required to reproduce $\geq 5\%$ of dust polarization and the high grain alignment efficiency detected inside Class 0/I YSOs by ALMA (\citealt{Valdivia_2019}, \citealt{Valentin_2020}, \citealt{Valentin_2023b}, \citealt{Giang_et_al_2024}). The finding in \cite{Valentin_2023b} implies protostellar grains have compact or composite structures to survive against RATD while propagating inside the jet and outflow. However, how VLGs with high tensile strength can form remains unclear. 

Besides, when estimating the critical sizes for rotational disruption, \cite{Hoang_2021_polarization_hole} and \cite{Valentin_2023b} adopted an analytical formula from \cite{Hoang_2019_nature}, which uses the average radiation field strength and the mean wavelength to estimate the net RAT acting on grains. Given the strong dependence of RAT efficiency on grain sizes, using the analytical formula may underestimate the RATD strength acting on grains. Besides, the transportation of dust grains along the jet and outflow has not yet been considered in the study. For outflowing grains propagating with a few km/s, results may not change significantly because the disruption can happen before grains migrate to larger distances where RATD becomes inefficient. However, for grains propagating with a few tens to a few hundred km/s inside the outflow, the fast weakening of RATs by decreasing radiation field strength may deactivate the effect of RATD on moving grains. Furthermore, since RATD is an important effect that accompanied the RAT alignment but not yet implemented in POLARIS, we will first incorporate this effect and study how it affects polarized dust emission remains unclear. Note that RATD cannot remove $100\%$ of the dust grains larger than the critical disruption size (\citealt{Hoang_2019_nature}), and only a fraction $f_{\rm high-J}$ of grains aligning with magnetic fields at high-J attractors can be fragmented by RATD. Given the strong dependence of $f_{\rm high-J}$ on the grain magnetic properties (\citealt{Lazarian_Hoang_2007b}, \citealt{Hoang_Lazarian_2016_mrat}) in protostellar environments (\citealt{Hoang_2022}, \citealt{Hoang+2022}, \citealt{Giang_2023a}, \citealt{Giang_Hoang_2024}, and \citealt{Giang_et_al_2024}), it indicates iron inclusions to control RATD activities inside Class 0 protostars. Therefore, an in-depth RATD study is required to accurately understand whether RATD suppresses the migration of dust grains from the inner disk to the inner envelope. Secondly, we aim to quantify the \cite{Valentin_2023b}'s argument on the effect of RATD on the properties of polarized dust emission observed in low/intermediate-mass protostars by ALMA. 

Our paper is organized as follows. We first describe the theoretical calculations of the grain velocity inside the jet/outflow and the rotational disruption by RATD in Section \ref{sec:theory}. The simultaneous modeling of RATD along with the grain transportation as a function of time will be shown in the follow-up Section \ref{sec:vgrain_omegat}. The incorporation and validity of RATD in POLARIS will be shown in Section \ref{sec:RATD_POLARIS}. We then present the effect of the new dust population by RATD on polarized dust emission in Section \ref{sec:ratd_polarization}. Further discussion about the impact of RATD on dust population, dust polarization, and the summary of our study will be in Sections \ref{sec:discussion} and \ref{sec:summary}, respectively.

\section{Theory of grain transportation and rotational disruption}\label{sec:theory}

\subsection{Grain acceleration by the drag force}\label{sec:vgrain}
We use the simple numerical model of \cite{Wong_2016} to get the grain velocity after they entrain the jet and outflow base. Considering the acceleration by the drag force from the gas-grain drifting and the deacceleration from gravity, they showed that grains could quickly gain the terminal velocity $v_{\rm grain}$ after leaving the uplifted point a few au. Assuming no deacceleration during the propagation of dust grains inside the jet and outflow, the velocity of grains with effective radius $a_{\rm eff}$\footnote{The effective grain size $a_{\rm eff}$ is defined as the radius of a spherical grain that shares the same volume with a considered irregular dust grain.} at distance $r$ to the center can be derived as:
\bea 
v_{\rm grain}(r,a_{\rm eff}) = v_{\rm gas}(r) - \sqrt{\frac{4 G M(\leq r) a_{\rm eff} \rho_{\rm grain}}{3 \mu n_{\rm H}(r) m_{\rm H} r^{2}}}, \label{eq:vgrain}
\ena
where $M(\leq r)$ is the total gas mass inside the inner region of radius $r$, $\rho_{\rm grain}$ is the volume grain mass density, and $\mu = 2$ is the atomic mass number. Generally, dust grains will move slower than gas when propagating outward, and the gas-grain decoupling happens stronger for larger grain sizes. The general difference between gas and grain velocity is below $< 3$ km/s (similar to finding in \citealt{Wong_2016}).

\subsection{Grain Spinup and Rotational Disruption by RATs}\label{sec:omega_t}
Besides being accelerated by the drag force, dust grains can also be spun up by RAdiative Torques (RATs, \citealt{Dolginov_1976}, \citealt{Draine_Weingartner_1996} \citealt{Lazarian_Hoang_2007a}, \citealt{Hoang_Lazarian_2008}). Radiative torque is formed from an anisotropic radiation field of unbalanced radiation forces between the right- and left-hand circularly polarized scattering and absorptions components. The spinning process by RATs is accompanied by the rotational damping by gas-grain collisions and thermal dust emission. Following \cite{Hoang_2019_nature}, the angular velocity $\Omega(t)$ of grains with effective radius $a_{\rm eff}$ achieved after moving $t$ time can be described as:
\begingroup\makeatletter\def\f@size{8}\check@mathfonts
\bea 
\Omega(t,a_{\rm eff}) = \frac{\tau_{\rm damp}}{I} \Bigg[\Gamma_{\rm RAT} - \Bigg(\Gamma_{\rm RAT} - \frac{I \Omega(t_{0})}{\tau_{\rm damp}}\Bigg)e^{-\frac{\rm t}{\tau_{\rm damp}}}\Bigg]\label{eq:omega_t},
\ena
\endgroup
where $\Gamma_{\rm RAT}$ is the net radiative torques of the radiation field acting on grains \footnote{Note that we are using the RAT calculation of compact grains (\citealt{Lazarian_Hoang_2007a}, \citealt{Hoang_Lazarian_2008}, \citealt{Herranen_2021}) for aggregate grains there (see Appendix \ref{sec:appen_RAT}). The recent study by \cite{Jager_2024} pointed out that aggregate and compact grains share similar behavior of RAT efficiency $Q_{\Gamma}$ with the ratio $\lambda/a_{\rm eff}$. However, $Q_{\Gamma}$ of aggregate grains is smaller by $\sim 10-100$ times when $\lambda/a_{\rm eff} \sim 1-100$, leading aggregate grains to be spun up by RATs slower than compact grains by twice times.}, $\tau_{\rm damp}$ is the typical grain damping timescale, $I$ is the grain inertia moment, and $\Omega(t_{0})$ is the initial angular velocity of dust grains. The detailed RAT calculations are in Appendix \ref{sec:appen_RAT}.
 
Dust grains will be disrupted by RATD when the induced centrifugal force exceeds the Van der Waals force that holds grain monomers together (\citealt{Hoang_2019_nature}, \citealt{Hoang_2020}). The maximum angular velocity of grains before being fragmented by RATD is called the disruption threshold  $\Omega_{\rm disr}$. The dependence of $\Omega_{\rm disr}$ on the maximum tensile strength $S_{\rm max}$ is given by 
 
\bea 
\Omega_{\rm disr} = \frac{2}{a_{\rm eff}} \Bigg(\frac{S_{\rm max}}{\rho_{\rm grain}}\Bigg)^{1/2}.\label{eq:omega_disr}
\ena
The maximum tensile strength depends strongly on the grain structure, composition, and grain sizes (i.e., larger grain sizes, lower maximum tensile strength). It can vary from high $S_{\rm max} \sim 10^{11}\erg\cm^{-3}$ for diamonds, $S_{\rm max} \sim 10^{9} -10^{10}\erg\cm^{-3}$ for polycrystalline bulk solid (\citealt{Burke_Silk_1974}, \citealt{Draine_Salpeter_1979}), to lower $S_{\rm max} \sim 10^{6} - 10^{9}\erg\cm^{-3}$ for composite grains (\citealt{Hoang_2019}), or $S_{\rm max} \sim 10^{3} - 10^{5}\erg\cm^{-3}$ for large, fluffy grains formed via the grain-grain coagulation in prosotellar disks and protoplanetary disks (\citealt{Li_Greenberg_1997}, \citealt{Seizinger_2013}, \citealt{Tatsuuma_2019}, \citealt{Hoang_2019}, \citealt{Kimura_2020}).

\begin{figure*}
\centering
    \includegraphics[width=\textwidth,height=\textheight,keepaspectratio]{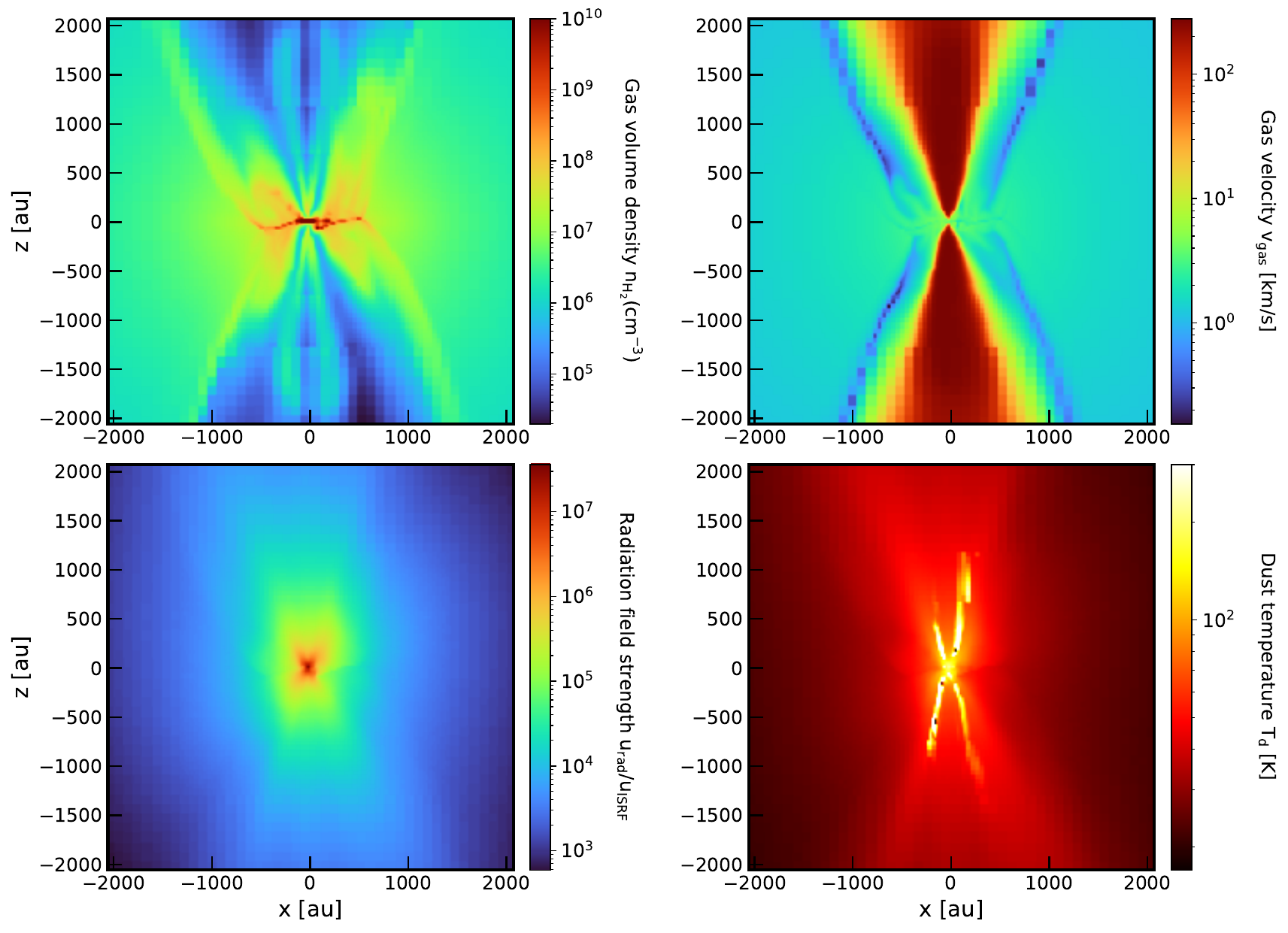}
    \caption{Upper row: spatial distribution of the gas volume density $n_{\rm H_{2}}$ and gas velocity $v_{\rm gas}$ on the slice containing the sink particle. The object is seen with edge$-$on direction. The protostellar core shows a symmetric structure around the vertical z$-$direction. The jet propagation shapes the outflow cavity, forming the high-velocity domain with an opening angle of $12^{\circ}$ where gas propagates with $v_{\rm gas} > 30$ km/s and low-velocity domain where $v_{\rm gas} \sim 3-30$ km/s. Lower panel: distribution of the radiation field strength $u_{\rm rad}/u_{\rm ISRF}$ and dust temperature $T_{\rm d}$ obtained after post-processing the MHD simulation with POLARIS. The net radiation field energy density $u_{\rm rad}$ is calculated by integrating over the electromagnetic spectrum from $\lambda_{\rm min} = 0.1\mum$ to $\lambda_{\rm max} = 2$ mm, that most of energy comes from thermal dust emission at infrared wavelengths. The radiation field strength decreases outward by the dust reddening effect. Grains can be heated to $\sim 200$ K in the center,  $T_{\rm d} \sim 80$ K inside the outflow, and lower $T_{\rm d}\sim 20$ K inside the envelope and the equatorial midplane.} 
     \label{fig:distribution_mcrt}
\end{figure*}

\section{Simultaneous modelling of grain motion and rotational disruption}\label{sec:vgrain_omegat} 
To understand the effect of RATD during the transportation of dust grains inside the jet and outflow, we first briefly describe our MHD simulation of the collapsing protostellar core in Section \ref{sec:MHD+radiation_field}. The variation of the grain angular velocity with grain sizes will be shown in Section \ref{sec:omega_grain_size}. The map of the disruption size as a function of time will be presented in Section \ref{sec:adisr_dynamic}. And the typical disruption size inside the jet and outflow during RATD period will be shown in Section \ref{sec:adisr_nH_vgas}.
  
\subsection{RAMSES MHD simulation}\label{sec:MHD+radiation_field}
Continuing from our previous study, we use the same simulation of the intermediate-mass Class 0 protostar adopted in \cite{Valentin_2023b} to model the effect of RATD on dust population. The MHD simulation is post-processed with the POLARIS code (\citealt{Reissl_2016}) to obtain the distribution of the radiation field strength and dust temperature inside the core. For a quick summary, our non-ideal MHD simulation models the star formation inside the intermediate-mass core having an initial uniform magnetic field along the rotation z-direction (\citealt{Valentin_2023b}). The core has no initial turbulent, the ambipolar diffusion is taken into account to describe the decoupling of magnetic fields and infalling gas (\citealt{Masson_2012}), and a jet is implemented by hand  (\citealt{Verliat_2022}) right after a sink particle forms (\citealt{Krumholz_2004}, \citealt{Bleuler_2014}).

To perform the radiative transfer with POLARIS, we use the snapshot at the moment when the core evolves $38.52$ kyr, the sink particle has a mass of $M_{\rm star} = 1.2M_{\odot}$, a size of $R_{\rm star} = 1.58R_{\odot}$, and a luminosity of $L_{\rm star} = 0.58L_{\odot}$. Following \cite{Valentin_2023b} and \cite{Giang_et_al_2024}, we consider the sink particle as a black body with a luminosity of $100L_{\odot}$ to maximize the effect of RATD on dust population and dust polarization in the intermediate protostar. The radiation spectrum extends from the minimum wavelength $\lambda_{\rm min} = 0.1\mum$ that corresponds to the lower cutoff of Hydrogen ionization, to the maximum wavelength $\lambda_{\rm max} = 3$mm where grain-radiation interactions become negligible. For the dust model, we adopt the uniform distribution of dust grains in the entire protostellar core with the typical dust-to-gas mass ratio $\eta = 0.01$ found in ISM. Dust grains are considered to have an oblate shape with the axial ratio $s = 1/2$, and to be a mixture of $67.5\%$ of silicate and $37.5\%$ of graphite, giving the grain mass density $\rho_{\rm grain} = 3.3\rm g cm^{-3}$ (Equation \ref{eq:vgrain}). We consider dust grains to follow the standard MRN distribution with $dn/da = n_{\rm H_{2}} C a^{-3.5}$ (\citealt{Mathis_1977}), with $C$ the normalization constant of the size distribution \footnote{The normalization constant $C$ can be calculated from the dust-to-gas mass ratio $\eta = 0.01$, giving $C = \frac{m_{\rm H_{2}} \eta}{\int_{\rm a_{\rm min}}^{\rm a_{\rm max}} 4/3 \pi \rho_{\rm grain} s a^{-0.5} da}$}. We consider the wide grain size from the minimum size $a_{\rm min} = 5$ nm to the maximum size $a_{\rm max} = 50\mum$.

As illustrated in \cite{Giang_et_al_2024}, dust grains in low/intermediate-mass Class 0 YSOs must be superparamagnetic material (SPM) to explain the high polarization fraction and the efficient grain alignment inferred by ALMA. Therefore, we consider grains to be SPM with  $\sim 30\%$ of iron embedding inside dust grains under the cluster form, represented by the volume filling factor of iron clusters $\phi_{\rm sp} = 0.1$. Each cluster contains a moderate number of iron atoms of $N_{\rm cl} = 10^{3}$. In addition, we consider dust grains to be highly inelastic material with $\mu Q = 3\times 10^{9}\erg\cm^{-3}$ ($\mu$ the shear modulus and $Q$ the Q-factor). Given the magnetic properties and the inelasticity of dust grains, we use the updated POLARIS developed by \cite{Giang_2023a} and \cite{Giang_et_al_2024} to model in detail the magnetic alignment of SPM grains in protostellar environments. We consider the joint action of super-Barnett and Inelastic relaxation to model the internal alignment of SPM grains. And we use the Larmor precession condition and the theory of RATs/MRAT (Magnetically enhanced RAdiative Torque mechanism) to model the external alignment between dust grains with magnetic fields. Detailed descriptions of our grain alignment process with POLARIS are shown in \cite{Giang_et_al_2024}. The summary of our chosen parameters in the study is listed in Table \ref{tab:parameter}.

We show in the upper row of Figure \ref{fig:distribution_mcrt} the gas volume density and gas velocity distribution in the 4000 au scale box. Without the initial density perturbation by turbulence, the core collapses isotropically, forming a central disk with gas density of $n_{\rm H_{2}} \sim 10^{10}\cm^{-3}$ and a symmetric protostellar outflow along the z direction. The outflow cavity has a width of $\sim 1000$ au at height $z \sim 1000$ au, with the density of $n_{\rm H_{2}} \sim 10^{5} - 10^{6}\cm^{-3}$ (upper left panel) and gas velocity of $v_{\rm gas} \sim 3 - 300$ km/s (upper right panel). The outflow cavity is broadening outward, reaching beyond $\geq 2000$ au in width at $\geq 2000$ au in height. Gas density increases and gas velocity decreases toward the outflow cavity wall, reaching $n_{\rm H_{2}} \sim 10^{7}-10^{8}\cm^{-3}$ and $v_{\rm gas} \sim 1-3$ km/s. We note that in our MHD simulation, we cannot resolve the sub-au scale around the protostar where the jet and outflow are launched (\citealt{Valentin_2023b}). Therefore, we inject the jet by hand right after the sink particle forms, with an initial velocity of $v_{\rm gas} \sim 277$ km/s (see the detailed of the jet implementation in \citealt{Verliat_2022}, \citealt{Valentin_2023b}, and \citealt{Lebreuilly_2024}). The jet propagation swept up material above and below the protostar, which shapes the outflow cavity structure. It explains why the large portion of outflowing gas can propagate outward with high velocity $v_{\rm gas} \sim 30-300$ km/s over the cone with the opening angle of $\sim 12^{\circ}$. Observationally, the high gas velocity of several ten to several hundred km/s is only detected inside the narrow jet with the typical opening angle of a few degrees ( \citealt{Jhan_2016}, \citealt{Antoniucci_2008}, \citealt{Elle_2013}, \citealt{Podio_2021}, \citealt{Dutta_2024}). The corresponding jet-mass loss rate is about $\dot{M}_{\rm j} \sim 10^{-6}  - 10^{-7}M_{\odot}/\rm yr$. Gas propagating inside the outflow cavity has lower velocity of $\sim 3-30$ km/s and smaller mass-loss rate of $\dot{M}_{\rm o} \sim 10^{-7} - 10^{-9}M_{\odot}/yr$ (\citealt{Cacciapuoti_2023b}). In our simulation, although the density and velocity structure inside the outflow cavity are not consistent with observations in low/intermediate-mass Class 0/I YSOs, the mass-loss rate inside the high-velocity domain (where $v_{\rm gas} \sim 30-300$km/s) and low-velocity domain (where $v_{\rm gas} \sim 3-30$ km/s) \footnote{To determine the mass-loss rate at each height $z$ inside the protostellar outflow, we define cells having gas velocity of $\sim 10\%$ of maximum gas velocity $v_{\rm gas,max}(z)$ being the jet boundary (the gas velocity range inside the jet will be $\sim 30-300$ km/s). Cells having $3 \leq v_{\rm gas} \leq 30$ km/s and density below $n_{\rm H_{2}} \leq 10^{7}\cm^{-3}$ is classified as the outflow cavity. The jet mass-loss rate $\dot{M}_{\rm j}$ and the outflow mass-loss rate $\dot{M}_{\rm o}$ at distance $r$ to the center are determined by integrating the product of $v_{\rm gas}$ and $M_{\rm gas} = m_{\rm H_{2}} n_{\rm H_{2}}$ over the outflow surface area. We obtain $\dot{M}_{\rm j} \sim 10^{-6}M{\odot}/\rm yr$ inside the jet base and $\sim 10^{-7}-10^{-8}M_{\odot}/\rm yr$ inside the jet lobe. The outflow mass-loss rate is $\dot{M}_{\rm o} \sim 10^{-8}-10^{-9}M_{\odot}/\rm yr$.} are among the typical observed values found inside the jet and outflow cavity. Therefore, we can classify the two above regions as jet and outflow and use them to understand how RATD modifies the migration of dust grains inside these areas.
 
    \begin{figure*}
\centering
    \includegraphics[width=\textwidth,height=\textheight,keepaspectratio]{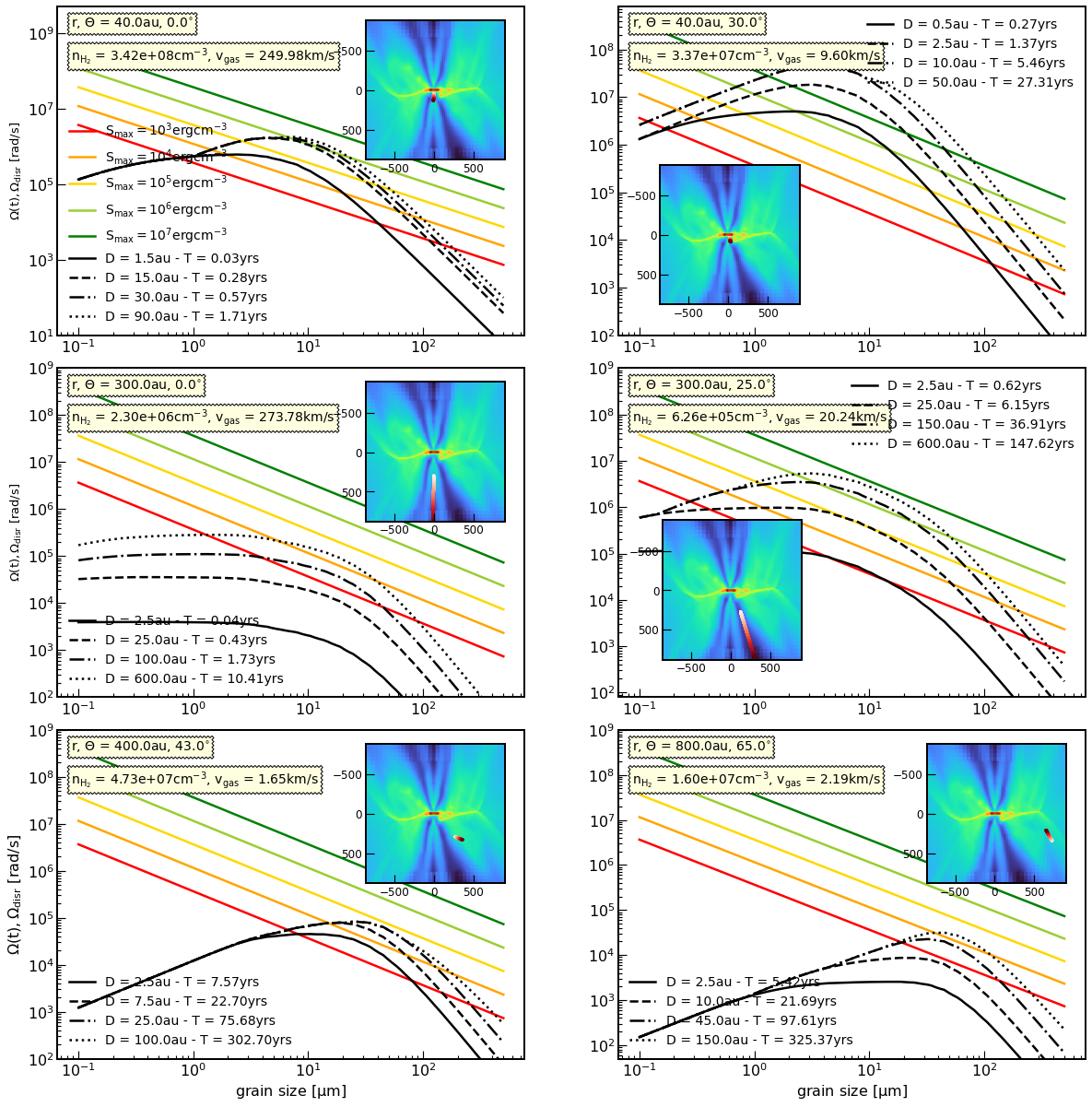}
    \caption{Variation of the grain angular velocity $\Omega(t,a)$ as a function of grain sizes with times (black lines). The corresponding displacement after each moving time is denoted in the legend box. Information of the starting point $[r,\Theta]$, with $r$ the distance to the protostar, and $\Theta$ the angle to the z$-$direction, $n_{\rm H_{2}}$, and $v_{\rm gas}$ are denoted in the lower left corner of each panel. The trajectories of dust grains are marked on the density map of size $800\times 800$ au placed in the corner of each panel. The gradient from white to red illustrates the moving direction of dust grains with time. We show in the same panel the variation of the disruption threshold $\Omega_{\rm disr}$ as a function of grain size for different maximum tensile strength $S_{\rm max}$, from $S_{\rm max} = 10^{3}\erg\cm^{-3}$ to $10^{7}\erg\cm^{-3}$. Grains will be considered to be destroyed by RATD when $\Omega(t) \geq \Omega_{\rm disr}$. Upper row: results for dust grains inside the jet base (left panel) and outflow base (right panel). Middle row: results for grains at $z = 300$ au moving with a few hundred km/s along z$-$direction (left panel) and a few tens km/s inside the outflow cavity (right panel). Lower row: results for grains propagating inside the outflow cavity wall (left panel) and infalling inside the inner envelope (right panel). RATD can quickly destroy micron-sized grains, VLGs, and submillimeter grains inside the jet/outflow base just after a few yr owning to the intense radiation field there. The disruption timescale increases to a few tens to a few hundred yr for grains at hundreds au scale. RATD happens weaker for grains propagating outward with a few hundred km/s and having higher maximum tensile strength. Additionally, the rotational disruption is mostly deactivated inside the dense outflow cavity wall and inner envelope, except for grains having $S_{\rm max} \sim 10^{3}-10^{4}\erg\cm^{-3}$. } 
     \label{fig:wt_grain_size}
\end{figure*}

The spatial distribution of the radiation field strength $U = u_{\rm rad}/u_{\rm ISRF}$ and dust temperature $T_{\rm d}$ obtained from the radiative transfer simulation with POLARIS are shown in the lower row of Figure \ref{fig:distribution_mcrt}. The radiation field strength and dust temperature decrease outward due to the dust reddening effect. Dust grains inside the outflow cavity are heated to higher temperatures ($T_{\rm d} \sim 60$ K) than grains inside the envelope ($T_{\rm d} \sim 20$ K) because of lower gas density and higher radiation field strength there. 

 \subsection{Numerical modeling the RATD activities during the transportation of grains inside jets and outflow cavities}\label{sec:omega_grain_size}
To model the variation of the grain angular velocity $\Omega(t)$ as a function of time, we solve Equation (\ref{eq:omega_t}) with a time step of $\Delta t = 0.5$ au$/v_{\rm grain}$, assuming grains initially are at rest with $\Omega(t_{0}) = 0$ rad/s. Given the smallest resolution in our datacube of 5 au, dust grains can receive the constant RAT and gas damping timescale when moving within 0.5 au, which validates the calculation of $\Omega(t)$ within $\Delta t$ time. After grains propagate in 0.5 au, we determine the new position of dust grains based on the gas velocity direction (assuming outflowing grains are well coupled with outflowing gas) and repeat the above steps, using the gas density, radiation field strength, and gas temperature in the new position to determine the new $\Omega(t)$. The process will repeat until the considered time $t$. The validity of the above assumptions will be discussed in detail in Section \ref{sec:discuss_limitation}.

We illustrate in Figure \ref{fig:wt_grain_size} the variation of the grain angular velocity $\Omega(t)$ as a function of grain sizes (black lines), from $0.05\mum$ to $500\mum$ with time \footnote{\cite{Tsukamoto_ashfall_2021}, \cite{lebreuilly_2023}, \cite{Tsukamoto_2023_coevolution} found that $500\mum$ grans can form inside the protostellar disk after the protostar forms $\sim 6\times 10^{3}$ yr}. The displacement distance and the corresponding traveling timescale of each black line are denoted in the label box. The upper row shows results for grains inside the jet base (left panel) and the outflow base (right panel). The middle row shows results for grains being at 300 au to the protostar. The left column illustrates grains moving with a few hundred km/s along the z$-$ direction, while the right column illustrates grains propagating inside the outflow cavity with a few to few tens km/s. The initial position of dust grains $[r,\Theta]$ ($r$ the distance to the protostar and $\Theta$ the angle to the vertical z-direction), the gas density $n_{\rm H_{2}}$, and gas velocity $v_{\rm gas}$ at the starting point are denoted in the lower left corner of each panel. The subplot in the upper corner of each panel shows the grain trajectory overplotted in the 2D slice of gas density distribution in the $800\times 800$ au scale box. The gradient of the trajectory from white to red shows the propagation direction of $10\mum$ grains. The vertical axis of the subplot shows the z-direction and the horizontal axis shows the x-direction. To understand how RATD disrupts dust grains, we show in each panel the variation of the disruption threshold $\Omega_{\rm disr}$ as a function of grain sizes (Equation \ref{eq:omega_disr}). The red, orange, yellow, and yellowgreen illustrate the disruption threshold of aggregate grains with $S_{\rm max} = 10^{3} - 10^{6}\erg\cm^{-3}$ and the green line corresponding to composite grains with $S_{\rm max} = 10^{7}\erg\cm^{-3}$ (Section \ref{sec:omega_t}).
 
Generally, the maximum angular velocity driven by RATs increases with increasing sizes of sub-micron grains, terminates at $\sim 2-4\mum$, and decreases with increasing sizes of micron grains, VLGs, and submillimeter grains. The rising feature of $\Omega(a_{\rm eff})$ for sub-micron grains is induced by the increased interaction cross-section $\pi a_{\rm eff}^{2}$ and the enhanced RAT efficiency between UV-optical photons with larger sub-micron grains (see Appendix \ref{sec:appen_RAT}). The RAT efficiency saturates for large grains beyond $>2\mum$, making $\Omega(a)$ decrease with increasing grain size as increasing the grain inertia moment (i.e., $I \sim a_{\rm eff}^{5}$, Equation \ref{eq:domega/dt}). Dust grains will be disrupted by RATD when $\Omega \geq \Omega_{\rm disr}$. The variation of $\Omega$ with grain sizes leads to two disruption thresholds, starting at the minimum size $a_{\rm disr,dynamic}$ and ending at the maximum size $a_{\rm disr,max,dynamic}$\footnote{The subscript $_{\rm dynamic}$ denotes the disruption size found in the dynamic approach. We use it to distinguish with $a_{\rm disr,POLARIS}$ and $a_{\rm disr,max,POLARS}$ found from POLARIS in the following Section \ref{sec:RATD_POLARIS}}. RATD does not appear if $\Omega < \Omega_{\rm disr}$ for all grain sizes. The disruption range will be broadened by decreasing the maximum tensile strength $S_{\rm max}$ and increasing the moving time $t$ (Equation \ref{eq:omega_t}). The detailed evolution of $\Omega$ with time during the transportation of $10\mum$ grains is shown in Appendix \ref{sec:appen_omega_t}.

Considering the grain propagation inside the jet and outflow, one can see that RATD less affects dust grains at $r > 300$ au before RATD happens. Besides, grains moving with higher velocity have higher possibilities of escaping from RATD owing to the quick weakening of RATs with decreasing the radiation field strength. For example, at $r = 300$ au, RATD cannot destroy micron-sized and VLGs moving with $\geq 200$ km/s, except grains have very low $S_{\rm max} = 10^{3}-10^{4}\erg\cm^{-3}$ (middle left panel). In contrast, if grains propagate with $v\sim 20$ km/s inside the outflow cavity (middle right panel), RATD can remove $10\mum$ grains with $S_{\rm max} = 10^{5}\erg\cm^{-3}$ after 6.15 yr (yellow and black dashed lines). The disruption range will extend to $a_{\rm disr,dynamic} \sim 1\mum$ and $ a_{\rm disr,max,dynamic} \sim 100\mum$ after they move in the next 125 au (or after 36.91 yr, yellow and black dashed dot lines).

In contrast, RATD only takes a few years to destroy dust grains inside the jet/outflow base due to the intense stellar radiation field there (upper row). For aggregate grains with $S_{\rm max} = 10^{5}\erg\cm^{-3}$ and $v_{\rm grain} \geq 100$ km/s (upper left panel), one gets $a_{\rm disr,dynamic} \sim 2\mum$ and $a_{\rm disr,max,dynamic} \sim 30\mum$ after grains move in 15 au within 0.28 yr. The disruption range is slightly extended to $a_{\rm disr,max,dynamic} \sim 50\mum$ after grains move in the next 75 au after 1.71 yr. RATD works stronger for grains moving with $\sim 6$ km/s inside the outflow cavity (upper right panel). One can obtain $a_{\rm disr,dynamic} \sim 1\mum$ and $a_{\rm disr,max,dynamic} \sim 30 \mum$ after grains with $S_{\rm max} = 10^{5}\erg\cm^{-3}$ propagating in 0.5 au after 0.27 yr (black solid line), and $a_{\rm disr,dynamic} \sim 0.4\mum$, $a_{\rm disr,max,dynamic} \sim 200\mum$ after they propagating 10 au after $5.46$ yr (black dashed dot line). For grains with higher $S_{\rm max} > 10^{7}\erg\cm^{-3}$ (green line), RATD cannot affect grains moving with $> 100$ km/s inside the jet base, but can remove grains inside the outflow base from $a_{\rm disr,dynamic} \sim 2\mum$ to $a_{\rm disr,max,dynamic} \sim 20\mum$ after 1.37 yr.

 \begin{figure*}
\centering
    \includegraphics[width=0.97\textwidth,height=0.97\textheight,keepaspectratio]{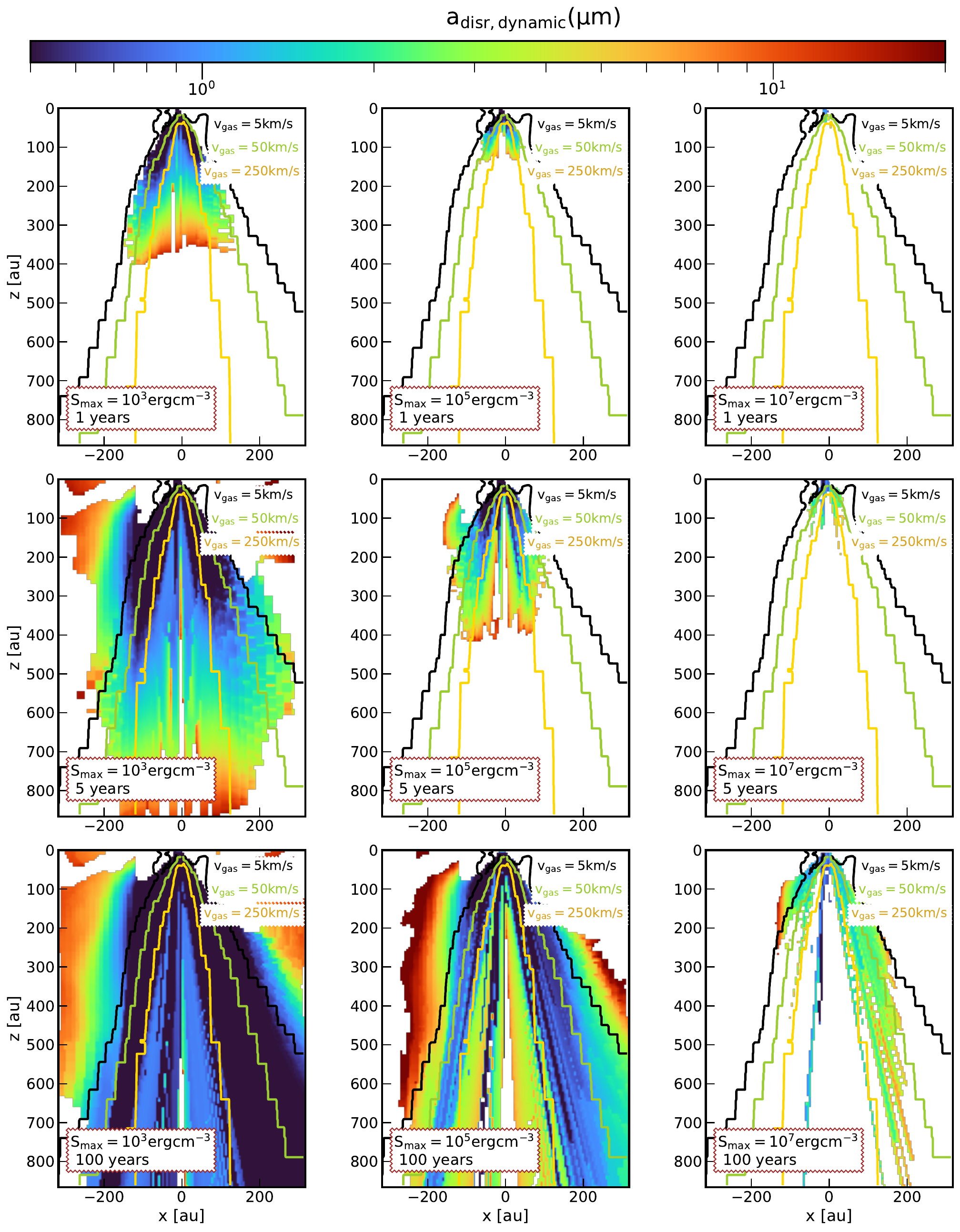}
    \caption{Spatial distribution of the minimum disruption size $a_{\rm disr,dynamic}$ over $600\times800$ au below the protostar after 1 yr (first row), 5 yr (middle row), and 100 yr (lower row). The left, middle, and right columns show results for grains with $S_{\rm max} = 10^{3}, 10^{5}, 10^{7}\erg\cm^{-3}$. The contours show the gas velocity distribution. The colorbar starts at $0.5\mum$ and ends at $20\mum$. We empty regions where grains are infalling toward the center, and regions without RATD. Small grains that survived from RATD inside the jet/outflow base will become the dominant dust population inside the jet and outflow following the spread out of material with time. The minimum disruption size is higher inside the high-velocity domain with $v_{\rm gas} > 250$ km/s and higher for grains with higher $S_{\rm max}$.   }
     \label{fig:adisr_min_dynamic}
\end{figure*}

 \begin{figure*}
\centering
    \includegraphics[width=0.97\textwidth,height=0.97\textheight,keepaspectratio]{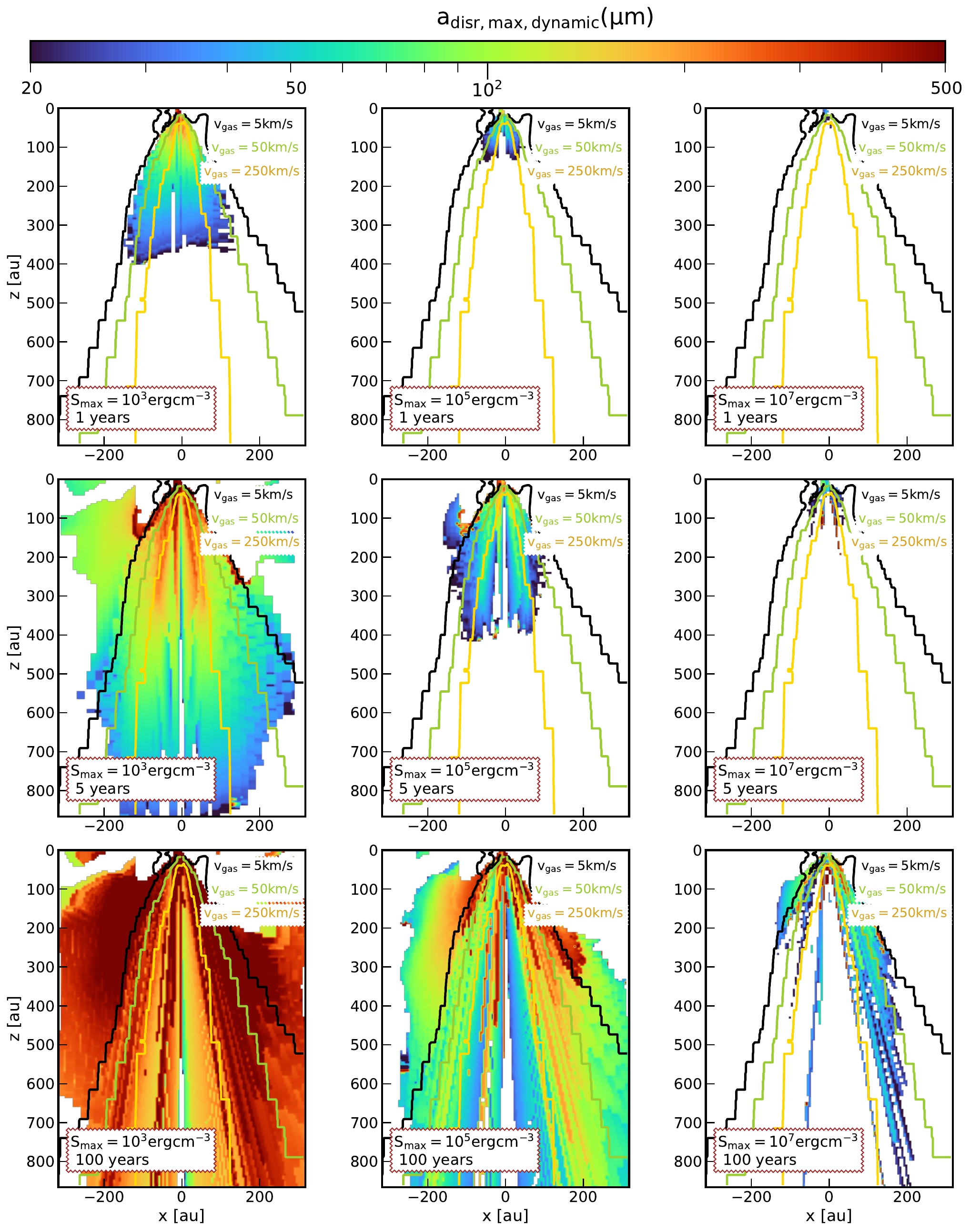}
    \caption{Similar to Figure \ref{fig:adisr_min_dynamic} but for the maximum disruption size $a_{\rm disr,max,dynamic}$. The colorbar starts from $20\mum$ and ends at $500\mum$. Similar to the evolution of $a_{\rm disr,dynamic}$, more VLGs and submillimeter grains will disappear from the jet and outflow following the propagation of dust grains with time. Particularly, VLGs and submillimeter grains up to $\sim 500\mum$ with $S_{\rm max} = 10^{3}\erg\cm^{-3}$ (left column) will totally disappear from the jet and outflow after 100 yr. In contrast, VLGs with $S_{\rm max} = 10^{7}\erg\cm^{-3}$ are only destroyed in some small areas having $v_{\rm gas} < 200$ km/s.}
     \label{fig:adisr_max_dynamic}
\end{figure*} 

We show in the lower row of Figure \ref{fig:wt_grain_size} the evolution of the grain angular velocity $\Omega(t,a)$ for grains propagating inside the outflow cavity wall (left panel) and grains infalling \footnote{We consider the well-coupling between gas and grains, or $v_{\rm grain} = v_{\rm gas}$} inside the inner envelope (right panel). In contrast to grains propagating inside the jet and outflow cavity, only VLGs having $S_{\rm max} \sim 10^{4}\erg\cm^{-3}$, or VLGs and submillimeter grains having $S_{\rm max} \sim 10^{3}\erg\cm^{-3}$ can be fragmented by RATD in such dense regions due to their low disruption threshold. The disruption timescale starts from 7 yr inside the outflow cavity wall and 22 yr inside the inner envelope for grains with $S_{\rm max} = 10^{3}\erg\cm^{-3}$. It lasts longer, reaching 22 yr and 100 yr for grains with $S_{\rm max} = 10^{4}\erg\cm^{-3}$. The inefficiency of RATD inside the outflow cavity wall and inner envelope majorly comes from the high gas density beyond $n_{\rm H_{2}} \sim 10^{7}\cm^{-3}$, which strongly attenuates the radiation field from the center and strengthens the gas randomization effect (see Appendix \ref{sec:appen_RAT}).

 \subsection{Evolution of the active region of RATD inside the jet and outflow}\label{sec:adisr_dynamic}
Given the different disruption timescale of grains propagating inside the jet and outflow (Figure \ref{fig:wt_grain_size}), we show in this section the 2D map of the disruption size range in the time series. We consider the region of $x \times z = 600 \times 800$ au below the protostar. The jet/outflow base is located at 20 au below the sink particle. The initial disruption size range inside the jet and outflow is $a_{\rm disr} = a_{\rm dirs,max} = 500\mum$ (no disruption activities). We start the calculation for grains at $z = 600$ au, using the same method described in Section \ref{sec:omega_grain_size} to get the disruption size range at each moving time step $\Delta t$. The disruption state inside the jet and outflow is updated continuously following the grain trajectory. We repeat the calculations for grains at lower $z < 600$ au until grains inside the jet and outflow base complete their movement within the considered $t$ time. 
 
We show the time-series of the minimum disruption size distribution $a_{\rm disr,dynamic}$ in Figure \ref{fig:adisr_min_dynamic}. The upper row shows the disruption picture after RATD happens 1 yr, the middle row shows results after 5 yr, and the lower row corresponds to the disruption picture after 100 yr. The left column illustrates the disruption size map for grains having $S_{\rm max} = 10^{3}\erg\cm^{-3}$, the middle and right columns illustrate for grains having $S_{\rm max} = 10^{5}\erg\cm^{-3}$ and $10^{7}\erg\cm^{-3}$, respectively. We overplot the gas velocity contour with $v_{\rm gas} = 5$ km/s, $50$ km/s, and $250$ km/s in black, green, and gold in each panel. Regions where grains are infalling toward the center, and regions where no disruption happens are replaced by empty cells. The colorbar starts from $0.05\mum$ and ends at $20\mum$. Cells exposed to stronger RATD will have smaller $a_{\rm disr,dynamic}$.
 
After 1 yr since RATD begins (upper row), only grains with $S_{\rm max} = 10^{3}\erg\cm^{-3}$ within $z < 400$ au (left panel) and grains with $S_{\rm max} = 10^{5}\erg\cm^{-3}$ within $z < 100$ au (middle panel) are destroyed by RATD due to their low disruption threshold. Dust grains are destroyed strongest inside the jet and outflow base, with $a_{\rm disr,dynamic} \sim 0.5\mum$ at $z < 100$ au, and weaker at larger $z$, with $a_{\rm disr,dynamic} \sim 10\mum$ at $z \sim 400$ au for grains with $S_{\rm max} = 10^{3}\erg\cm^{-3}$. After 5-100 yr (middle and lower rows), the active region of RATD expands from $z \sim 400$ au (for grains with $S_{\rm max} = 10^{3}\erg\cm^{-3}$) or $z \sim 100$ au (for grains with $S_{\rm max} = 10^{5}\erg\cm^{-3}$) to $z > 800$ au. The expansion of RATD region is caused by the spread out of small grains below $a_{\rm disr,dynamic}$ inside the jet and outflow, and also by the disruption of large grains at high $z$ after few tens yr (Figure \ref{fig:wt_grain_size}, middle right panel). In particular, after RATD happens $\sim 100$ yr, one gets $a_{\rm disr,dynamic} \sim 0.05-0.1\mum$ at $z = 800$ au for grains with $S_{\rm max} = 10^{3}\erg\cm^{-3}$, and $a_{\rm disr,dynamic} \sim 1-4\mum$ for grains with $S_{\rm max} = 10^{5}\erg\cm^{-3}$.

Opposite to the strong disruption picture of grains with $S_{\rm max} = 10^{3}-10^{5}\erg\cm^{-3}$, RATD takes $\sim 5$ yr to start destroying grains with $S_{\rm max} = 10^{7}\erg\cm^{-3}$ in the jet/outflow base (middle right panel). After 100 yr, small grains surviving from RATD propagate to $z \sim 500$ au (lower right panel). But opposite to the full dominance of small grains inside the jet and outflow as grains with $S_{\rm max} \leq 10^{5}\erg\cm^{-3}$, small grains with $S_{\rm max} = 10^{7}\erg\cm^{-3}$ only become the major dust population in some parts where $v_{\rm gas} < 250$ km/s. The inefficient RATD for composite grains majorly comes from their high disruption threshold, which is further amplified by the transportation of dust grains inside the jet and outflow (Figure \ref{fig:wt_grain_size}).

\begin{figure*}
\centering
    \includegraphics[width=\textwidth,height=\textheight,keepaspectratio]{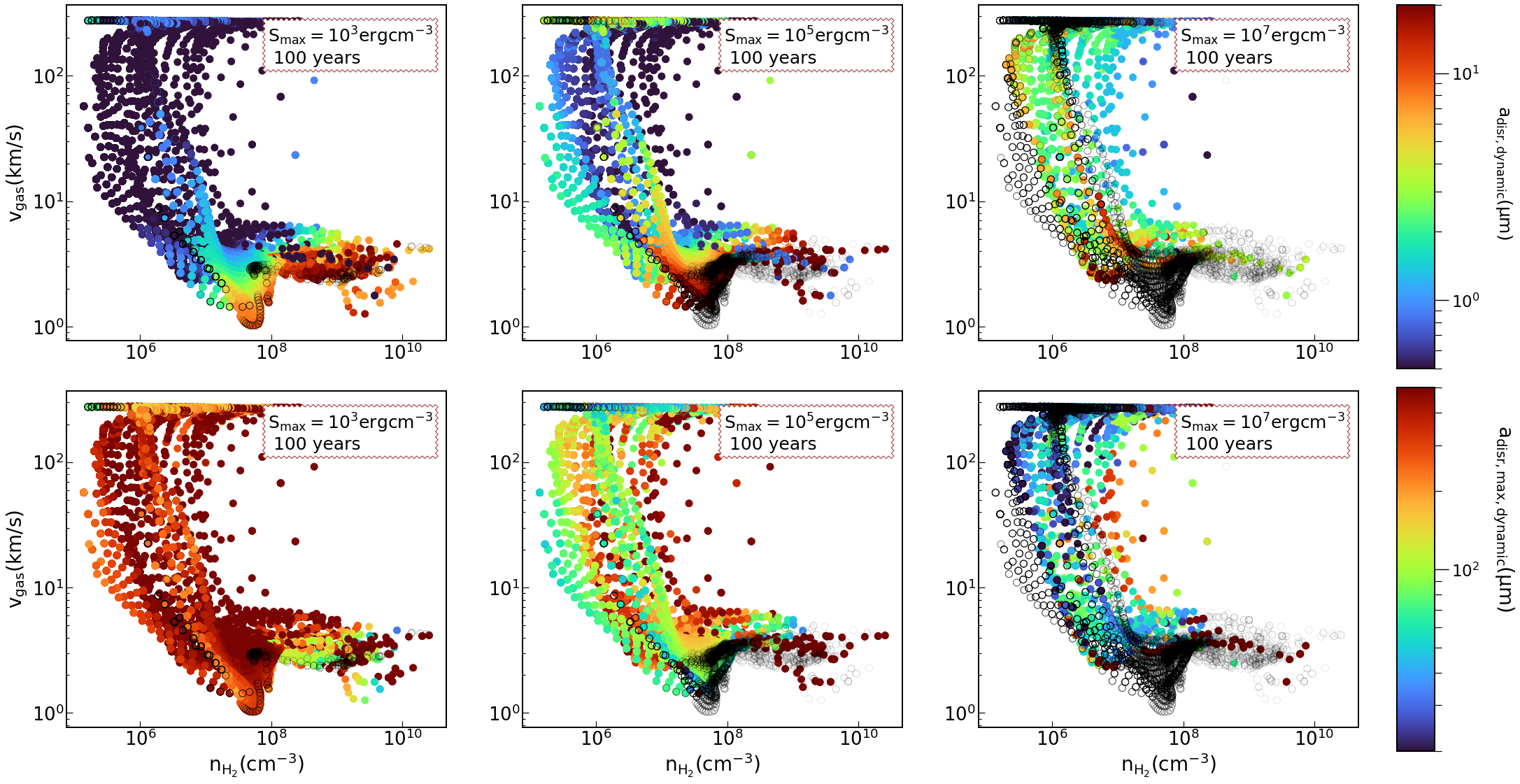}
    \caption{Variation of the minimum disruption size $a_{\rm disr,dynamic}$ (upper row) and maximum disruption size $a_{\rm disr,max,dynamic}$ (lower row) within $600\times 800$ au inside the outflow after 100 yr as a function of gas volume density $n_{\rm H_{2}}$ and gas velocity $v_{\rm gas}$. Color points show the lower and upper threshold of the disruption size range, while empty black circles illustrate regions unaffected by RATD. We show results for grains with $S_{\rm max} = 10^{3}, 10^{5}, 10^{7}\erg\cm^{-3}$ from left to right, respectively. The strongest disruption activities happens inside the outflow cavity where $n_{\rm H_{2}} \sim 10^{5} - 10^{7}\cm^{-3}$, $v_{\rm gas} \sim 3-30$ km/s, and inside the low-velocity jet where $v_{\rm gas} \sim 30-200$ km/s. The disruption size range decreases toward the high-velocity jet where $v_{\rm gas} > 200$ km/s and to the outflow cavity wall where $n_{\rm H_{2}} \geq 10^{7}\cm^{-3}$, $v_{\rm gas} \sim 1-3$ km/s. The maximum grain size constrained by RATD inside the jet and outflow is $< 0.5\mum$ for grains with $S_{\rm max} = 10^{3}\erg\cm^{-3}$. In contrast, almost grains with $S_{\rm max} = 10^{7}\erg\cm^{-3}$ are safe from RATD inside the jet and outflow, regardless of their sizes. }
    \label{fig:adisr_nH_vgas}
\end{figure*} 

We show in Figure \ref{fig:adisr_max_dynamic} the evolution of the maximum disruption size distribution $a_{\rm disr,max,dynamic}$ inside the jet and outflow from 1 yr to 100 yr, for grains with $S_{\rm max} = 10^{3}, 10^{5}, 10^{7}\erg\cm^{-3}$. The colorbar starts at $20\mum$ and ends at $500\mum$. Similar to the evolution of $a_{\rm disr,dynamic}$ shown in Figure \ref{fig:adisr_min_dynamic}, more VLGs and submillimeter grains will disappear from the jet and outflow by RATD with time. And the destruction of large grains happens stronger for grains with lower $S_{\rm max}$. In particular, for grains with $S_{\rm max} = 10^{3}\erg\cm^{-3}$ (left column), RATD can remove large grains within $\sim 30-150\mum$ from $z < 400$ au after 1 yr, and it will remove all VLGs and submillimeter grains up to $\sim 500\mum$ within $z \sim 800$ au after 100 yr. For grains with $S_{\rm max} = 10^{5}\erg\cm^{-3}$ (middle column), VLGs up to $\sim 50-100\mum$ can be destroyed from the $600\times 800$ au box scale after RATD happens 100 yr (lower middle panel). In contrast, only VLGs of $\sim 30\mum$ with $S_{\rm max} = 10^{7}\erg\cm^{-3}$ can disappear from some small areas inside the low-velocity region where $v_{\rm gas} < 250$ km/s (lower right panel). The large remaining part of the jet and outflow is still filled with micron-sized grains, VLGs, and submillimeter grains unaffected by RATD.

  \begin{figure*}
\centering
    \includegraphics[width=0.96\textwidth,height=0.96\textheight,keepaspectratio]{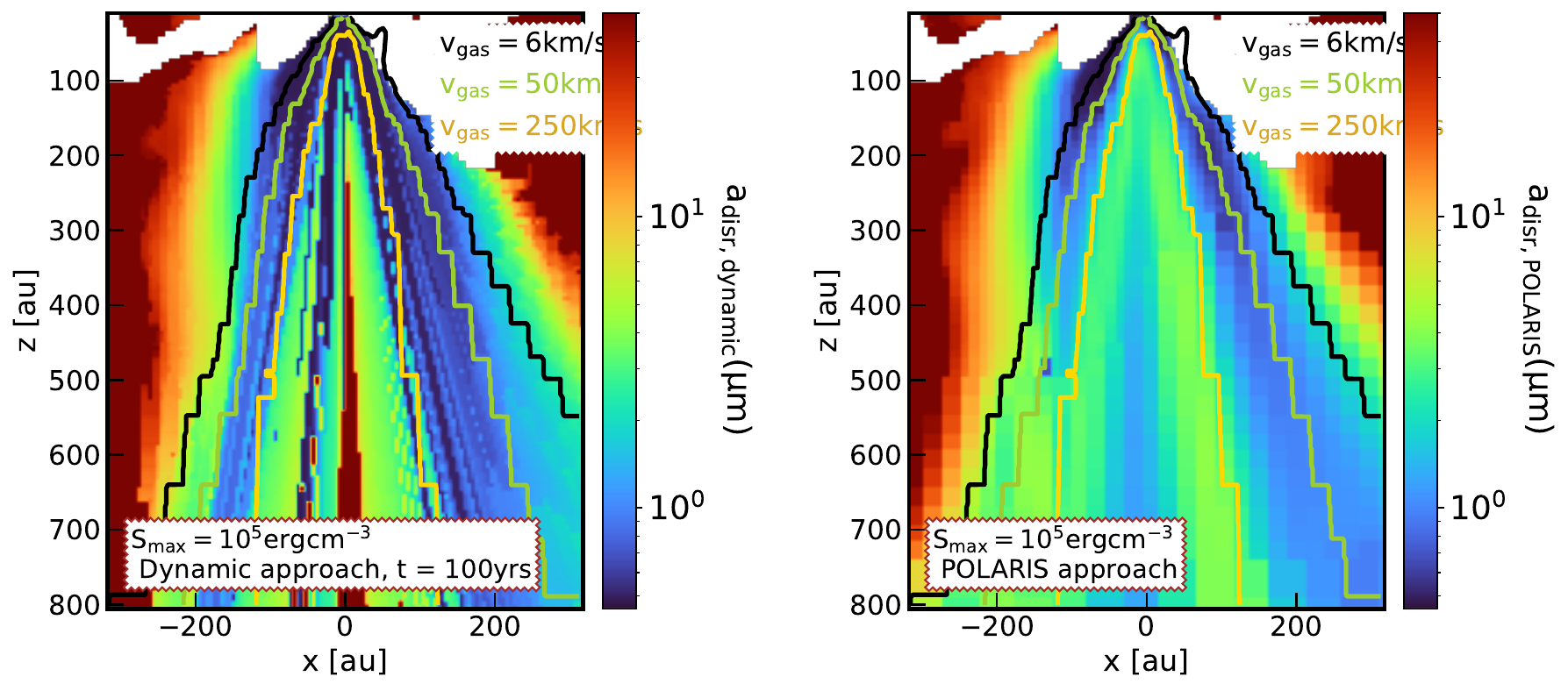}
    \includegraphics[width=0.97\textwidth,height=0.97\textheight,keepaspectratio]{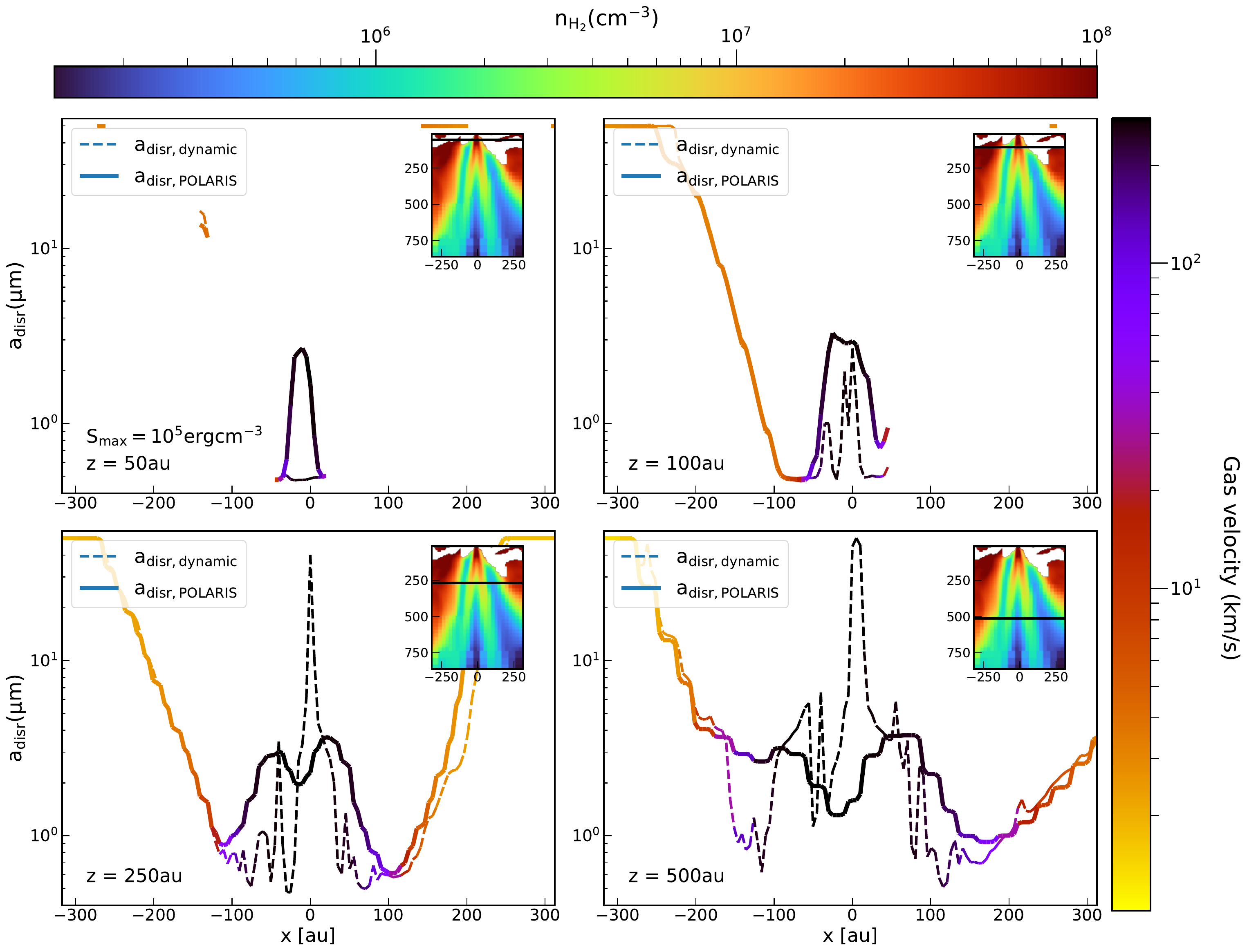}
    \caption{Upper row: comparison of the minimum disruption size distribution obtained from the dynamic approach after 100 yr $a_{\rm disr,dynamic}$ (left panel) and POLARIS approach $a_{\rm disr,POLARIS}$ (right panel), considering aggregate$-$type grains with $S_{\rm max} = 10^{5}\erg\cm^{-3}$. The map shows the $600\times 800$ au below the protostar. Middle and lower rows: Variation of $a_{\rm disr,dynamic}$ (dashed lines) and $a_{\rm disr,POLARIS}$ (solid lines) along the horizontal cut shown in the subpanel displaying the 2D slice of gas volume density $n_{\rm H_{2}}$. The solid and dashed lines are color-coded based on the gas velocity distribution placed on the right of the figure, while the colorbar of the density distribution is placed on top. We choose different cuts at $z = 50, 100, 250, 500$ au, and show results following the clockwise direction. Generally, POLARIS approach induces the similar disruption size map as found after 100 yr from the dynamic approach due to the low disruption threshold of dust grains. However, since POLARIS cannot take the transportation of small grains surviving from RATD and the escape of fast-moving grains into account, it may overestimate the minimum disruption size in some parts of the outflow, especially in the region where grains propagating with $v_{\rm gas} > 200$ km/s.}
  \label{fig:compare_Smax1e5}
     \end{figure*}

\subsection{Grain size distribution constrained by RATD}\label{sec:adisr_nH_vgas}
To summary the disruption range inside the jet and outflow, we show in Figure \ref{fig:adisr_nH_vgas} the variation of $a_{\rm disr,dynamic}$ (upper row) and $a_{\rm disr,max,dynamic}$ (lower row) shown in Figures \ref{fig:adisr_min_dynamic} and \ref{fig:adisr_max_dynamic} after 100 yr as a function of $n_{\rm H_{2}}$ and $v_{\rm gas}$. As discussed in Section \ref{sec:MHD+radiation_field}, the region where gas propagates with $v_{\rm gas} \sim 3-30$ km/s can be considered as the outflow cavity, and the region where gas propagates with $v_{\rm gas} \sim 30-300$ km/s can be considered as the jet. The outflow cavity wall has higher $n_{\rm H_{2}} > 10^{7}\cm^{-3}$ and lower $v_{\rm gas} \sim 1-4$ km/s, with the high $n_{\rm H_{2}} \sim 10^{8}-10^{10}\cm^{-3}$ indicating the cavity wall part near the equatorial midplane. Color circles show the disruption range inside the active region of RATD, while the empty black circles illustrate regions without being affected by RATD (empty cells inside Figures \ref{fig:adisr_min_dynamic} and \ref{fig:adisr_max_dynamic}).

As shown in Figures \ref{fig:adisr_min_dynamic} and \ref{fig:adisr_max_dynamic}, more dust grains will be destroyed by RATD, i.e., lower $a_{\rm disr,dynamic}$ and larger $a_{\rm disr,max,dynamic}$, with decreasing $S_{\rm max}$ (Figures \ref{fig:adisr_min_dynamic} and \ref{fig:adisr_max_dynamic}). But to be more detailed, the disruption size range is largest inside the outflow cavity where $n_{\rm H_{2}} < 10^{7}\cm^{-3}, v_{\rm gas} \sim 3-30$ km/s, and the low-velocity jet where $v_{\rm gas} < 200$ km/s. It is narrower in the high-velocity jet where $v_{\rm gas} > 200$ km/s and inside the outflow cavity wall where $n_{\rm H_{2}} > 10^{7}\cm^{-3}$. For grains with $S_{\rm max} = 10^{5}\erg\cm^{-3}$ (left column), the maximum grain size constrained by RATD inside the outflow cavity and low-velocity jet is $\sim 0.5\mum$. The disruption size range is narrower of $\sim 1 - 200\mum$ inside the high-velocity jet, and is beyond $> 2-5\mum$ inside the outflow cavity wall. Some VLGs near the equatorial midplane also can be removed by RATD. In contrast, for grains with $S_{\rm max} = 10^{5}\erg\cm^{-3}$ (middle column), the disruption size range reduces from $\sim 1-200\mum$ inside the outflow cavity and low-velocity jet to $\sim 3-100\mum$ inside the high-velocity jet. Almost no disruption can happen inside the outflow cavity wall. 

\section{Modelling RATD in POLARIS}\label{sec:RATD_POLARIS}
Given the distribution of the minimum and maximum disruption size by RATD in Section \ref{sec:vgrain_omegat}, theoretically, we can use them as the input dust model for POLARIS to model the effect of RATD on polarized dust emission. However, this work meets difficulty in converting the distribution of $a_{\rm disr,dynamic}$ and $a_{\rm disr,max,dynamic}$ to the POLARIS format. We also cannot adopt the dynamic approach (Section \ref{sec:vgrain_omegat}) in POLARIS because of the lack of tools to solve the evolution of the equation of grain motion. Therefore, to model the effect of RATD on dust population and dust emission polarization using POLARIS, we assume grains to be at rest with $v_{\rm grain} = 0$ km/s to solve the equation of grain motion. Their disruption picture is thus independent of grain velocity as in the dynamic approach (Section \ref{sec:vgrain_omegat}), and is controlled only by the distribution of the gas density, radiation field strength, and dust temperature shown in Figure \ref{fig:distribution_mcrt}. From the dynamic view, this assumption may work for grains propagating with a few km/s inside the outflow cavity (Figure \ref{fig:wt_grain_size}, right column), in which RATD can destroy grains before they migrate to larger distances where RATD is deactivated. However, given the complex distribution of gas velocity inside the jet and outflow (Figure \ref{fig:distribution_mcrt}, upper right panel), we need to check the validation of the POLARIS approach before using them to study the RATD impact on dust polarization. We first describe our incorporation of RATD in POLARIS in Section \ref{sec:polaris_adisr}. The comparison of RATD picture obtained from the POLARIS and dynamic approach is shown in Section \ref{sec:adisr_polaris_dynamic}. Finally, we will describe the calculation for the new size distribution of small grains enhanced by RATD in Section \ref{sec:polaris_eta}.

\subsection{New incorporation in POLARIS}\label{sec:polaris_adisr}
When grains are at rest, they will receive the constant RATs and be spun up continuously until reaching the maximum angular velocity $\Omega_{\rm RAT}$. By replacing $t = \inf$, Equation (\ref{eq:omega_t}) becomes:
\bea 
\Omega_{\rm RAT} = \frac{\Gamma_{\rm RAT} \tau_{\rm damp}}{I}.
\label{eq:omega_rat}
\ena

Giving the maximum tensile strength $S_{\rm max}$, we can determine the disruption angular velocity $\Omega_{\rm disr}$ for every grain size  (Equation \ref{eq:omega_disr}). The minimum and maximum disruption size found by the POLARIS approach, $a_{\rm disr,POLARIS}$ and $a_{\rm disr,max,POLARIS}$, inside each cell is determined by comparing $\Omega_{\rm RAT}$ with $\Omega_{\rm disr}$ over the grain size distribution. However, before further being spun up and disrupted by RATD, dust grains must be able to maintain their orientation in space, i.e., they must be magnetically aligned. Therefore, we need to compare the disruption range $a_{\rm disr,POLARIS}-a_{\rm disr,max,POLARIS}$ with the alignment range $a_{\rm align} - a_{\rm max,JB}^{\rm Lar}$ \footnote{$a_{\rm align}$ is the minimum alignment size determined by suprathermal condition by RATs, and $a_{\rm max,JB}^{\rm Lar}$ is the maximum alignment size determined by Larmor precession condition, see our \cite{Giang_et_al_2024}} to determine the realistic range size that RATD can destroy. In particular, no dust grains can be affected by RATD if $a_{\rm disr,POLARIS} > a_{\rm max,JB}^{\rm Lar}$. If $a_{\rm align} < a_{\rm disr,POLARIS} < a_{\rm max,JB}^{\rm Lar}$, the maximum disruption size will be assigned to $a_{\rm max,JB}^{\rm Lar}$ if $a_{\rm disr,max,POLARIS} > a_{\rm max,JB}^{\rm Lar}$. We keep $a_{\rm disr,max,POLARIS}$ if $a_{\rm disr,max,POLARIS} < a_{\rm max,JB}^{\rm Lar}$.

\subsection{Disruption size obtained from the POLARIS and dynamic approach}\label{sec:adisr_polaris_dynamic}
To understand how the POLARIS approach changes the disruption picture inside the protostellar jet and outflow, we compare their disruption size map with results using the dynamic approach after 100 yr (lower row of Figure \ref{fig:adisr_min_dynamic}). As we consider the single maximum grain size of $a_{\rm max} = 50\mum$ inside the protostellar core to simulate the RT in POLARIS, the minimum and maximum disruption size from the dynamic approach will be adjusted to $50\mum$ if $a_{\rm disr,dynamic} > 50\mum$ and $a_{\rm disr,max,dynamic} > 50\mum$. We compare results obtained for aggregate$-$type grains in Section \ref{sec:compare_aggregate_grains} and for composite grains in Section \ref{sec:compare_composite_grains}, respectively.

\subsubsection{For aggregate$-$type grains}\label{sec:compare_aggregate_grains}
We show in the upper row of Figure \ref{fig:compare_Smax1e5} the spatial distribution of $a_{\rm disr,dynamic}$ after 100 yr (left panel) and $a_{\rm disr,POLARIS}$ (right panel) in the  $600\times 800$ au scale box below the protostar, considering aggregate$-$type grains with $S_{\rm max} = 10^{5}\erg\cm^{-3}$. The black, green, and gold contours indicate the gas velocity distribution with $v_{\rm gas} = 5$ km/s, $50$ km/s, and $250$ km/s, respectively. Empty cells at the top of each panel indicate regions where gas and grains are infalling toward the center. For aggregate$-$type grains, the disruption picture obtained from the POLARIS and dynamic approach is quite similar from the outflow to the low- and high-velocity jet regions. The similar disruption picture obtained in the two approaches comes from the low dust disruption threshold, which allows RATD to destroy aggregate grains before they escape to the inefficient RATD region. However, as the dynamic approach can account for both the migration of small surviving grains and the escape of fast-moving grains, it induces much more complicated distribution of $a_{\rm disr,dynamic}$ inside the jet and outflow than the POLARIS approach.
 
To look into the difference in detail, we provide in the middle and lower rows of Figure \ref{fig:compare_Smax1e5} the horizontal-cut of $a_{\rm disr,POLARIS}$ (solid lines) and $a_{\rm disr,dynamic}$ (dashed lines) at different height $z = 50,100,250,500$ au following the clockwise direction. The position of the cut is marked by the black line in the 2D slice of the gas density distribution placed in the upper right corner of each panel. The density map focuses on the same $600\times 800$ au region shown in the upper row, with the color scale positioned at the top of the figure. The solid and dashed lines are color-coded by the gas velocity distribution placed on the right of the figure. We note that the gap in the variation of $a_{\rm disr,dynamic}$ and $a_{\rm disr,POLARIS}$ along the horizontal direction comes from the empty region in the upper row, where grains are infalling toward the central region.

   \begin{figure*}
\centering
    \includegraphics[width=0.96\textwidth,height=0.96\textheight,keepaspectratio]{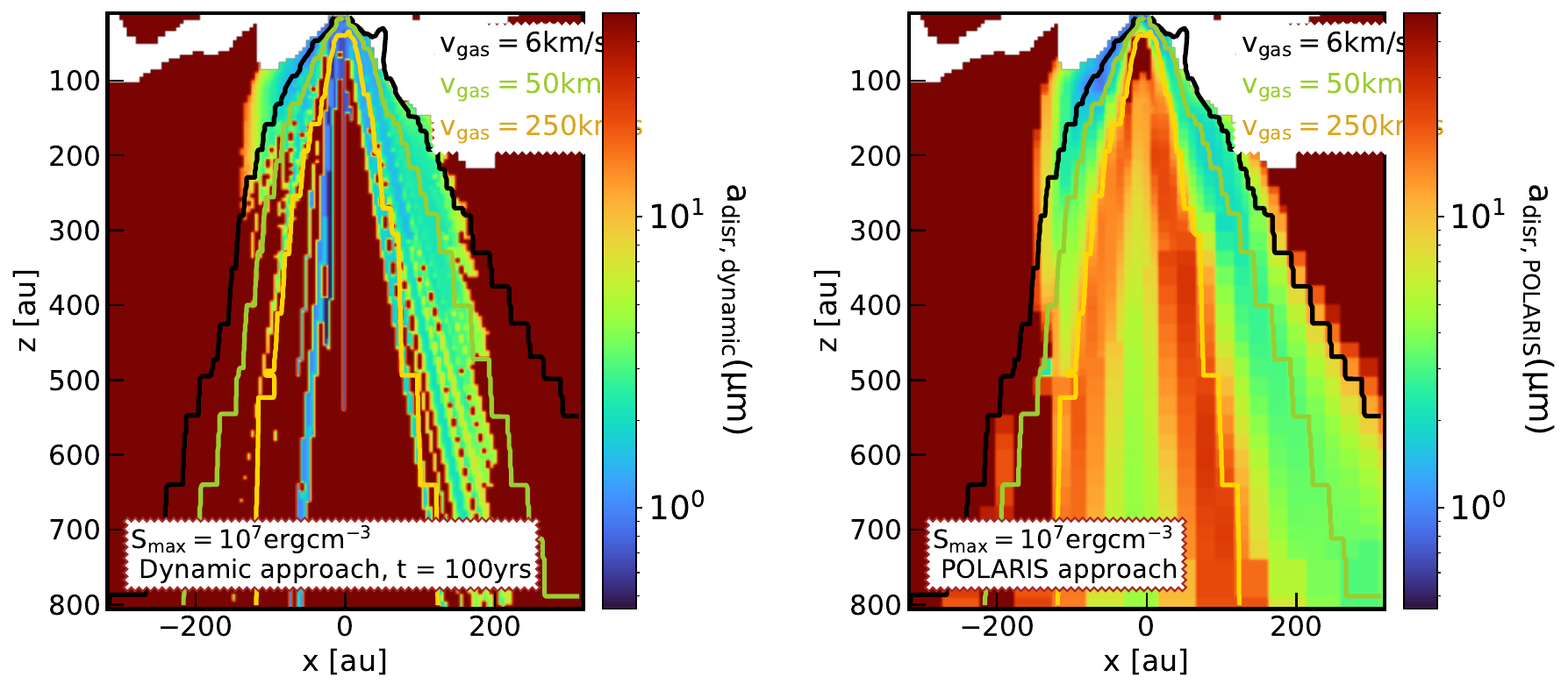}
    \includegraphics[width=0.97\textwidth,height=0.97\textheight,keepaspectratio]{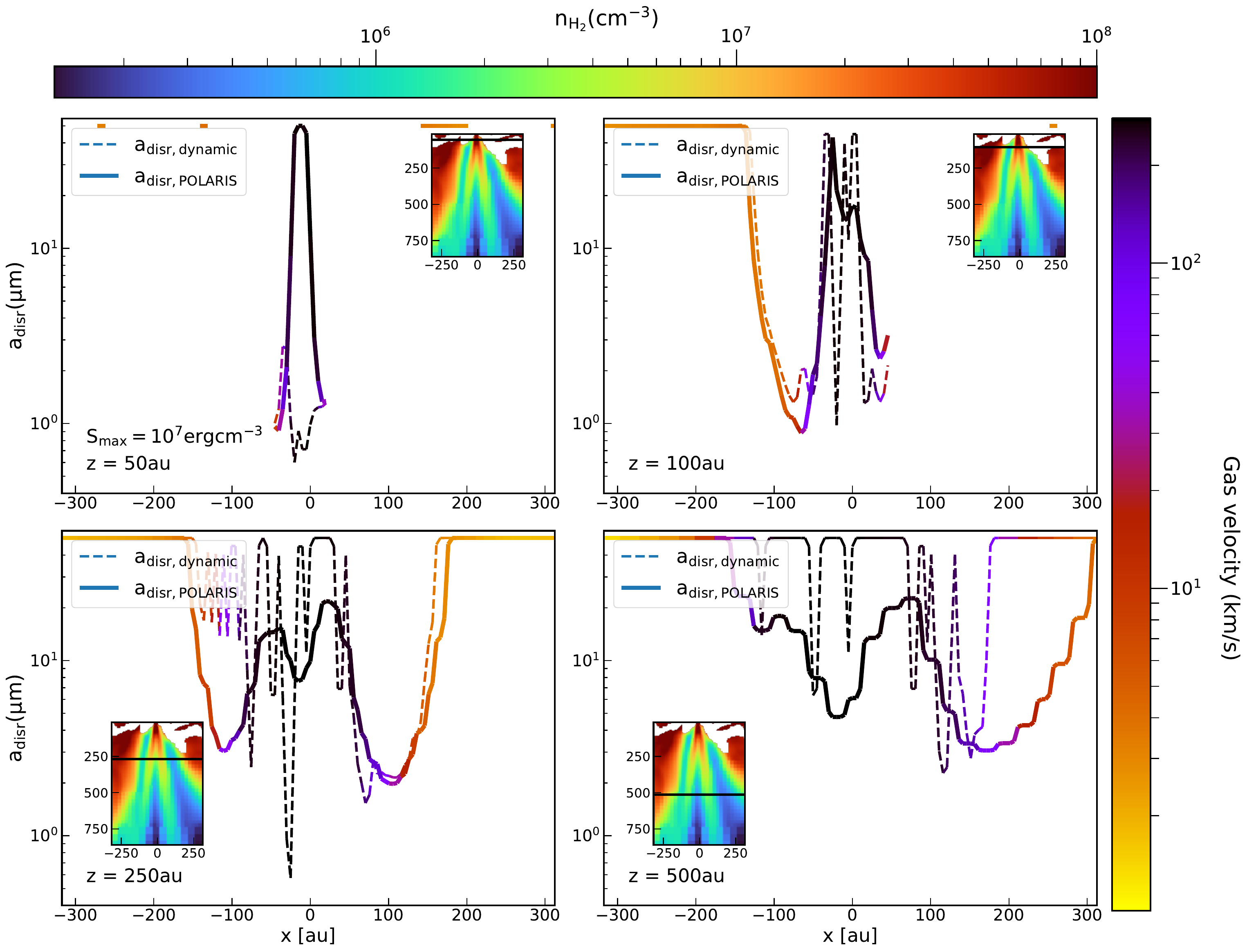}
    \caption{Similar results to Figure \ref{fig:compare_Smax1e5} but for composite grains with $S_{\rm max} = 10^{7}\erg\cm^{-3}$. In contrast to aggregate$-$type grains, the high disruption threshold of composite grains makes the POLARIS approach (assuming grains are at rest) nearly overestimate the realistic disruption picture for grains with high maximum tensile strength. The POLARIS approach predicts the minimum disruption size of $a_{\rm disr,POLARIS} \sim 4\mum$ inside the outflow cavity and low-velocity jet, and higher $\sim 20\mum$ in high-velocity jet region. In reality, only some small parts inside the jet and outflow are affected by RATD, with $a_{\rm disr,dynamic} \sim 3\mum$ there.}
     \label{fig:compare_Smax1e7}
\end{figure*}

At each height $z$, $a_{\rm disr,POLARIS} \approx a_{\rm disr,dynamic}$ inside the outflow where $v_{\rm gas} < 20$ km/s. Inside the jet where $v_{\rm gas} \sim 20-300$ km/s, while POLARIS provides $a_{\rm disr,POLARIS} \sim 3\mum$ at $z = 50$ au (middle left panel), the dynamic approach shows $a_{\rm disr,dynamic} \sim 0.5\mum$ owing to the occupation of surviving sub-micron grains migrating from the jet base (upper row of Figure \ref{fig:compare_Smax1e5}). At $z \sim 100$ au (middle right panel), one gets $a_{\rm disr, POLARIS} \sim 3\mum$, while $a_{\rm disr,dynamic}$ varies among $\sim 0.5\mum$, $1\mum$, $2\mum$, and $3\mum$ owing to different migration histories of dust grains from the inner region. At $z = 250 - 500$ au (lower row), while POLARIS provides a smooth distribution of $a_{\rm disr,POLARIS} \sim 2 - 3\mum$ across the x$-$direction, the dynamic approach shows more complicated distribution, including both sub-micron grains of $0.5\mum$ and micron-sized grains of $2\mum$ migrated from the inner region, and large micron-sized grains of $\sim 7\mum$ and $50\mum$ escaped from RATD (Figure \ref{fig:wt_grain_size}, lower row, left panel).

\begin{figure*}
\centering
    \includegraphics[width=\textwidth,height=\textheight,keepaspectratio]{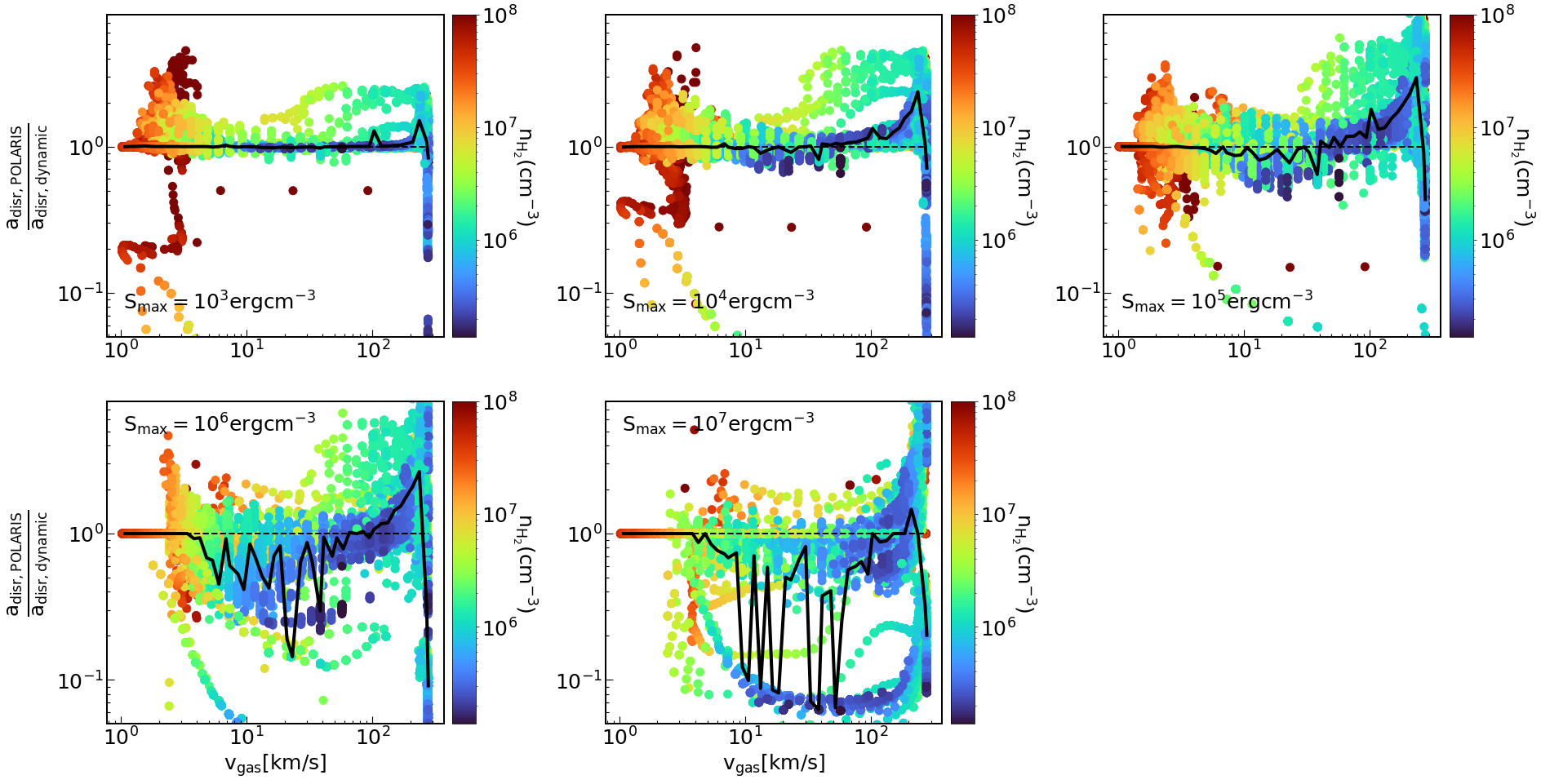}
    \caption{Variation of the ratio $a_{\rm disr,POLARIS}/a_{\rm disr,dynamic}$ as the function of gas density $n_{\rm H_{2}}$ and gas velocity $v_{\rm gas}$ (colorcode), for grains having $S_{\rm max} = 10^{3}-10^{7}\erg\cm^{-3}$. The POLARIS approach can reproduce well the minimum disruption size found from the dynamic approach for grains with $S_{\rm max} \leq 10^{5}\erg\cm^{-3}$ propagating with $v_{\rm gas} < 60$ km/s, but overestimates the disruption picture for grains having $S_{\rm max} \geq 10^{6}\erg\cm^{-3}$ in this velocity domain. For grains propagating with $v_{\rm gas} > 60$ km/s, POLARIS underestimates the RATD strength on grains having $S_{\rm max} \leq 10^{5}\erg\cm^{-3}$ ($a_{\rm disr,POLARIS} > a_{\rm disr,dynamic}$) as they cannot simulate the spread out of small grains surviving from RATD inside the jet and outflow. In contrast, it overestimates the disruption state for grains having $S_{\rm max} \geq 10^{6}\erg\cm^{-3}$ because they cannot account for the suppression of RATD by the fast-moving composite grains. The median difference in the disruption size from the two approaches is roughly twice for grains having $S_{\rm max} \geq 10^{5}\erg\cm^{-3}$.}
     \label{fig:ratio}
\end{figure*}

\subsubsection{For composite grains}\label{sec:compare_composite_grains}
We then next examine the disruption of composite grains with $S_{\rm max} = 10^{7}\erg\cm^{-3}$. The spatial distribution of the disruption size is shown in the upper row and the comparison between $a_{\rm disr,dynamic}$ and $a_{\rm disr,POLARIS}$ at different heights are shown in the middle and lower rows of Figure \ref{fig:compare_Smax1e7}. Difference from the similarity of $a_{\rm disr,POLARIS}$ and $a_{\rm disr,dynamic}$ obtained for aggregate grains in Figure \ref{fig:compare_Smax1e5}, the dynamic approach induces much weaker disruption picture than the POLARIS approach because composite grains can quickly escape from RATD when propagating inside the jet and outflow (upper row). Particularly, while almost dust grains beyond $a_{\rm disr,POLARIS} \geq 4-10\mum$ inside the jet and outflow can be destroyed by RATD in the POLARIS approach, the dynamic approach shows dust destruction to happen only inside some areas of jet and outflow where $v_{\rm gas} < 250$ km/s, with $a_{\rm disr,dynamic} \sim 4\mum$. Besides, while RATD in the POLARIS approach can affect dust grains in the entire $600\times 800$ au of the jet and outflow, RATD in the dynamic approach only can affect grains up to $z \sim 500$ au after 100 yr. The large area of the jet and outflow is still filled with micron-sized and VLGs, i.e., $a_{\rm disr,dynamic} \sim 50\mum$ (lower row).

\subsubsection{On the difference between the disruption range using POLARIS and dynamic approach.}
We show in Figure \ref{fig:ratio} the variation of the ratio $a_{\rm disr,POLARIS}/a_{\rm disr,dynamic}$ as the function of $v_{\rm gas}$ obtained inside $600\times800$ au below the protostar. We show results for grains having $S_{\rm max} = 10^{3}, 10^{4}, 10^{5}, 10^{6}, 10^{7}\erg\cm^{-3}$ following the clockwise direction. Data is color-coded by the gas volume density $n_{\rm H_{2}}$. The black lines show the median variation of $a_{\rm disr,POLARIS}/a_{\rm disr,dynamic}$ with $v_{\rm gas}$. 
 
One can clearly see that for grains with $S_{\rm max} \leq 10^{5}\erg\cm^{-3}$ (upper row), $a_{\rm disr,POLARIS}/a_{\rm disr,dynamic} \sim 1$ inside the outflow and low-velocity jet where $v_{\rm gas} < 60$ km/s. For grains propagating with $v_{\rm gas} > 60$ km/s, the POLARIS approach slightly underestimates the realistic minimum disruption size, i.e., larger $a_{\rm disr,POLARIS}$, because they cannot account for the migration of small micron-sized grains surviving from RATD inside the jet and outflow (Figure \ref{fig:compare_Smax1e5}). The difference reaches $\sim 1.5$ times for grains having $S_{\rm max} = 10^{4}\erg\cm^{-3}$, and $\sim 2$ times for grains having $S_{\rm max} = 10^{5}\erg\cm^{-3}$. 

For grains with $S_{\rm max}$ to $\geq 10^{6}\erg\cm^{-3}$, $a_{\rm disr,POLARIS}/a_{\rm disr,dynamic}$ decreases to below 1 inside the outflow cavity and low-velocity jet where $v_{\rm gas} < 60$ km/s (lower row). The overestimation of RATD from POLARIS, i.e., lower $a_{\rm disr,POLARIS}$, in these regions comes from the jet and outflow part beyond 300 au, i.e., the boundary of RATD obtained from the dynamic approach after 100 yr (Figure \ref{fig:compare_Smax1e7}). Theoretically, $a_{\rm disr,POLARIS}/a_{\rm dynamic}$ will approach $\sim 1$ for the time being. However, considering RATD to happen within a few to a few hundred yr (the typical timescale of accretion burst event when RATD is predicted to operate, \citealt{Lee_2007}, \citealt{Jhan_2016}, \citealt{Contreras_2024}), grains propagating with $v_{\rm gas} < 60$ km/s may not have enough time to fully distribute over the jet and outflow cavity of hundred au across, giving smaller ratio $a_{\rm disr,POLARIS}/a_{\rm disr,dynamic}$ during RATD lifetime. For grains propagating inside the jet with $v_{\rm gas} > 60$ km/s, $a_{\rm disr,POLARIS}/a_{\rm disr,dynamic} > 1$ for grains with $S_{\rm max} \sim 10^{6}\erg\cm^{-3}$ (similar as grains with $S_{\rm max} \leq 10^{4}\erg\cm^{-3}$), but either larger or smaller than 1 for grains with $S_{\rm max} = 10^{5}\erg\cm^{-3}$. Such complex behavior of $a_{\rm disr,POLARIS}/a_{\rm disr,dynamic}$ is caused by different RATD strength acting on moving grains. The difference between the two approaches is around twice for grains having $S_{\rm max} \geq 10^{6}\erg\cm^{-3}$. Given such a small difference for grains experiencing weak RATD efficiency, it is acceptable to continue using POLARIS results to understand how RATD modifies the grain size distribution and properties of polarized dust emission.

\subsection{Fraction of grains destroyed by RATD and the new grain size distribution}\label{sec:polaris_eta}
As RATD converts large grains to smaller sizes, we expect to obtain more small grains below $a_{\rm disr,POLARIS}$ inside the active region of RATD. We call $f_{\rm disr}$ as the fraction of grains within $a_{\rm disr,POLARIS}-a_{\rm disr,max,POLARIS}$ that RATD destroys. As grains aligning with $\B$ at high-\textit{J} can have stable orientation in space, they are easier to be spun up by RATs and destroyed by RATD than grains aligning with $\B$ at low-\textit{J} attractors with thermal rotation. Therefore, $f_{\rm disr}$ can be described by $f_{\rm high-J}$, the fraction of grains having magnetic alignment by RATs/MRAT mechanism at high-\textit{J} attractors (see \citealt{Giang_et_al_2024} for detailed description). One expects $25\%, 50\%$, and $100\%$ of grains will be removed by RATD if this grain size has $f_{\rm high-J} = 0.25$, $0.5$, and $1$, respectively. 

We assume that $f_{\rm disr}$ of grains within $a_{\rm disr,POLARIS} - a_{\rm disr,max,POLARIS}$ will be totally destroyed to smaller sizes below $a_{\rm disr,POLARIS}$. Given the disruption range size and the disruption fraction for each disrupted grain size $f_{\rm disr}(a)$, we can calculate the new grain size distribution of small grains below $a_{\rm disr,POLARIS}$. Considering small grains enhanced by RATD to follow the MRN distribution but with the new power index $dn/da \sim a^{\alpha}$, the value of $\alpha$ can be determined by the mass conservation, given by:
\begingroup\makeatletter\def\f@size{7.5}\check@mathfonts
\bea 
\int_{\rm a_{\rm min}}^{\rm a_{\rm disr,POLARIS}} a^{-0.5} da &+& \int_{\rm a_{\rm disr,POLARIS}}^{\rm a_{\rm disr,max,POLARIS}} a^{-0.5} f_{\rm disr}(a) da \\ \nonumber
&=& \int_{a_{\rm min}}^{\rm a_{\rm disr,POLARIS}} a^{3+\alpha} da. \label{eq:eta}
\ena
\endgroup
The first term of the left-hand side describes the total mass of all dust grains below $a_{\rm disr,POLARIS}$ which RATD does not destroy, while the second term describes the mass of destroyed grains within $a_{\rm disr,POLARIS} - a_{\rm disr,max,POLARIS}$. The right-hand side describes the total mass of small grains below $a_{\rm disr,POLARIS}$ after RATD happens. Grains above $ a_{\rm disr,max,POLARIS}$ are not affected by RATD, so they do not contribute to the calculation of $\alpha$. The spatial distribution of the new size distribution $\alpha$ under RATD effect is shown in Figure \ref{fig:eta_midplane}. Generally, more small grains will be enhanced by RATD, i.e., larger negative $\alpha$, with decreasing $S_{\rm max}$ (which induces larger disruption range) and increasing the grain magnetic susceptibility (which induces larger $f_{\rm disr}$). For example, sub-micron grains with $N_{\rm cl} = 10^{3}$ and $S_{\rm max} = 10^{5}\erg\cm^{-3}$ inside the outflow cavity can follow the steeper slope with $\alpha \sim -3.64$ owing to the strong reduction of grains beyond $\geq 0.5\mum$ by RATD there (Figure \ref{fig:adisr_nH_vgas}).

\begin{table}
\centering
\caption{Setup for the radiation field and dust model in POLARIS}
\scalebox{0.8}{
\begin{tabular}{lll}
\hline
Quantity & Symbol & Value \\   
\hline
\multicolumn{3}{c}{\textbf{ Radiation sources}}\\
\hline 
Stellar radius              & $R_{\rm star}$       & $1.23R_{\odot}$ \\ 
Effective temperature       & $T_{\rm star}$       & 16457 K  \\
Central luminosity          & $L_{\rm center}$       & $100 L_{\odot}$ \\

\hline
\multicolumn{3}{c}{\textbf{Dust model}}\\
\hline
Grain axial ratio           & $s$                    & 0.5\\
Dust-to-gas mass ratio      &                       & 0.01\\
Initial size distribution   & $\rm dn/da$              & C$ a^{-3.5}$\\
Minimum grain size          & $a_{\rm min}$        & 5\AA  \\ 
Maximum grain size          & $a_{\rm max}$        & $50\mu m$  \\
Fraction of silicate        &                      & $67.5\%$  \\
Fraction of graphite        &                      & $32.5\%$  \\

\hline
\multicolumn{3}{c}{\textbf{Grain structure}}\\
\hline
Maximum tensile strength & $S_{\rm max}$ &  $10^{5},10^{5},10^{7},10^{8}\erg\cm^{-3}$ \\

\hline
\multicolumn{3}{c}{\textbf{Grain magnetic properties (for SPM grains)}}\\
\hline
Iron atom/cluster & $N_{\rm cl}$ & $ 10^{2}, 10^{3}, 10^{4}$ \\
Volume filling factor  & $\phi_{\rm sp}$ & 0.1 \\
of iron clusters & & \\

\hline 
\multicolumn{3}{c}{\textbf{Elasticity of grains}}\\
\hline
Elasticity & $\mu Q$ & $3\times10^{9}\erg\cm^{-3}$ \\

 \hline
\multicolumn{3}{c}{\textbf{Internal alignment degree }}\\
\hline
\multicolumn{3}{c}{Grains with fast internal relaxation}\\
\hline
High-$J$ attractors & $Q_{\rm X}^{\rm high-J}$ & 1 \\
Low-$J$ attractors & $Q_{\rm X}^{\rm low-J}$ & TE Boltzmann distribution\\
\hline
\multicolumn{3}{c}{Grains with slow internal relaxation}\\
\hline
High-$J$ attractors & $Q_{\rm X}^{\rm high-J}$ & 0.15 \\
Low-$J$ attractors & $Q_{\rm X}^{\rm low-J}$ & 0.05\\

 \hline
\multicolumn{3}{c}{\textbf{External alignment degree}}\\
\hline
High-$J$ attractors & $Q_{\rm J}^{\rm high-J}$ &  1 \\
Low-$J$ attractors & $Q_{\rm J}^{\rm low-J}$ & 1 \\

\hline
\label{tab:parameter}
\end{tabular}}   
\end{table}

 \begin{table*}
  \centering
         \caption{Setup for model RATA and RATA$-$RATD. The alignment model in RATA and RATA$-$RATD is similar to model RATA$-$INELASTIC in \cite{Giang_et_al_2024}: the alignment size range is controlled by the Larmor precession rate; the internal alignment is controlled by the join action of super-Barnett and inelastic relaxation; and the external alignment mechanism followed by the fraction of grains at high-\textit{J} attractors $f_{\rm high-J}$ is determined by the magnetic relaxation ratio $\delta_{\rm m}$ (Section \ref{sec:MHD+radiation_field}).}
  \begin{tabular} {cccc}
  \hline 
  \textbf{Model name} & \textbf{RATD effect} & \textbf{Disruption range} & \textbf{$f_{\rm disr}$} \\
  \hline 

RATA & No   & --    &  --   \\  
RATA$-$RATD & Yes   &  $[a_{\rm disr,POLARIS} -  a_{\rm disr,max,POLARIS}]$$^{a}$ & $f_{\rm high-J}$$^{b}$ \\ 
  
  \hline 
    \label{tab:model}
    \end{tabular}\\
     \footnotesize{($^a$): Disruption range is examined again with the alignment range.}\\
     \footnotesize{($^b$): Fraction of disrupted grains $f_{\rm disr}$ depends on the values of $f_{\rm high-J}$, which depends on the grain size, grain magnetic susceptibility, gas damping timescale, and magnetic field strength.}\\
\end{table*}
 
\section{Impact of RATD on polarized dust emission}\label{sec:ratd_polarization}
 
\subsection{Treatment of polarized radiative transfer of Stokes parameters under RATD}\label{sec:polaris_stokes}
Knowing the grain alignment efficiency (see detailed in \citealt{Giang_et_al_2024}), the disruption range, and the new size distribution of small grains below $a_{\rm disr,POLARIS}$, we can determine the new coefficients used to solve the polarized radiative transfer of Stokes parameters. {\footnote{Indeed, the change of $a_{\rm disr,POLARIS}$ and $\alpha$ affect the dust extinction with stellar and thermal dust radiation field, which leads to future changes in radiation field strength, dust temperature, and disruption sizes by RATD. Therefore, we must loop the Monte-Carlo radiative transfer and RATD calculation until the change in all parameters becomes insignificant. However, due to the cost of such a simulation, we only use results from one round of calculation for analysis in the rest of this paper. The new alignment and disruption size distribution from the second round of MCRT simulation is shown in Appendix \ref{sec:appen_RATD_more_step}. }  Theoretically, the polarization state of photons can be described by the Stokes vectors \textbf{S} $ = (I ~ Q ~ U ~ V)^{\rm T}$, with Stokes I describes the intensity, Stokes Q and U describe the linear polarization, and Stokes V describes the state of circular polarization. At each wavelength, the general formula of this equation in the dusty environments is given by (\citealt{Martin_1974}, \citealt{Reissl_2016}):

\begin{gather}
    \frac{d}{ds} 
    \begin{pmatrix}
     I\\
     Q\\
     U\\
     V
    \end{pmatrix}
=
-\begin{pmatrix}
    \alpha_{\rm I} && \alpha_{\rm Q} && 0 && 0 \\
    \alpha_{\rm Q} && \alpha_{\rm I} && 0 && 0 \\
    0 && 0 && \alpha_{\rm I} && \kappa_{\rm Q}  \\
    0 && 0 && -\kappa_{\rm Q} && \alpha_{\rm I} \\
\end{pmatrix}
\begin{pmatrix}
 I\\
 Q\\
 U\\
 V
\end{pmatrix}
+ 
\begin{pmatrix}
 j_{\rm I} \\
 j_{\rm Q} \\
 0 \\
0
 \label{eq:polarized_radiative_transfer}
\end{pmatrix},
\end{gather}

where $\alpha_{\rm I}$, $\alpha_{\rm Q}$, $\kappa_{\rm Q}$ are the extinction coefficient, linear polarization coefficient, and circular polarization coefficient integrated over grain size distribution. $j_{\rm I}$ and $j_{\rm Q}$ are the emissivity coefficient of non-aligned and aligned dust grains, respectively.

The calculation of the new $\alpha_{\rm I}, \alpha_{\rm Q}, \kappa_{\rm Q}, j_{\rm I}$, and $j_{\rm Q}$ under RATD effect is given in Appendix \ref{sec:appen_stokes}. Here, we summarize the final changes in their formulas:
\begingroup\makeatletter\def\f@size{8}\check@mathfonts
\bea
\alpha_{\rm I}^{\rm new}  &=& n_{\rm H_{2}} C ~ \Bigg [\int_{\rm a_{\rm min}}^{\rm a_{\rm disr,POLARIS}} C_{\rm ext}  a^{\alpha} da \nonumber \\ 
&+& \int_{\rm a_{\rm disr,POLARIS}}^{\rm a_{\rm max}} C_{\rm ext}^{\rm new} (1-f_{\rm disr}(a)) a^{-3.5} da\Bigg],\label{eq:alpha_I_new}
\ena
\endgroup

\begingroup\makeatletter\def\f@size{8}\check@mathfonts
\bea
\alpha_{\rm Q}^{\rm new}  &=& n_{\rm H_{2}} C ~\Bigg [\int_{\rm a_{\rm min}}^{\rm a_{\rm disr,POLARIS}} C_{\rm ext}^{\rm pol}  a^{\alpha} da \nonumber \\
&+& \int_{\rm a_{\rm disr,POLARIS}}^{\rm a_{\rm max}} C_{\rm ext}^{\rm pol,new} (1-f_{\rm disr}(a)) a^{-3.5} da\Bigg],\label{eq:alpha_Q_new}
\ena
\endgroup

\begingroup\makeatletter\def\f@size{8}\check@mathfonts
\bea
\kappa_{\rm Q}^{\rm new}  &=& n_{\rm H_{2}} C ~\Bigg [\int_{\rm a_{\rm min}}^{\rm a_{\rm disr,POLARIS}} C_{\rm circ} a^{\alpha} da \nonumber \\
&+& \int_{\rm a_{\rm disr,POLARIS}}^{\rm a_{\rm max}} C_{\rm circ}^{\rm new} (1-f_{\rm disr}(a)) a^{-3.5} da\Bigg],\label{eq:alpha_Q_new}
\ena
\endgroup

\begingroup\makeatletter\def\f@size{7.5}\check@mathfonts
\bea
j_{\rm I}^{\rm new}  &=& n_{\rm H_{2}} C  ~\Bigg[\int_{\rm a_{\rm min}}^{\rm a_{\rm disr,POLARIS}} C_{\rm abs}  B_{\lambda}(T_{\rm d}(a)) a^{\alpha} da \nonumber \\
&+& \int_{\rm a_{\rm disr,POLARIS}}^{\rm a_{\rm max}} C_{\rm abs}^{\rm new} (1-f_{\rm disr}(a)) B_{\lambda}(T_{\rm d}(a)) a^{-3.5} da\Bigg] \label{eq:j_I_new}
\ena
\endgroup

and 
\begingroup\makeatletter\def\f@size{7.5}\check@mathfonts
\bea
j_{\rm Q}^{\rm new}  &=& n_{\rm H_{2}} C ~ \Bigg[\int_{\rm a_{\rm min}}^{\rm a_{\rm disr,POLARIS}} C_{\rm abs}^{\rm pol}  B_{\lambda}(T_{\rm d}(a)) a^{\alpha} da \nonumber \\
&+& \int_{\rm a_{\rm disr,POLARIS}}^{\rm a_{\rm max}} C_{\rm abs}^{\rm pol, new}  (1-f_{\rm disr}(a)) B_{\lambda}(T_{\rm d}(a)) a^{-3.5} da\Bigg].\label{eq:j_Q_new}
\ena
\endgroup

In above equations, $C_{\rm ext}, C_{\rm exp}^{\rm pol}, C_{\rm circ}, C_{\rm abs}, C_{\rm abs}^{\rm pol}$ are the extinction, linear polarization extinction, circular polarization, absorption, and polarization absorption cross-section of aligned dust grains before RATD happens. The similar terms but with denotation $^{\rm new}$ are the new cross-sections modified by RATD (see Appendix \ref{sec:appen_stokes}). $C$ is the normalization constant of the dust size distribution. The term $a^{-3.5}$ describes the distribution of remaining grains surviving from RATD. And $a^{\alpha}$ describes the distribution of small grains below $a_{\rm disr,POLARIS}$ (Section \ref{sec:polaris_eta}).

Then, by plugging $\alpha_{\rm I}^{\rm new}$, $\alpha_{\rm Q}^{\rm new}$, $\kappa_{\rm Q}^{\rm new}$, $j_{\rm I}^{\rm new}$, and $j_{\rm Q}^{\rm new}$ into Equation (\ref{eq:polarized_radiative_transfer}), we can solve the polarized radiative transfer of Stokes parameters to get the Stokes I, Q, U maps under the effect of RATD. We use similar setups as in \cite{Giang_et_al_2024}, with the plane detector placed at 300 pc from the object to observe 8000 au of the protostellar core. The plane detector contains 1000$\times$1000 pixels, which resolves the core at 8 au. We observe the core at 1.3mm. Given the synthetic map of Stokes I, Q, and U, the polarization fraction in the unit of percentage will be:

\bea 
p(\%) = \frac{\sqrt{Q^{2} + U^{2}}}{I} \times 100\%.
\ena

 \begin{figure*}
\centering
 \includegraphics[width=\textwidth,height=\textheight,keepaspectratio]{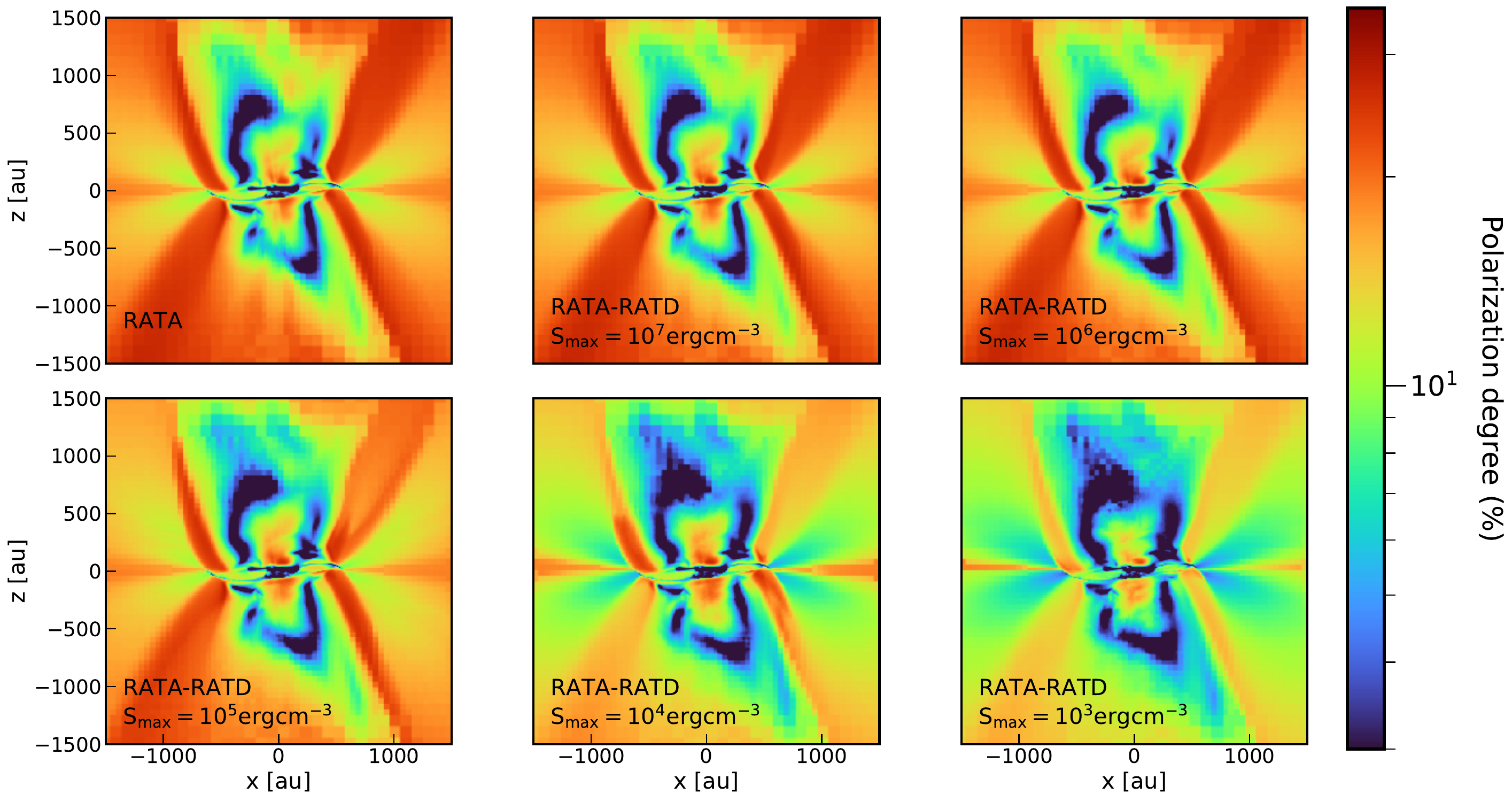}

    \caption{Comparison of the polarization degree map obtained at 1.3mm between model RATA and model RATA$-$RATD of SPM grains with $N_{\rm cl} = 10^{3}$, assuming $S_{\rm max} = 10^{3} - 10^{7}\erg\cm^{-3}$. Without RATD (upper left panel), one expects to see the exceeded polarization degree along the outflow cavity wall owing to the efficient magnetic alignment of micron-sized and VLGs by inelastic relaxation and MRAT mechanism (\citealt{Giang_et_al_2024}). When RATD happens, the polarization degree unchanges for grains having $S_{\rm max} \geq 10^{5}\erg\cm^{-3}$, but it reduces about twice times for grains having $S_{\rm max} \leq 10^{4}\erg\cm^{-3}$. The latter case happens because RATD can destroy aligned dust grains inside the outflow cavity wall and inner envelope due to their low disruption threshold (see Figure \ref{fig:wt_grain_size}, lower row for the dynamic approach, and Figure \ref{fig:adisr_midplane} for the POLARIS approach).} 
     \label{fig:RATD_visibility}
\end{figure*}

\subsection{Dust polarization degree map}
To understand the impact of RATD on polarized dust emission, we consider model RATA, which does not include the RATD effect, and model RATA$-$RATD, which takes RATD into account. Detailed setups of each model are summarized in Table \ref{tab:model}.

We show in Figure \ref{fig:RATD_visibility} the comparison of the polarization degree map obtained from model RATA (upper left panel) and model RATA$-$RATD, assuming grains to have $S_{\rm max}=10^{3}-10^{7}\erg\cm^{-3}$ and $N_{\rm cl} = 10^{3}$. In model RATA, the polarization fraction peaks along the outflow cavity wall, of $p \sim 25\%$, owing to the efficient magnetic alignment of micron-sized and VLGs driven by the joint action of Inelastic relaxation and MRAT mechanism (see \citealt{Giang_et_al_2024} for detailed explanation). The value of $p(\%)$ is smaller of $p \sim 15\%$ inside the inner envelope and the equatorial midplane. It reaches $< 5\%$ inside the disk owing to the reduction of the internal alignment and MRAT efficiency in high gas density regions. By taking RATD into account, the polarization degree remains the same for grains with $S_{\rm max} \geq 10^{6}\erg\cm^{-3}$. As shown in Figure \ref{fig:adisr_midplane}, RATD can remove micron-sized and VLGs with $S_{\rm max}\geq 10^{6}\erg\cm^{-3}$ inside the jet and outflow cavity. However, the dust polarization radiating from these regions is subdominant to emission from foreground envelope grains that are unaffected by RATD. That explains why the disruption of grains with $S_{\rm max} \geq 10^{6}\erg\cm^{-3}$ does not affect the net observed degree of polarization. In contrast, for grains with $S_{\rm max} \sim 10^{3}-10^{5}\erg\cm^{-3}$, RATD can remove both large grains inside the jet, outflow cavity, outflow cavity wall, and inner envelope (Figure \ref{fig:adisr_midplane}). The removal of large grains beyond the outflow cavity reduces polarized dust emission obtained along the outflow cavity wall and inner envelope clearer. $p(\%)$ inside the outflow cavity wall decreases from $\sim 25\%$ to $p \sim 15-20\%$, and $p(\%)$ near the equatorial midplane decreases from $\sim 15\%$ to $\sim 6\%$ when taking the effect of RATD on aggregate grains with $S_{\rm max} \sim 10^{3}-10^{5}\erg\cm^{-3}$ into account.

\section{Discussion}\label{sec:discussion}

In this section, we will discuss more about the effect of RATD on the migration of dust grains from the protostellar disk to the inner envelope proposed by \cite{Wong_2016} and \cite{Tsukamoto_ashfall_2021}. Further discussion about the impact of RATD on dust polarization, stellar feedback, dust dynamics inside the jet and outflow will also be carried out.

\subsection{Dust grain population in RATD period}\label{sec:discuss_disruption} 
As shown in \cite{Valentin_2023b} and Appendix \ref{sec:appen_Lstar}, RATD can start affecting aggregate$-$type grains with low $S_{\rm max} \leq 10^{4}\erg\cm^{-3}$ when the central luminosity of YSOs exceeds $> 5L_{\odot}$. During this period, their micron-sized grains, VLGs, and submillimeter grains uplifted to the jet/outflow base will be quickly fragmented to smaller sizes just after a few yr.(Figures \ref{fig:adisr_min_dynamic}, \ref{fig:adisr_max_dynamic}, \ref{fig:RATD_20Lsun} and \ref{fig:RATD_5Lsun})\footnote{We note that this disruption timescale is not universal for low/intermediate Class 0 YSOs. Grains will take a longer time to be destroyed by RATD if the jet/outflow base is denser and more clumpy than our adopted model}. Small grains surviving from RATD will migrate outward, replacing VLGs and submilimeter grains to be the major dust population inside the jet and outflow. We found the maximum grain size inside this region to be constrained at $\sim 3, 1, 0.5\mum$ by RATD for $L_{\rm center} = 5, 20, 100L_{\odot}$. The dominance of sub-micron and small micron-sized grains inside the jet and outflow can maintain a few to few hundred yr (i.e., the typical lifetime of the accretion burst expected to activate RATD, \citealt{Jhan_2022}, \citealt{Contreras_2024}), until $L_{\rm center}$ reduces to below the threshold for RATD being activated. When RATD turns off, large aggregate$-$type grains uplifted from the disk surface start to replace small grains to be the major dust popularization inside the jet and outflow. One expects this area to be fully occupied by large grains again after the last small grains generation surviving from RATD reaches beyond a few thousand au scale. Besides the strong dust disruption inside the jet and outflow, VLGs with low $S_{\rm max} \leq 10^{4}\erg\cm^{-3}$ infalling inside the inner envelope can also be destroyed by RATD due to their low disruption threshold.

Grains having higher $S_{\rm max} \sim 10^{5}-10^{6}\erg\cm^{-3}$ need to be exposed to stronger radiation field strength to be affected by RATD, i.e., $L_{\rm center} = 100L_{\odot}$ (Figure \ref{fig:adisr_min_dynamic}, middle column). But in this case, only micron-sized and VLGs are removed from the jet and outflow cavity by RATD. Submillimeter grains can freely migrate toward the inner envelope, and VLGs/submillimeter grains inside the outflow cavity wall and inner envelope are safe from RATD (Figure \ref{fig:adisr_max_dynamic}, middle column). The situation is worse for composite grains with $S_{\rm max} \geq 10^{7}\erg\cm^{-3}$ (Figure \ref{fig:adisr_nH_vgas}, right column). Even for $L_{\rm center} = 100L_{\odot}$, only some small areas inside the jet and outflow where $v_{\rm gas} < 250$ km/s can be occupied by small grains enhanced by RATD. The major dust population inside the jet and outflow still be VLGs and submillimeter grains because of the inefficient RATD on grains having high disruption threshold.

\subsubsection{Whether RATD totally suppresses the propagation of VLGs and submillimeter grains inside the jet and outflow?}\label{sec:discuss_RATD_propagation}
Given that RATD can strongly affect aggregate$-$type grains having $S_{\rm max} \leq 10^{6}\erg\cm^{-3}$ since they are uplifted to the jet/outflow base, it turns to the second question that whether RATD can totally suppress the propagation of large grains inside the jet and outflow cavity. As discussed in Section \ref{sec:polaris_eta}, the fraction of grains destroyed by RATD $f_{\rm disr}$ depends on how much large grains can be stably aligned with $\B$ at high-\textit{J} attractors, i.e., the fraction $f_{\rm high-J}$. To understand the disruption fraction of RATD acting on protostellar grains, we show in Figure \ref{fig:f_disr} the maximum size for grains having $f_{\rm high-J} = 0.5$ ($a_{\rm max,JB}^{\rm DG,0.5}$, upper row) and $f_{\rm high-J} = 1$ ($a_{\rm max,JB}^{\rm DG,1}$, lower row), considering $a_{\rm max} = 500\mum$, and $L_{\rm center} = 100L_{\odot}$. Given the strong dependence of $f_{\rm high-J}$ on the grain magnetic properties, we show results for SPM grains with $N_{\rm cl} = 10^{2}$, $10^{3}$, and $10^{4}$, from left to right, respectively.
 
 \begin{figure*}
\centering
    \includegraphics[width=\textwidth,height=\textheight,keepaspectratio]{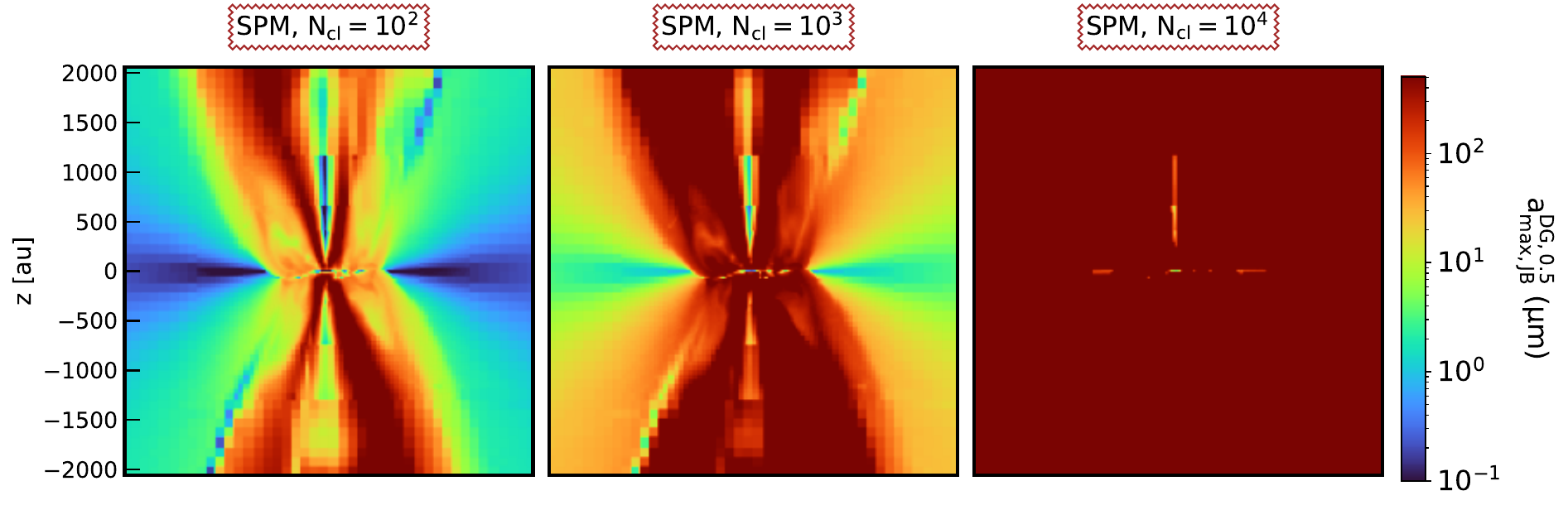}
    \includegraphics[width=\textwidth,height=\textheight,keepaspectratio]{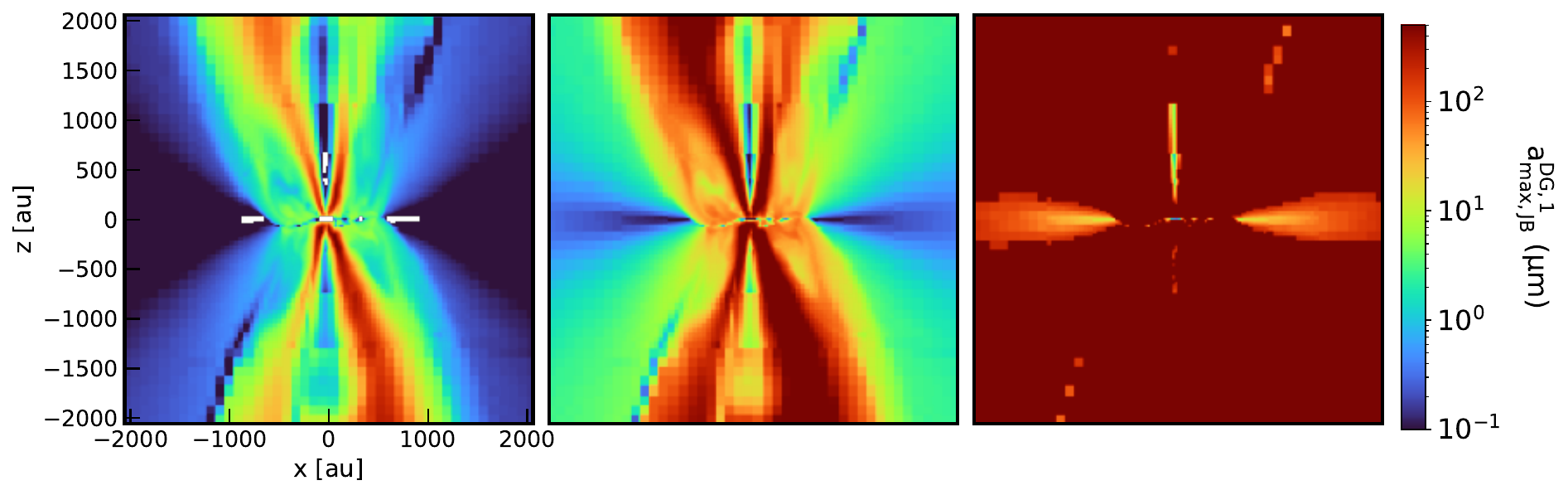}
    \caption{Spatial distribution of $f_{\rm disr}$ - the fraction of grains destroyed by RATD, representing via the maximum grain size which can be aligned with $\B$ by MRAT alignment with $f_{\rm high-J} = 0.5$ ($a_{\rm max,JB}^{\rm DG,0.5}$ (upper row) and $f_{\rm high-J} = 1$ ($a_{\rm max,JB}^{\rm DG,1}$ (lower row) (Section \ref{sec:polaris_eta}). By increasing the number of iron inclusions locked inside dust grains, more micron-sized and VLGs can have perfect magnetic alignment by MRAT owing to the enhanced magnetic relaxation efficiency. Grains can achieve perfect magnetic alignment with $f_{\rm high-J} = 1$ regardless of their positions inside the protostar if they have big iron clusters with $N_{\rm cl} = 10^{4}$. } 
     \label{fig:f_disr}
\end{figure*} 
  
One can see that for the protostar with $L_{\rm center} = 100L_{\odot}$, if SPM grains have $N_{\rm cl} \geq 10^{3}$ (middle and right column), $\sim 100\%$ of grains having $S_{\rm max} \leq 10^{6}\erg\cm^{-3}$ can be destroyed by RATD while propagating inside the jet and outflow cavity, i.e., $a_{\rm max,JB}^{\rm DG, 1} \sim 500\mum$. The disruption fraction is slightly smaller inside the narrow jet lobe and the outflow cavity wall due to the reduced MRAT alignment efficiency by high gas density there (Figure \ref{fig:distribution_mcrt}, upper left panel). The perfect disruption still happens for micron-sized below $< 10\mum$ in these areas, but $f_{\rm disr}$ is only $\sim 25-50\%$ for VLGs and submillimeter grains there. The disruption fraction for aggregate$-$type grains inside the jet and outflow cavity reduces with decreasing the grain magnetic susceptibility. For example, if grains have $N_{\rm cl} = 10^{2}$ (left column), only $25-50\%$ of VLGs and submillimeter grains rotating suprathermally are disrupted to smaller sizes by RATD while propagating inside the jet and outflow cavity. This disruption fraction can increase with decreasing $L_{\rm center}$ because MRAT alignment becomes more efficient due to the enhanced magnetic susceptibility of grains having lower temperatures.

As discussed in \cite{Valentin_2023b}, grains aligning with $\B$ at thermal rotation, i.e., low-\textit{J} attractors, can still be destroyed by RATD if they can be uplifted to high-\textit{J} attractors via the gas bombardment (\citealt{Hoang_Lazarian_2008}, \citealt{Lazarian_Hoang_2021}). Considering the jet base with $n_{\rm H_{2}} \sim 10^{8}\cm^{-3}$ and $T_{\rm d} \sim 100$ K, the gas damping timescale is approximately $\tau_{\rm gas} \sim 2$ yrs, yielding the typical timescale of  $5-10\tau_{\rm gas} \sim 10 - 20$ yr for grains being uplifted to high-\textit{J} point. However,  grains moving with a hundred km/s can reach $r \sim 400-800$ au to the center after $\sim 10 - 20$ yrs, beyond the region where RATD can affect fast-moving grains (Figure \ref{fig:wt_grain_size}, lower row, left panel). For grains propagating along the outflow cavity, they experience weaker gaseous damping due to lower gas density and smaller gas temperature there. Consequently, outflowing grains can even take a few hundred yr to change their alignment state from low to high-\textit{J} attractors. During this time, large thermal grains may already migrate beyond the active region of RATD. Thus, "slow disruption" cannot help thermal aggregate grains to be destroyed by RATD while propagating inside the jet and outflow cavity.

\subsubsection{Whether RATD totally suppresses the existence of VLGs inside the outflow cavity wall and inner envelope?}\label{sec:discuss_RATD_migration}

As discussed in Section \ref{sec:discuss_RATD_propagation}, if aggregate grains with $S_{\rm max} \leq 10^{6}\erg\cm^{-3}$ have $N_{\rm cl} \leq 10^{2}$, $\sim 50-75\%$ of thermal grains can freely propagate inside the jet and outflow without being destroyed by RATD. As shown in \cite{Tsukamoto_ashfall_2021}, such large grains can decouple from outflowing gas and migrate to the outflow cavity wall and inner envelope by centrifugal forces (\citealt{Tsukamoto_ashfall_2021}). Except $< 10\%$ of them will be destroyed by grain-grain shattering after entering these areas (\citealt{Wong_2016}), $> 90\%$ of thermal grains can stably exist inside the cavity wall and inner envelope without being affected by RATD.

The situation is different for their suprathermal grains that alrealdy exist beyond the outflow cavity. Except for grains with $S_{\rm max} \sim 10^{5}-10^{6}\erg\cm^{-3}$, we found that for $L_{\rm center} = 100L_{\odot}$, suprathermal VLGs with $S_{\rm max} = 10^{4}\erg\cm^{-3}$ inside the outflow cavity wall, and suprathermal VLGs/submillimeter grains with $S_{\rm max} = 10^{3}\erg\cm^{-3}$ inside the cavity wall/inner envelope still be destroyed by RATD due to their low disruption threshold (Figure \ref{fig:wt_grain_size}, lower row). The disruption fraction can reach $\sim 100\%$ if grains hold $N_{\rm cl} \sim 10^{4}$, but $\sim 25\%$ if grains have lower $N_{\rm cl} \leq 10^{3}$ (Figure \ref{fig:f_disr}). For the latter case, $\sim 75\%$ of suprathermal grains are safe from RATD inside the outflow cavity wall and inner envelope, and they can freely reflux to the outer disk as expected in \cite{Tsukamoto_ashfall_2021}. Such strong disruption picture for aggregate grains with $S_{\rm max} \leq 10^{4}\erg\cm^{-3}$ beyond the outflow cavity may only appear around the high-luminous protostar with $L_{\rm center} = 100L_{\odot}$. For protostars with lower $L_{\rm center} < 100L_{\odot}$, RATD may not strong enough to affect their existence inside the outflow cavity wall and inner envelope.

\subsubsection{Validation of our calculation}\label{sec:discuss_limitation}
In this section, we will discuss more about the effect of our assumption made in Section \ref{sec:vgrain_omegat} on the disruption picture shown in Section \ref{sec:adisr_dynamic}. The first one is about the gas-grain coupling used to describe the grain propagation direction inside the jet and outflow. As shown in \cite{Tsukamoto_ashfall_2021}, while sub-micron and micron-sized grains below $< 10\mum$ can maintain their well coupling with outflowing gas and distribute over few hundreds au of the outflow, VLGs beyond $> 10\mum$ and submillimeter grains tends to decouple from outflowing gas at $z \geq 50$ au and migrate toward the inner envelope (see Figure 1 in \citealt{Tsukamoto_ashfall_2021}). This decoupling lowers the amount of VLGs and submillimeter grains inside the outflow but enriches the large dust population inside the outflow cavity wall and inner envelope (Figure 5 in \citealt{Tsukamoto_ashfall_2021}).  

Above realistic gas-grain decoupling does not affect the disruption of sub-micron and micron-sized grains inside the jet and outflow base as shown in Figure \ref{fig:adisr_min_dynamic}. Similarly, it does not alter the dominance of small micron-sized grains with $S_{\rm max} \leq 10^{6}\erg\cm^{-3}$ inside the jet and outflow during RATD period as discussed in Section \ref{sec:discuss_disruption}. For large grains beyond $>10\mum$ that break the gas-grain coupling condition, Figure \ref{fig:wt_grain_size} shows that RATD only takes $< 5.46$ yr to destroy VLGs (for all considered values $S_{\rm max} = 10^{3} - 10^{7}\erg\cm^{-3}$) and submillimeter grains (for $S_{\rm max} = 10^{3}-10^{4}\erg\cm^{-3}$) inside the outflow base (upper right panel). The corresponding displacement that grains propagating with $\sim 10$ km/s take during $\sim 5.46$ yr is about $10$ au. Given such a small displacement, the gas-grain coupling may still be held until VLGs and submillimeter grains are destroyed by RATD, which validates the maximum disruption size found inside the outflow base in Figure \ref{fig:adisr_max_dynamic} (upper row, left panel). Such fast disruption in this area will clear out the appearance of VLGs and submillimeter grains inside the outflow for the time being, similar to what we found in the lower row of Figure \ref{fig:adisr_max_dynamic}. 

For large grains at $z > 50$ au before RATD happens, they may be migrating toward the outflow cavity wall and inner envelope instead of still ejecting outward along the outflow cavity as our assumption. However, for aggregate grains having $S_{\rm max} \leq 10^{4}\erg\cm^{-3}$, RATD can still destroy both VLGs and submillimeter grains either they are on the path toward the inner envelope or already outside the outflow cavity due to their short disruption timescale (Figure \ref{fig:wt_grain_size}, middle right panel and lower row). Therefore, the gas-grain decoupling does not alter the role of RATD in preventing the migration of such grains from the disk to the inner envelope. However, the presence of the decoupling effect may let VLGs and submillimeter grains inside the inner envelope to quickly settle down the dense midplane (\citealt{Lebreuilly_2020}, \citealt{Marchand_2023}), that reduces the impact of RATD on infalling large aggregate grains. Further studies including RATD in the dust evolution are required to accurately understand whether RATD suppresses the Ashfall motion of large aggregate grains inside the protostellar core. For aggregate grains with $S_{\rm max} \sim 10^{5}-10^{6}\erg\cm^{-3}$, their VLGs of $\sim 10-20\mum$ may still be destroyed by RATD before reaching the inner envelope due to their short disruption timescale of $\sim 6-10$ yr (which corresponds to the displacement of $\sim 25-40$ au, Figure \ref{fig:wt_grain_size}, middle right panel). But their larger grains may safely enter the outflow cavity wall and inner nevelope without being removed by RATD. Similarly, more VLGs and submillimeter composite grains with $S_{\rm max} \geq 10^{7}\erg\cm^{-3}$ can enrich the inner envelope by this decoupling feature. 
  
Another assumption in our study is the fixed dust-to-gas mass ratio $\eta = 0.01$ typically found inside ISM. If the Ashfall dust model happens before RATD is activated, one expects the dust-to-gas mass ratio distribution to form the U$-$shape, with lower $\eta \sim 0.004$ inside the outflow cavity and higher $\eta \sim 0.012$ inside the outflow cavity wall and inner envelope (\citealt{Tsukamoto_ashfall_2021}, \citealt{Marchand_2023}). One also expects to see more VLGs and submillimeter grains inside the enhanced $\eta$ region than inside the outflow. Under this condition, one expects RATD to be stronger, i.e., smaller $a_{\rm disr}$ and larger $a_{\rm disr,max}$, inside the outflow cavity and weaker inside the outflow cavity wall and inner envelope due to smaller/larger dust reddening effect in these regions. Given the strong conversion from large grains to small grains inside the outflow base, the operation of RATD not only makes sub-micron and small micron-sized grains to be the major dust population inside the outflow cavity, but also increases again the total grain mass in this area (or increasing $\eta$). The enhanced amount of small grains inside the outflow cavity accelerates the reprocessing of stellar UV-optical photons which weakens the RAT disruption acting on sub-micron and micron-sized grains, i.e., $a_{\rm disr}$ increases. In contrast, more IR photons from dust emission can escape from the dust reddening effect, which further enhances the RAT disruption acting on large grains. More VLGs and submillimeter grains with low $S_{\rm max} \leq 10^{5}\erg\cm^{-3}$ at $z > 50$ au can be fragmented to smaller sizes before they enter the outflow cavity wall and inner envelope. This change further reduces the amount of large grains entering the shell surrounding the outflow but further enhances the small dust population in this area. As small micron-sized grains can be well coupled with infalling gas, it can slow down the dust settling and the increase of $\eta$ inside the disk midplane as predicted in model that VLGs and submillimeter grains are enriched inside the inner envelope (\citealt{Lebreuilly_2020}, \citealt{Vorobyov_2019}, \citealt{Tsukamoto_ashfall_2021}, \citealt{Tsukamoto_2023_coevolution}, \citealt{Marchand_2023}). Further studies are required to better understand the impact of RATD on the dust evolution around the protostars.
 
In Section \ref{sec:vgrain_omegat}, we assume grains initially to be at rest with $\Omega(t_{0}) = 0$ rad/s. As found in \cite{Giang_et_al_2024}, grains beyond $2\mum$ inside the disk may partly have suprathermal rotation by RATs, while the remaining can have thermal rotation due to the gas-grain collisions. To understand how the disruption picture changes if we assign thermal or suprathermal rotation to the initial angular velocity $\Omega(t_{\rm 0})$, we adopt two cases of $\sim 5\mum$ grains having $S_{\rm max} = 10^{5}\erg\cm^{-3}$ and $50\mum$ grains having $S_{\rm max} = 10^{3}\erg\cm^{-3}$ inside the outflow base with typical $T_{\rm g} = 100$ K. The disruption threshold are $\Omega_{\rm disr} = 7\times 10^{5}$ rad/s and $7\times 10^{3}$ rad/s (Equation \ref{eq:omega_disr}). The corresponding thermal rotation for $5\mum$ and $50\mum$ grains will be $\Omega_{\rm th} = \sqrt{k T_{\rm g}/I_{\parallel}} = 7$ rad/s and $0.02$ rad/s, respectively. The suprathermal rotation for alignment ($\Omega = 3\Omega_{\rm ther}$) will be $21$ rad/s and $0.06$ rad/s. Given the huge difference in $\Omega_{\rm disr}$ and thermal/suprathermal rotation of dust grains, adopting $\Omega(t_{\rm 0}) = 0$ rad/s does not weaken RATD activities shown in Figures \ref{fig:adisr_min_dynamic} and \ref{fig:adisr_max_dynamic}.
   
 \subsection{Iron inclusions and RATD: which one is more important in controlling dust polarization fraction in protostellar environments?} 
As discussed in \cite{Valentin_2023b}, protostellar VLGs should have composite or compact structure with $S_{\rm max} \geq 10^{7}\erg\cm^{-3}$ to preserve their existence inside the protostellar outflow required to explain $\geq 5\%$ of dust polarization observed in low-intermediate mass Class 0/I YSOs by ALMA (\citealt{Valdivia_2019}, \citealt{Valentin_2023b}). But as shown in \cite{Seizinger_2013}, \cite{Kataoka_2017_dust}, the grain coagulation inside protostellar disks tends to produce highly porous VLGs with the typical low $S_{\rm max} \sim 10^{3} - 10^{5} \erg \cm^{-3}$ (\citealt{Tatsuuma_2019}, \citealt{Kimura_2020}). Yet how large grains with high $S_{\rm max}$ can form still be unclear.
 
As shown in Figure \ref{fig:RATD_visibility}, the observed polarization degree within 1500 au around the protostar is similar to results without RATD if grains have $S_{\rm max} \geq 10^{5}\erg\cm^{-3}$. In contrast, for grains having $S_{\rm max} \leq 10^{4}\erg\cm^{-3}$, RATD can destroy both VLGs and submillimeter grains while they propagate inside the outflow cavity wall and infall inside the inner envelope. Such large removal of large grains beyond the outflow cavity visible the suppression of RATD on polarization fraction as expected in \cite{Hoang_2021} and \cite{Valentin_2023b}. The observed polarization degree along the outflow cavity wall and inner envelope is reduced about twice when RATD happens. Indeed, for SPM grains with $N_{\rm cl} = 10^{3}$ shown in Figure \ref{fig:RATD_visibility}, RATD only can remove $\sim 25\%$ and $\sim 50\%$ of VLGs from the inner envelope and outflow cavity. The grain disruption fraction can increase to one if grains have $N_{\rm cl} = 10^{4}$, inducing stronger suppression of RATD on $p(\%)$. But regardless of the disruption fraction, RATD cannot remove grains to below $\sim 10\mum$ (Figures \ref{fig:wt_grain_size} and \ref{fig:adisr_midplane}) - the minimum size required to reproduce the observed ALMA polarization degree in the inner 500 au region (\citealt{Giang_Hoang_2024}, Figure 8). In contrast, even with the presence of VLGs, we cannot reproduce $p > 1\%$ there if grains have $N_{\rm cl} < 10^{2}$ and be aligned with $\B$ by RATs only (\citealt{Giang_Hoang_2024}, Figure 7). MRAT mechanism must be taken into account, letting iron inclusions be more important than RATD in controlling the observed polarization degree in Class 0/I protostars. The observed polarization degree can be smaller by twice if RATD happens inside the core, but grains do not need to have composite or compact structure to explain ALMA dust polarization observations as argued in \cite{Valentin_2023b}.

\subsection{Origin of dust polarization inside protostellar outflow in RATD period: B$-$RAT or k$-$RAT?}\label{sec:discuss_krat}
One of the topics raised in \cite{Valentin_2023b} and \citealt{Giang_et_al_2024} is about the origin of dust polarization observed inside the protostellar outflow. Different from the envelope and inner envelope, dust grains inside outflows can either have the magnetic alignment (B$-$RAT, by, i.e., RATs/MRAT) or radiative alignment (k$-$RAT, or alignment with radiation field by RATs, \citealt{Lazarian_Hoang_2007a}) owing to the intense radiation field in this low density region. By comparing the Larmor precession and radiative precession, \cite{Valentin_2023b} found that k$-$RAT can become the alignment direction of VLGs beyond $\geq 10\mum$ at low-\textit{J} attractor inside outflows. However, if the MRAT mechanism is the origin behind the grain alignment that would produce ALMA dust polarization of $p\geq 2-30\%$ beyond the disk, dust grains will be SPM with $N_{\rm cl} \geq 10^{2}$. Considering $a_{\rm max} = 500\mum$ inside outflows, Figure \ref{fig:f_disr} showed that if grains have $N_{\rm cl} \sim 10^{2}$, $\sim 50\%$ of VLGs beyond $\geq 20\mum$ and $\sim 25\%$ of submillimeter grains beyond $\geq 100\mum$ will be aligned with $\B$ at high-\textit{J} attractors. As k$-$RAT prefers grains rotating thermally (\citealt{Hoang+2022}), radiative alignment can become the alignment axis of $50-75\%$ of grains at low-\textit{J} as indicated in \cite{Valentin_2023b}. However, grains beyond $\geq 1\mum$ at low-\textit{J} always have slow internal alignment by slow internal relaxation in protostellar environments (\citealt{Giang_et_al_2024}). As a result, one cannot detect the polarization signal from radiative alignment inside outflows because of the dominant dust emission from outflowing grains aligning with $\B$ at high-\textit{J} attractors. In the case of SPM grains with higher $N_{\rm cl} \geq 10^{3}$, $\sim 100\%$ of grains inside outflows are driven to have the magnetic alignment at high-\textit{J} points, that suppresses the possibility for k$-$RAT operating there (\citealt{Valentin_2023b}).

Taking RATD into account, k$-$RAT can become the alignment direction for thermal grains that are not destroyed by RATD inside outflows. However, similar to the period without RATD, radiative alignment is hard to become the origin behind the observed polarization signal inside the protostellar outflow because of their slow internal relaxation. Their emission
can be suppressed by polarization signals from magnetically aligned dust grains from foreground envelopes. Moreover, one may expect to observe the polarization signal from k$-$RAT near the outflow base. However, we do not expect ALMA to detect the imprint of radiative alignment there because of the obscuration from magnetically aligned grains inside the foreground equatorial midplane.

Besides k$-$RAT, Mechanical Torques (MET) can also operate inside the outflow due to the decoupling of gas and large grains (\citealt{Lazarian_Hoang_2007a}, \citealt{Hoang_2018}). As discussed in \cite{Hoang+2022} and \cite{Giang_et_al_2024}, MET can help to enhance the magnetic alignment for outflowing large grains by exerting more spinning torques on grains. It also can drive mechanical alignment (alignment of grains with gas flows, or v$-$MET, \citealt{Hoang_2018}) when the mechanical precession is faster than Larmor precession and gas damping timescale. But similar to k$-$RAT, v$-$MET prefers to drive the alignment for thermal grains (\citealt{Hoang+2022}). Consequently, it cannot replace magnetic alignment to be the origin behind dust polarization observed inside outflows, regardless of the presence of RATD.

\subsection{Another impact of RATD inside the protostellar core}

The first consequence of the dominance of sub-micron and small micron-sized grains inside the outflow by RATD is the shift of the dust reddening effect toward the UV-optical range. As discussed in Section \ref{sec:discuss_limitation}, this change weakens the alignment and disruption of small grains, but strengthens these activities for large micron-sized grains, VLGs, and submillimeter grains (Appendix \ref{sec:appen_RATD_more_step}). Besides being less aligned and disrupted, sub-micron and small micron-sized grains also feel weaker radiative pressure than the moment without RATD. In contrast, large grains will be pushed outward stronger. However, the net radiative pressure acting on all grain sizes will be reduced $\sim 3$ times if the mean maximum grain size decreases from $\sim 50\mum$ to $< 0.5\mum$  (\citealt{Hoang_2021}).  
 
The impact of RATD on stellar feedback may be more obvious in massive protostellar cores. Unlike low/intermediate-mass protostars, high-mass protostars can rapidly evolve to main-sequence OB stars while still embedded inside the massive collapsing core. The intense UV radiation from OB stars develops an expanding bubble driven by the strong stellar wind, surrounded by the HII and photodissociation regions (\citealt{Hollenbach_1997}). Recent photometric and polarimetric observations of massive star-forming regions are significantly increased by JCMT and ALMA facility (BISTRO, MagMAR project), aiming to resolve the long-standing question about the formation, evolution, and feedback of massive stars to surrounding environments (\citealt{Lyo_2021}, \citealt{Fernández-López_2021}, \citealt{Jihye_2022}, \citealt{Chung_2023}). Similar to studies of low/intermediate star-forming regions, it also allows for constraining the dust physical properties, tracking the grain dynamic dynamics of dust grains, and testing RATD physics in such high-dynamic regions (i.e., as for clouds and filaments using SOFIA, JCMT, \citealt{Ngoc_2021}, \citealt{Thuong_2022}, \citealt{Tram_Hoang_2022}, \citealt{Ngoc_2023}, or Orion bar by ALMA, \citealt{Valentin_2023c}).

Another side effects of RATs is rotational desorption (\citealt{Hoang_Tram_2020}) acting on the ice mantle covered on the grain surface. Traditionally, ice mantles will be sublimated and release COMs back to the gas phase when the dust temperature exceeds $\geq 100$K (the ice sublimation threshold, \citealt{Herbst_2009}), denoted as the snow line. Taking the spinning of dust grains by RATs into account, \cite{Hoang_Tram_2020} found that ice mantles can be separated from the grain surface whenever the tensile stress induced by the centrifugal force exceeds the adhesive strength of grain-ice mantle connection. Icy mantles will be fragmented into smaller ones and rapidly sublimated to the gas phase, releasing a bulk of water vapour and COMs even outside the snow line with $T_{\rm d} < 100$ K. The rotation of dust grains also reduces the binding energy between COMs and icy grain surface, allowing COMs to return to the gas phase faster. This mechanism is named as ro-thermal desorption (\citealt{Hoang_Tung_2019}. Applying rotational desorption and ro-thermal desorption into the protoplanetary disk of Herbig Ae/Be stars and some T-Tauri stars, \cite{Tung_Hoang_2020} found that the snow line can be extent beyond the ice sublimation threshold, which may explain the exceed detection of PAHs/nanoparticles in the
disk surface layer (\citealt{Siebenmorgen_2010}).

RATD can also affect the dust destruction and molecule abundances inside the jet. As shown in \cite{Gusdorf_2008a}, \cite{Gusdorf_2008b}, the propagation of shocks along the jet can destroy grains via thermal sputtering. Grain-grain shattering also can happen in this area (\citep{Guillet_2007,Guillet_2009,Guillet_2011,Guillet_2013}), which can be followed by vaporisation of ice mantles if the colliding energy heats them beyond the sublimation threshold (\citealt{Guillet_2007}, \citealt{ Guillet_2009}, \citealt{Guillet_2011}, \citealt{Guillet_2013}). Grain-grain shattering with vaporisation works more efficiently than thermal sputtering if the shock velocity is below $v_{\rm sh} <30$ km/s (\citealt{Guillet_2009}). Inside faster shocks, grain-grain shattering can happen inside the shock front where the temperature exceeds thousands Kelvin. The enhancement of dust cross-section by shattering strengthens thermal sputtering, which happens later in the post-shock where the temperature is several hundred Kelvin (\citealt{Guillet_2011}). However, studies of Gufdoff et al. and Guillet et al. use the standard MRN distribution of $dn/da \sim a^{-3.5}$ with the maximum size $a_{\rm max} \sim 0.3\mum$. Given the enhanced population of sub-micron grains inside the jet by RATD (i.e., larger negative $\alpha \sim -3.6$, Figure \ref{fig:eta_midplane}), shattering and sputtering may be more effective inside the shock region. More SiO (from the vaporisation of ice mantles) and atomic Si (from thermal sputtering, that will react with OH and $\rm O_{\rm 2}$ to become SiO) will go back to the gas phase. Accompanied by the increased shock temperature by enhancing small dust abundance, one may expect to obtain the higher intensity of SiO and other molecules from highly-excited rotational levels (\citealt{Guillet_2013}).

As discussed in Section \ref{sec:discuss_RATD_migration}, RATD enhances amount of sub-micron and micron-sized grains inside the outflow cavity wall and inner envelope if grains have low $S_{\rm max} \leq 10^{4}\erg\cm^{-3}$. Indeed, protostellar grains are not totally neutral, but are charged by absorpting electrons and ions (\citealt{Zhao_2016}). The enhanced amount of sub-micron and micron-sized grains strengthens the depletion of electrons from the outflow cavity wall and inner envelope. Consequently, the ambipolar diffusion resistivity may decrease because ions lose their counterparts for recombination (\citealt{lebreuilly_2023}, \citealt{Tsukamoto_2020_dust_model}, \citealt{Tsukamoto_2023_coevolution}), thus improves the coupling between infalling gas and magnetic fields. As discussed in Section \ref{sec:discuss_limitation}, dust grains inside the inner envelope may be coupled with infalling gas better when RATD happens (\citealt{Lebreuilly_2020}, \citealt{Marchand_2023}). RATD may thus slow the ashfall motion of large grains to the outer disk. But it enhances the abundance of sub-micron and micron-sized grains on the disk scale.

\section{Summary}\label{sec:summary}
In our paper, we aim to understand the effect of RATD on the transportation of dust grains from the protostellar disk to the inner envelope via the outflow channel proposed by \cite{Wong_2016} and \cite{Yusuke_2022}. The second question is about the effect of RATD on properties of dust polarization observed in intermediate Class 0/I protostars. Our major findings are summarized below:

\begin{enumerate} 
 
\item RATD starts to destroy aggregate grains having $S_{\rm max} \leq 10^{4}$ when the bolometric luminosity exceeds $L_{\rm center} > 5L_{\odot}$, with the short disruption timescale of a few yr inside the jet/outflow base. We found that if such aggregate grains are SPM with $N_{\rm cl} \geq 10^{3}$, RATD can totally remove VLGs and submillimeter grains up to $\sim 500\mum$ from the outflow, letting small grains below $\leq 3\mum, 1\mum, 0.5\mum$ to be the major dust population inside the outflow of protostars having $L_{\rm center} = 5,20,100L_{\odot}$. This action prevents the migration of VLGs and submillimeter grains from the inner disk toward the inner envelope and can also partly weaken the redistribution of large grains from the inner envelope toward the outer disk. For aggregate$-$type grains with lower $N_{\rm cl} \leq 10^{2}$, $\sim 25 - 50\%$ of large suprathermal grains will be destroyed by RATD while propagating inside the outflow and infalling inside the inner envelope. The remaining $50-75\%$ of thermal large grains with $S_{\rm max} \leq 10^{4}\erg\cm^{-3}$ can freely enrich the inner envelope without affected by RATD.

\item SPM grains having higher $S_{\rm max} \sim 10^{5}-10^{6}\erg\cm^{-3}$ need to exposed to stronger radiation field strength with $L_{\rm center} \sim 100L_{\odot}$ in order to be affected by RATD. Similar to grains having $S_{\rm max} \leq 10^{4}\erg\cm^{-3}$, RATD can remove $\sim 25-100\%$ of VLGs and submillimeter grains up to $\sim 200\mum$ propagating inside the outflow cavity, depending on $N_{\rm cl}$. However, RATD cannot affect large grains inside the outflow cavity wall and inner envelope due to their larger disruption threshold. This mechanism also not totally prevent the migration of large composite grains from the protostellar disk to the inner envelope, regardless of $L_{\rm center}$.

\item RATD can also prevent the propagation of dust grains having several tens to several hundreds km/s inside the jet. However, due to the suppression of RATD by the fast transportation of dust grains, submillimeter grains beyond $>100\mum$ with $S_{\rm max} \leq 10^{5}\erg\cm^{-3}$, and VLGs beyond $>50\mum$ with $S_{\rm max} \geq 10^{7}\erg\cm^{-3}$ can freely migrate outward without being rotationally disrupted. The disruption fraction inside the jet by RATD is around $\sim 25-50\%$, weaker than inside the outflow due to the higher density that reduces the efficient magnetic alignment of dust grains there.

\item  We show that the physics of RATD incorporated in POLARIS can describe well the dust population distribution of aggregate grains with $S_{\rm max} \leq 10^{5}\erg\cm^{-3}$ propagating with $v_{\rm gas} < 60$ km/s. For such aggregate grains propagating with $v_{\rm gas} > 60$ km/s, POLARIS underestimates the disruption picture for grains having $S_{\rm max} \leq 10^{5}\erg\cm^{-3}$ because it cannot take the migration of small grains surviving from RATD into account. For grains with $S_{\rm max} \geq 10^{6}\erg\cm^{-3}$, POLARIS overestimates the disruption picture because it cannot consider the suppression of RATD by fast-moving grains having high $S_{\rm max}$. The general discrepancy of the disruption size between the POLARIS and dynamic approach is around twice.

\item  Using the new dust population constrained by RATD in the POLARIS approach, we found that the observed polarization degree is reduced about twice time after VLGs having $S_{\rm max} \leq 10^{4}\erg\cm^{-3}$ are destroyed from the outflow cavity wall and inner envelope by RATD. However, RATD cannot remove grains below $10\mum$, i.e., the minimum size required to explain $p \sim 1-10\%$ observed in the inner envelope by ALMA. Therefore, grains do not need to have composite or compact structure as argued by \cite{Valentin_2023b} to explain the efficient grain alignment found by ALMA in low/intermediate-mass protostars.

\item We suggest that inside the outflow cavity, the radiative alignment (k$-$RAT) can happen for the remaining VLGs surviving from RATD at low-\textbf{J} attractors. However, it may be hard to detect the signature of the k-RAT alignment inside the protostellar outflow even with or without RATD. The reason is that k$-$RAT occurs o thermally rotating grains with inefficient internal alignment, whose polarized emission is dominated by magnetically aligned dust grains inside the outflow (in case of without RATD) and in the foreground envelope (in case of with RATD).

\end{enumerate}
\section*{Acknowledgements}
N.C.G and T.H. acknowledge the support from the main research project (No. 2025186902) from Korea Astronomy and Space Science (KASI).  

 
\bibliography{main}
\appendix
 
\section{Spinup process by RATs}\label{sec:appen_RAT}
We use the simplest case of an oblate grain with axial ratio $s=1/2$ to describe the net Radiative torques acting on dust grains. We define $a_{\rm eff}$ as an effective radius of an equivalent spherical grain that shares a similar volume with the oblate grain with the major axis $a$. Giving $a_{\rm eff} = a s^{1/3}$, one gets (\citealt{Hoang_2019}):

\bea 
\Gamma_{\rm RAT} = \int_{\lambda_{\rm min}}^{\lambda_{\rm max}} \pi a_{\rm eff}^{2} \gamma_{\lambda} u_{\lambda}Q_{\Gamma}(a_{\rm eff}, \lambda)\times \frac{\lambda}{2 \pi} d\lambda,\label{eq:gamma_rat}
\ena

where $u_{\lambda}$ and $\gamma_{\lambda}$ are the spectral energy density and the anisotropic degree of radiation at wavelength $\lambda$ obtained from the Monte-Carlo Radiative Transfer simulation with POLARIS (\citealt{Reissl_2014}, \citealt{Reissl_2016}). $Q_{\Gamma}(a_{\rm eff}, \lambda)$ is the RAT efficiency which is a function of wavelength $\lambda$ and effective size $a_{\rm eff}$. Following \cite{Hoang_2019}, $Q_{\Gamma}$ can be considered as a constant of $\approx 0.4$ for grains of size $a_{\rm eff} \ge \lambda /1.8$, and $Q_{\Gamma}$ decreases with decreasing grain sizes as $Q_{\rm RAT} \sim 0.4 (\lambda / (1.8 a_{\rm eff}))^{-3}$ for $a_{\rm eff} \leq \lambda / 1.8$. Equation (\ref{eq:gamma_rat}) will be integrated over the radiation spectrum from $\lambda_{\rm min} = 0.1\mum$ to $\lambda_{\rm max} = 3$ mm (Section \ref{sec:MHD+radiation_field}) to find the net RAT efficiency acting on dust grains.

The equation of rotational motion under RATs is given by (\citealt{Hoang_Lazarian_2014}):
\bea 
\frac{I d\Omega}{dt} = \Gamma_{\rm RAT} - \frac{I \Omega}{\tau_{\rm damp}}, \label{eq:domega/dt}
\ena
where $\Omega$ is the grain angular velocity, $I = 8/15 \rho_{\rm grain} \pi s a^{5}$ is the inertia moment of oblate grain, and $\tau_{\rm damp} = \tau_{\rm gas}/(1+\rm FIR)$ characterizes the damping timescale of rotating grain. The first term $\tau_{\rm gas}$ characterizes the damping timescale caused by gas-grain collisions followed by gas evaporation, while the second dimensionless parameter $\rm FIR$ characterizes the damping coefficient caused by thermal dust emission. In protostellar environments, the high gas density makes thermal dust emission negligible to gas-grain collisions (or $FIR << 1$) in reducing the grain rotational energy. Thus, grains are majorly slowed down by gas collisions, with the typical timescale of (\citealt{Hoang_2019}):
\bea 
\tau_{\rm gas} = \frac{3}{4 \sqrt{\pi}}\frac{I}{n_{\rm H_{2} m_{\rm H_{2}} v_{\rm th} a^{4} \Gamma_{\parallel}}},\label{eq:tau_gas}
\ena,
where $n_{\rm H_{2}}$ is the gas volume density of hydrogen molecules, $v_{\rm th} = \sqrt{k_{\rm B}T_{\rm gas}/I}$ is the thermal grain velocity with $k_{\rm B}$ the Boltzmann constant, $T_{\rm gas}$ the gas temperature, and $\Gamma_{\parallel}$ is the geometrical factor of oblate grains (see also \citealt{Hoang+2022}).

By solving Equation (\ref{eq:domega/dt}), the angular velocity after $t$ time, $\Omega(t)$, is given by (Section \ref{sec:omega_t}):

\bea 
\Omega(t) = \frac{\tau_{\rm damp}}{I} \Bigg[\Gamma_{\rm RAT} - \Bigg(\Gamma_{\rm RAT} - \frac{I \Omega(t_{0})}{\tau_{\rm damp}}\Bigg)e^{-\frac{\rm t}{\tau_{\rm damp}}}\Bigg] .
\ena 

\section{Grain dynamic under RATD activities}\label{sec:appen_omega_t}
To illustrate the spinup of RATs while grains are propagating inside jets and outflow cavities better (Figure \ref{fig:wt_grain_size}), we show in Figure \ref{fig:wt_10um} the variation of $\Omega(t)$ with time for $10\mum$ grains in solid black lines. The corresponding black dashed line illustrates the variation of $\Omega_{\rm rest}(t)$ if grains are at rest, i.e., $v_{\rm grain} = 0$ km/s. The new position of grains with time is marked on the right y-axis (blue dashed lines), and their trajectory (from white to black) is marked on the density map placed in the corner of each panel. The critical disruption thresholds $\Omega_{\rm disr}$ for different $S_{\rm max} = 10^{3}-10^{7}\erg\cm^{-3}$ are plotted in color lines, and grains will be disrupted by RATD when $\Omega(t) \geq \Omega_{\rm disr}$.  We show in the upper row the evolution of $\Omega(t)$ for  $10\mum$ grains inside the jet base (left panel) and outflow base (right panel). The middle row shows results for $10\mum$ grains propagating inside the jet with few hundreds km/s (left panel) and inside the outflow cavity with few tens km/s (right panel). And the lower row illustrates $\Omega(t)$ for $100\mum$ grains propagating inside the outflow cavity wall (left panel) and infalling inside the inner envelope (right panel). The initial distance of grains to the center $r$, the gas density, and the corresponding gas velocity and grain velocity are noted in the lower left corner.

One can see that in all panels, $\Omega(t)$ and $\Omega_{\rm rest}(t)$ share similar values when grains start moving outward owing to the high radiation field strength. For the time being, $\Omega(t)$ increases slower (and even decreases if grains enter dense clumps, as shown in the lower row \footnote{We note that $10\mum$ grains inside the jet and outflow base are already destroyed by RATD before $\Omega(t)$ decreasing, so this issue does not affect the map of $a_{\rm disr,dynamic}$ shown in Figure \ref{fig:adisr_min_dynamic}. But if the disruption threshold has not yet been met, the reduction of $\Omega(t)$ by increasing gas density may allow grains to survive from RATD}) owing to decreased radiation field strength with distances. In contrast to this trend, $\Omega_{\rm rest}(t)$ continues to increase until reaching the saturated values. That explains why RATD becomes weaker when we take the propagation of dust grains inside jets and outflow cavities into account (Section \ref{sec:adisr_polaris_dynamic}). 

Inside the jet/outflow base (upper row), RATD can immediately destroy $10\mum$ grains having $S_{\rm max} \leq 10^{6}\erg\cm^{-3}$ just after $< 0.5$ yrs. For grains that already uplifted to $z = 300$ au before RATD happens (middle row), $10\mum$ grains having $S_{\rm max} = 10^{3}$ and $10^{4}\erg\cm^{-3}$ takes $\sim 1$ yr and $\sim 5$ yr to be destroyed by RATD if they propagate with $v_{\rm gas} \sim 273$ km/s along the z$-$direction (left panel). Grains having $S_{\rm max} \geq 10^{5}\erg\cm^{-3}$ inside the jet are not affected by RATD because $\Omega(t)$ being saturated at values below their disruption threshold after $\sim 10$ yr (Figure \ref{fig:wt_grain_size}, middle left panel). In contrast, the angular velocity of grains propagating inside the outflow cavity (middle right panel) takes $\geq 100$ yr to be saturated because of lower $v_{\rm grain}$. That explain why RATD can affect $10\mum$ grains having $S_{\rm max} \sim 10^{3}-10^{6}\erg\cm^{-3}$ in this area (Figure \ref{fig:wt_grain_size}, middle right panel). However, in case $S_{\rm max} = 10^{5}\erg\cm^{-3}$, grains already move from $z = 300$ au to $z = 600$ au within $\sim 6$ yr before RATD affecting them. Taking the decoupling and redirection of VLGs by centrifugal forces into account, $10\mum$ may already enter the outflow cavity wall where RATD is deactivated for grains having $S_{\rm max} \geq 10^{5}\erg\cm^{-3}$. That explains why in Section \ref{sec:discuss_RATD_migration}, we state that RATD may not suppress the migration of this class of VLGs beyond the outflow base. 

RATD happens much weaker inside the outflow cavity wall (lower left panel) and the inner envelope (lower right panel) because of the weak radiation field strength and high gas damping there. Only $10\mum$ having $S_{\rm max} = 10^{3}$ and $10^{4}\erg\cm^{-3}$ can be destroyed by RATD. The disruption timescale will be $\sim 10$ and $\sim 50$ yr for $S_{\rm max} = 10^{3}\erg\cm^{-3}$, and is longer, $\sim 40$ and $300$ yr for $S_{\rm max} = 10^{4}\erg\cm^{-3}$.  

 \begin{figure*}
\centering
    \includegraphics[width=\textwidth,height=\textheight,keepaspectratio]{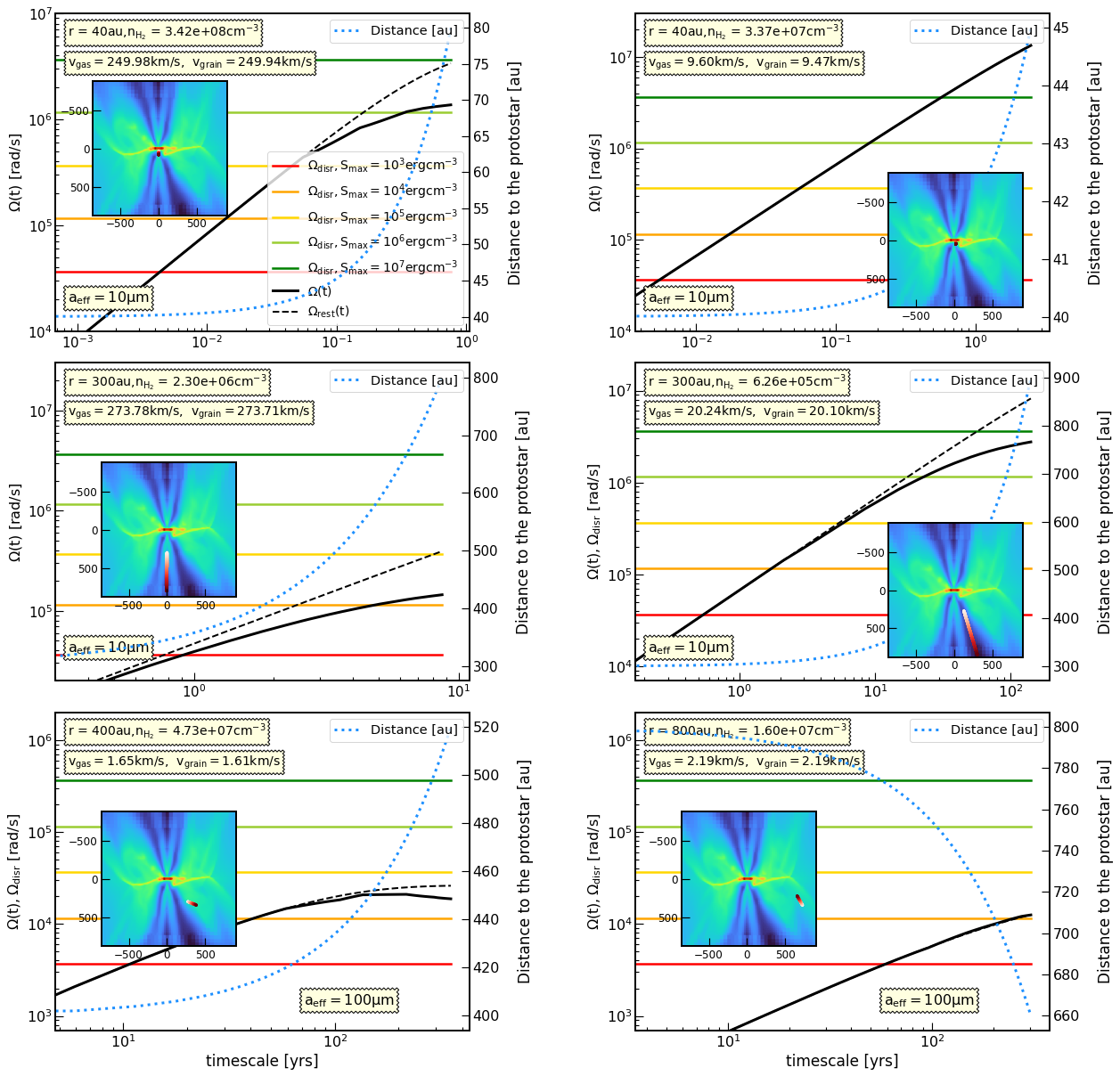}
    \caption{Evolution of the angular velocity $\Omega$ with times (black lines) for $10\mum$ grains along with their transportation (marked by the blue dashed line in the right y-axis) inside the outflow. The back dash lines show the variation of $\Omega_{\rm rest}$ with time, if grains are at rest in their initial position. The left column illustrates the dynamic of grains moving with few hundreds km/s inside the jet lobe. The right column relates to grains moving with few tens km/s near the outflow cavity wall. The upper row shows results for grains starting from the outflow/jet base within 50 au to the protostar, while the lower row presents results for grains starting at 500 au to the center. The trajectory of grains is marked on the density map placed in the upper left corner. The disruption threshold of $10\mum$ grains $\Omega_{\rm disr}$ for different values of $S_{\rm max}$ is marked by color lines. Grains will be disrupted by RATD at the moment when $\Omega(t) > \Omega_{\rm disr}$.} 
     \label{fig:wt_10um}
\end{figure*}

\section{Polarized radiative transfer of Stokes parameters under RATD activities}\label{sec:appen_stokes}
As discussed in Section \ref{sec:polaris_stokes}, the change in grain size distribution by RATD will change the interaction cross-section between dust grains and the radiation field. To understand the effect of RATD on polarized dust emission, we need to calculate the new extinction and absorption efficiency and also the emissivity of thermal grains and aligned dust grains. As shown in Equation (\ref{eq:polarized_radiative_transfer}, the polarized radiative transfer of Stokes parameter when photons propagate inside the dusty environments is given by:

\begin{gather}
    \frac{d}{ds} 
    \begin{pmatrix}
     I\\
     Q\\
     U\\
     V
    \end{pmatrix}
=
-\begin{pmatrix}
    \alpha_{\rm I} && \alpha_{\rm Q} && 0 && 0 \\
    \alpha_{\rm Q} && \alpha_{\rm I} && 0 && 0 \\
    0 && 0 && \alpha_{\rm I} && \kappa_{\rm Q}  \\
    0 && 0 && -\kappa_{\rm Q} && \alpha_{\rm I} \\
\end{pmatrix}
\begin{pmatrix}
 I\\
 Q\\
 U\\
 V
\end{pmatrix}
+ 
\begin{pmatrix}
 j_{\rm I} \\
 j_{\rm Q} \\
 0 \\
0
 \label{eq:polarized_radiative_transfer_appendix}
\end{pmatrix},
\end{gather}

where $\alpha_{\rm I}$, $\alpha_{\rm Q}$, $\kappa_{\rm Q}$ are the extinction coefficient, linear polarization coefficient, and circular polarization coefficient integrated over grain size distribution. $j_{\rm I}$ and $j_{\rm Q}$ are the emissivity of thermal grains and aligned dust grains, respectively.

The extinction coefficient $\alpha_{\rm I}$ at wavelength $\lambda$ before RATD changes the dust population is given by:
 \bea 
\alpha_{\rm I} = \int_{a_{\rm min}}^{a_{\rm max}} C_{\rm ext} \frac{dn}{da} da = n_{\rm H_{2}} C \int_{a_{\rm min}}^{a_{\rm max}}  C_{\rm ext}(a,\lambda) a^{-3.5} da,
\label{eq:alpha_I}
\ena
where $C_{\rm ext}$ is the extinction cross-section of grains of size $a$ with wavelength $\lambda$, which is:

\bea 
C_{\rm ext} = \frac{(4+3R\cos^{2}\psi-R) C_{\rm ext,\|}}{6} + \frac{(2 - 3\cos^{2}\psi + R)C_{\rm ext, \perp}}{6}.
\label{eq:Cext}
\ena

In above equation, $C_{\rm ext,\|}$ and $C_{\rm ext, \perp}$ are the extinction cross-section when the electric field component \textbf{E} of incoming photon oscillating parallel and perpendicular with the grain major axis, $\psi$ is the angle between the magnetic field direction and the observed LOS, and $R$ is the Rayleigh reduction factor that describes the net internal and external alignment degree of grains with magnetic fields. For grains within the alignment range $a_{\rm align} - a_{\rm max,JB}^{\rm Lar}$, the formula of $R$ following the RAT alignment theorem is given by:
\bea 
R = f_{\rm high-J} Q_{\rm X}^{\rm high-J} Q_{\rm J}^{\rm high-J} + (1-f_{\rm high-J})Q_{\rm X}^{\rm low-J} Q_{\rm J}^{\rm low-J},
\label{eq:R}
\ena
where $Q_{\rm X}^{\rm high-J}$ and $Q_{\rm J}^{\rm high-J}$ describe the internal and external alignment degree of grains aligning with $\B$ at high-\textit{J} attractors. While $Q_{\rm X}^{\rm low-J}$ and $Q_{\rm J}^{\rm low-J}$ describe the alignment degree of grains aligning with $\B$ at low-\textit{J} (our \citealt{Giang_et_al_2024}). For grains beyond the alignment range, $R$ will be set to zero to describe the random orientation of dust grains.

When RATD destroys a fraction of $f_{\rm high-J}$ of grains at high-\textit{J} attractors, the Rayleigh reduction factor will be reduced to:
\bea 
R_{\rm new} = (1-f_{\rm high-J})Q_{\rm X}^{\rm low-J} Q_{\rm J}^{\rm low-J},
\label{eq:R_new}
\ena
because only the remained $(1-f_{\rm high-J})$ of grains at low-\textit{J} contribute to produce polarized dust emission. By placing $R$ in Equation (\ref{eq:Cext}) by $R_{\rm new}$, one gets the new extinction cross-section $C_{\rm ext}^{\rm new}$, and then the new extinction efficiency under the change in dust population and grain alignment degree by RATD, which is given by:
\bea
\alpha_{\rm I}^{\rm new}  = n_{\rm H_{2}} C ~\Bigg [ \int_{\rm a_{\rm min}}^{\rm a_{\rm disr,POLARIS}} C_{\rm ext} a^{\alpha} da + \int_{\rm a_{\rm disr,POLARIS}}^{\rm a_{\rm max}} C_{\rm ext}^{\rm new} (1-f_{\rm disr}(a))  a^{-3.5} da \Bigg] .\label{eq:alpha_I_new}
\ena
The first term describes the extinction cross-section produced by small grains following the new size distribution $dn/da  \sim a^{\alpha}$ (Section \ref{sec:polaris_eta}. And the second term describes the extinction cross-section of the remaining large grains which are not affected by RATD.

For the linear polarization efficiency $\alpha_{\rm Q}$, their formula before RATD happens is given by:
\bea 
\alpha_{\rm Q}  = \int_{a_{\rm min}}^{a_{\rm max}} C_{\rm pol}^{\rm ext} \frac{dn}{da} da,
\label{eq:alpha_Q}
\ena
where $C_{\rm pol}^{\rm ext}$ is the linear polarized extinction cross-section, which is:
\bea 
C_{\rm pol}^{\rm ext} = \frac{(C_{\rm ext,\|} - C_{\rm ext,\perp}) R \sin^{2}\psi }{2}.
\label{eq:Cpol}
\ena

Following the similar method used to calculate $\alpha_{\rm I}^{\rm new}$, we replace $R_{\rm new}$ (Equation \ref{eq:R_new}) to Equation (\ref{eq:Cpol}) to get new $C_{\rm ext}^{\rm pol,new}$, then use it to replace $C_{\rm ext}^{\rm new}$ in Equation (\ref{eq:alpha_I_new}) to get the new linear polarization efficiency $\alpha_{\rm Q}^{\rm new}$. The new fomular of $\alpha_{\rm Q}^{\rm new}$ is given by:
\bea
\alpha_{\rm Q}^{\rm new}  = n_{\rm H_{2}} C \Bigg[\int_{\rm a_{\rm min}}^{\rm a_{\rm disr,POLARIS}} C_{\rm ext}^{\rm pol}  a^{\alpha} da + \int_{\rm a_{\rm disr,POLARIS}}^{\rm a_{\rm max}} C_{\rm ext}^{\rm pol,new} (1-f_{\rm disr}(a)) a^{-3.5} da\Bigg].\label{eq:alpha_Q_new}
\ena

The circular polarization efficiency $\kappa_{\rm Q}$ is given by:
\bea 
\kappa_{\rm Q}  = \int_{a_{\rm min}}^{a_{\rm max}} C_{\rm circ} \frac{dn}{da} da,
\label{eq:kappa_Q}
\ena
where $C_{\rm circ} = \frac{C_{\rm circ}' R \sin^{2}\psi }{2} $ is the circular polarization cross-section of aligned dust grains and $C_{\rm circ}'$ the grain intrinsic circular polarization cross-section. Following the similar method used to determine $\alpha_{\rm I}^{\rm new}$ and $\alpha_{\rm Q}^{\rm new}$, the new circular polarization efficiency $\kappa_{\rm Q}^{\rm new}$ during RATD period changes to:
\bea
\kappa_{\rm Q}^{\rm new}  = n_{\rm H_{2}} C \Bigg[\int_{\rm a_{\rm min}}^{\rm a_{\rm disr,POLARIS}} C_{\rm circ}  a^{\alpha} da + \int_{\rm a_{\rm disr,POLARIS}}^{\rm a_{\rm max}} C_{\rm circ}^{\rm new} (1-f_{\rm disr}(a)) a^{-3.5} da\Bigg],\label{eq:kappa_Q_new}
\ena
with $C_{\rm circ}^{\rm new} = \frac{C_{\rm circ}' R_{\rm new} \sin^{2}\psi }{2} $.

For the second matrix in Equation (\ref{eq:polarized_radiative_transfer_appendix}), the emissivity $j_{\rm I}$ and polarized emissivity $j_{\rm Q}$ without RATD effect is given by:
\bea 
j_{\rm I} =  \int_{a_{\rm min}}^{a_{\rm max}} C_{\rm abs} B_{\lambda}(T_{\rm d}(a)) \frac{dn}{da} da,
\label{eq:j_I}
\ena
and

\bea 
j_{\rm Q} =  \int_{a_{\rm min}}^{a_{\rm max}} C_{\rm abs}^{\rm pol} B_{\lambda}(T_{\rm d}(a)) \frac{dn}{da} da,
\label{eq:j_I}
\ena

where $C_{\rm abs}$ and $C_{\rm abs}^{\rm pol}$ is the absorption and polarization absorption cross-section, and $B_{\lambda}(T_{\rm d}(a)$ is the Planck emission of dust grains with temperature $T_{\rm d}(a)$ at wavelengths $\lambda$. $C_{\rm abs}$ and $C_{\rm abs}^{\rm pol}$ share similar formulas with $C_{\rm ext}$ and $C_{\rm ext}^{\rm pol}$ (Equations \ref{eq:Cext} and \ref{eq:Cpol}), with $C_{\rm ext,\parallel}$ and $C_{\rm ext,\perp}$ being replaced by $C_{\rm abs,\parallel}$ and $C_{\rm abs, \perp}$.

Following similar methods used to calculate $C_{\rm ext}^{\rm new}$ and $C_{\rm ext}^{\rm pol,new}$, one can get the new $C_{\rm abs}^{\rm new}$ and $C_{\rm abs}^{\rm pol,new}$. The new emissivity and polarized emissivity under RATD effect will be given by:
\bea
j_{\rm I}^{\rm new}  = n_{\rm H_{2}} C ~\Bigg[\int_{\rm a_{\rm min}}^{\rm a_{\rm disr,POLARIS}} C_{\rm abs}  B_{\lambda}(T_{\rm d}(a)) a^{\alpha} da
+ \int_{\rm a_{\rm disr,POLARIS}}^{\rm a_{\rm max}} C_{\rm abs}^{\rm new} (1-f_{\rm disr}(a)) B_{\lambda}(T_{\rm d}(a)) a^{-3.5} da\Bigg],\label{eq:j_I_new}
\ena
and 
\bea
j_{\rm Q}^{\rm new} = n_{\rm H_{2}} C ~\Bigg[\int_{\rm a_{\rm min}}^{\rm a_{\rm disr,POLARIS}} C_{\rm abs}^{\rm pol}  B_{\lambda}(T_{\rm d}(a)) a^{\alpha} da + \int_{\rm a_{\rm disr,POLARIS}}^{\rm a_{\rm max}} C_{\rm abs}^{\rm pol, new}  (1-f_{\rm disr}(a)) B_{\lambda}(T_{\rm d}(a)) a^{-3.5} da\Bigg].\label{eq:j_Q_new}
\ena

\section{Spatial distribution from POLARIS}\label{sec:appen_distribution_polaris}

Figure \ref{fig:adisr_midplane} shows the spatial distribution of the minimum disruption size $a_{\rm disr,POLARIS}$ within 2000 au on the midplane containing the sink particle. The distribution is a result from POLARIS approach (Section \ref{sec:polaris_adisr}), assuming the maximum tensile strength of grains of $S_{\rm max} = 10^{3}, 10^{4}, 10^{5}, 10^{6}$ and $10^{7}\erg\cm^{-3}$, from left to right, respectively. The maximum grain size shown in the figure is $a_{\rm max} = 50\mum$. Empty cells indicate the region where RATD does not destroy grains.

 \begin{figure*}
\centering
    \includegraphics[width=\textwidth,height=\textheight,keepaspectratio]{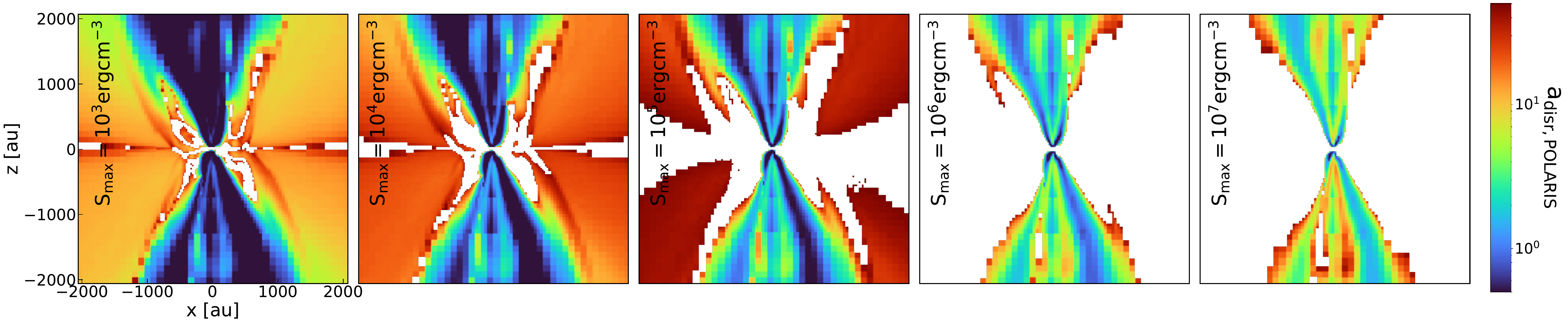}
    \caption{ Spatial distribution of the minimum disruption size $a_{\rm disr,POLARIS}$ obtained from POLARIS with different values of $S_{\rm max} = 10^{3}-10^{7}\erg\cm^{-3}$, from left to right, respectively. We show results on the midplane containing the sink particle, with the spatial scale of 2000 au around the protostar. The empty region illustrates the area without RATD activities. More sub-micron grains inside the jet and outflow cavity will be destroyed by RATD by decreasing $S_{\rm max}$. RATD can constrain the maximum grain size to $\sim 0.5\mum$ if grains have $S_{\rm max} \leq 10^{5}\erg\cm^{-3}$ (first, second, and third panel) and $\sim 3-4\mum$ if grains have $S_{\rm max} \sim 10^{7}\erg\cm^{-3}$ (fifth panel) (similar to results shown in Figure \ref{fig:adisr_nH_vgas}. VLGs beyond $10\mum$ can be removed from the outflow cavity wall, inner envelope, and envelope if the maximum tensile strength reduces to below $10^{5}\erg\cm^{-3}$. RATD does not affect dust grains on the equatorial midplane because of the stronger dust reddening effect along this direction and extra damping on the RAT efficiency by the perpendicular between radiation and magnetic field there.}
     \label{fig:adisr_midplane}
 
\centering
    \includegraphics[width=\textwidth,height=\textheight,keepaspectratio]{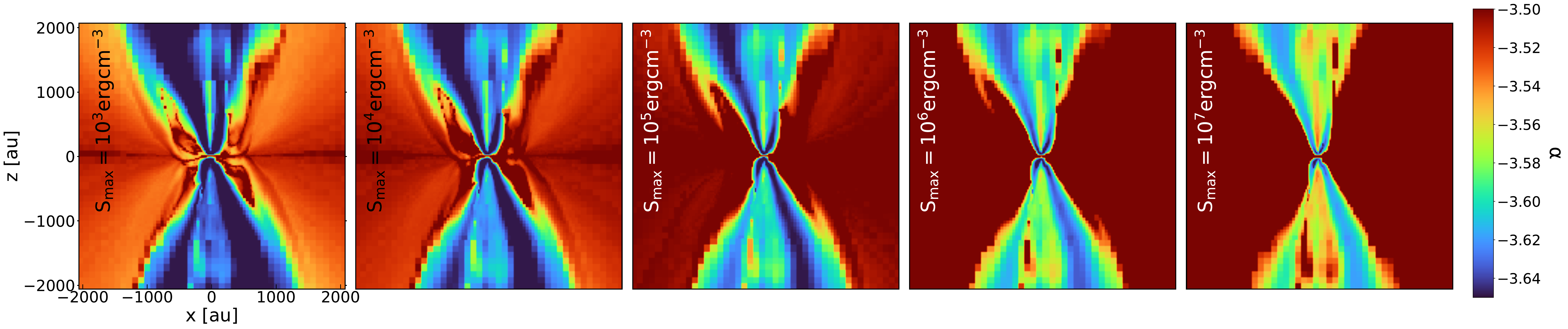}
    \caption{Grain size distribution of small grains below $a_{\rm disr,POLARIS}$, $\alpha$, constrained by RATD within 2000 au around the protostar. Grains inside the region without being affected by RATD (empty regions in Figure \ref{fig:adisr_midplane}) have the standard power index of MRN distribution of $\alpha = -3.5$, while grains below $a_{\rm disr,POLARIS}$ in the active region of RATD will have higher negative $\alpha$ due to the fragmentation of large grains beyond $a_{\rm disr,POLARIS}$ (shown in Figure \ref{fig:adisr_midplane}) to smaller sizes. For grains having $S_{\rm max} = 10^{3}\erg\cm^{-3}$ (first panel), sub-micron grains below $a_{\rm disr,POLARIS} \sim 0.5\mum$ inside the outflow cavity will follow the steep distribution with $\alpha \sim -3.64$, while micron-sized grains below $a_{\rm disr,POLARIS} \sim 10\mum$ inside the outflow cavity wall, inner envelope, and envelope will follow the distribution with $\alpha \sim -3.54$. The size distribution becomes shallower with increasing $S_{\rm max}$. For grains having $S_{\rm max} = 10^{7}\erg\cm^{-3}$ (fifth panel), only sub-micron and micron-sized grains below $a_{\rm disr,POLARIS} \sim 3-4\mum$ inside the outflow cavity will follow the steeper distribution with $\alpha \sim -3.6$. Dust grains in the remaining part still follow the standard MRN distribution with $\alpha = -3.5$.}
     \label{fig:eta_midplane}
\end{figure*}

Similar to the minimum disruption size obtained from the dynamic approach (Figures \ref{fig:adisr_min_dynamic} and \ref{fig:adisr_nH_vgas}, upper row), dust grains will be destroyed by RATD stronger when having lower maximum tensile strength. One gets $a_{\rm disr,POLARIS} \sim 0.05\mum$ inside the jet and outflow cavity for grains having $S_{\rm max} \sim 10^{3}-10^{4}\erg\cm^{-3}$. The minimum disruption size in these regions increases to $\sim 1\mum$ for grains having $S_{\rm max} \sim 10^{4}-10^{5}\erg\cm^{-3}$ and to $\sim 2-4\mum$ for grains having $S_{\rm max} = 10^{7}\erg\cm^{-3}$. Inside the outflow cavity wall, inner envelope, and envelope where $n_{\rm H_{2}} \geq 10^{7}\cm^{-3}$, VLGs beyond $10\mum$ can be removed by RATD if $S_{\rm max} \sim 10^{3}-10^{4}\erg\cm^{-3}$. RATD can destroy VLGs beyond $\geq 30\mum$ in the envelope beyond $500$ au if $S_{\rm max} = 10^{5}\erg\cm^{-3}$, but cannot touch to VLGs having $S_{\rm max} \geq 10^{6}\erg\cm^{-3}$. Dust grains inside the protostellar disk and equatorial midplane are not affected by RATD regardless of $S_{\rm max}$ because of the strong dust reddening effect and the RAT damping caused by the perpendicular between radiation and magnetic fields.

\begin{figure*}
\centering
    \includegraphics[width=\textwidth,height=\textheight,keepaspectratio]{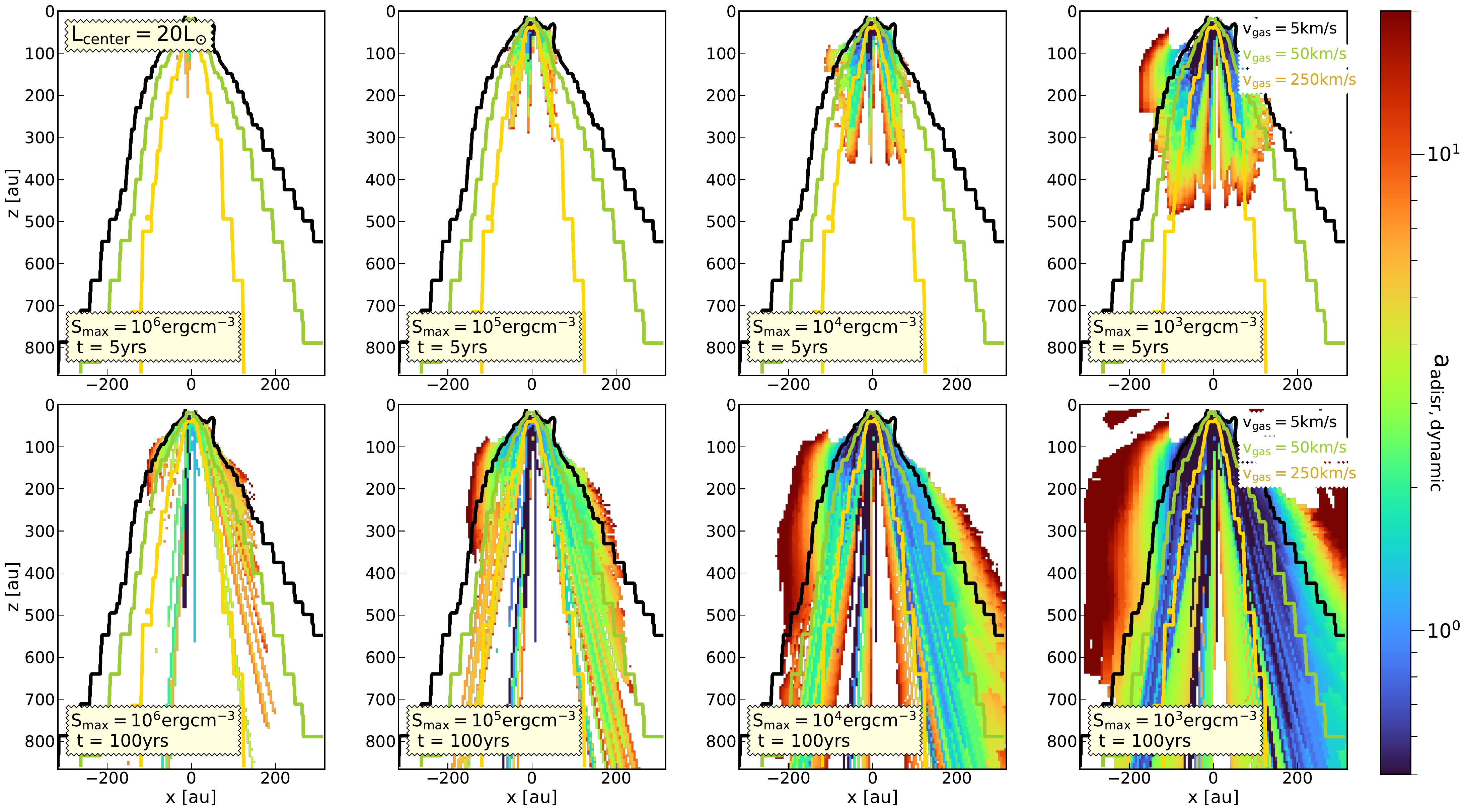}
    \caption{Evolution of the minimum disruption size obtained from the dynamic approach $a_{\rm disr,dynamic}$ inside the jet and outflow after RATD is turned on 5 yr and 100 yrs, for grains having $S_{\rm max} = 10^{3}-10^{6}\erg\cm^{-3}$. The adopted bolometric luminosity is $L_{\rm center} = 20L_{\odot}$. The contour shows the gas velocity distribution, and empty cells show regions unaffected by RATD. With lower $20L_{\odot}$, RATD do not prevent the migration of dust grains having $S_{\rm max} \geq 10^{5}\erg\cm^{-3}$ inside the jet and outflow. For grains having $S_{\rm max} = 10^{4}\erg\cm^{-3}$, RATD can destroy the presence of large grains beyond $2\mum$ from the outflow cavity, but can not touch to large grains propagating inside the jet with few hudnreds km/s. For grains having $S_{\rm max} = 10^{4}\erg\cm^{-3}$, one gets $a_{\rm disr,dynamic} \sim 1\mum$ inside the outflow cavity and $\sim 3\mum$ inside the jet.}
     \label{fig:RATD_20Lsun}
\centering
    \includegraphics[width=\textwidth,height=\textheight,keepaspectratio]{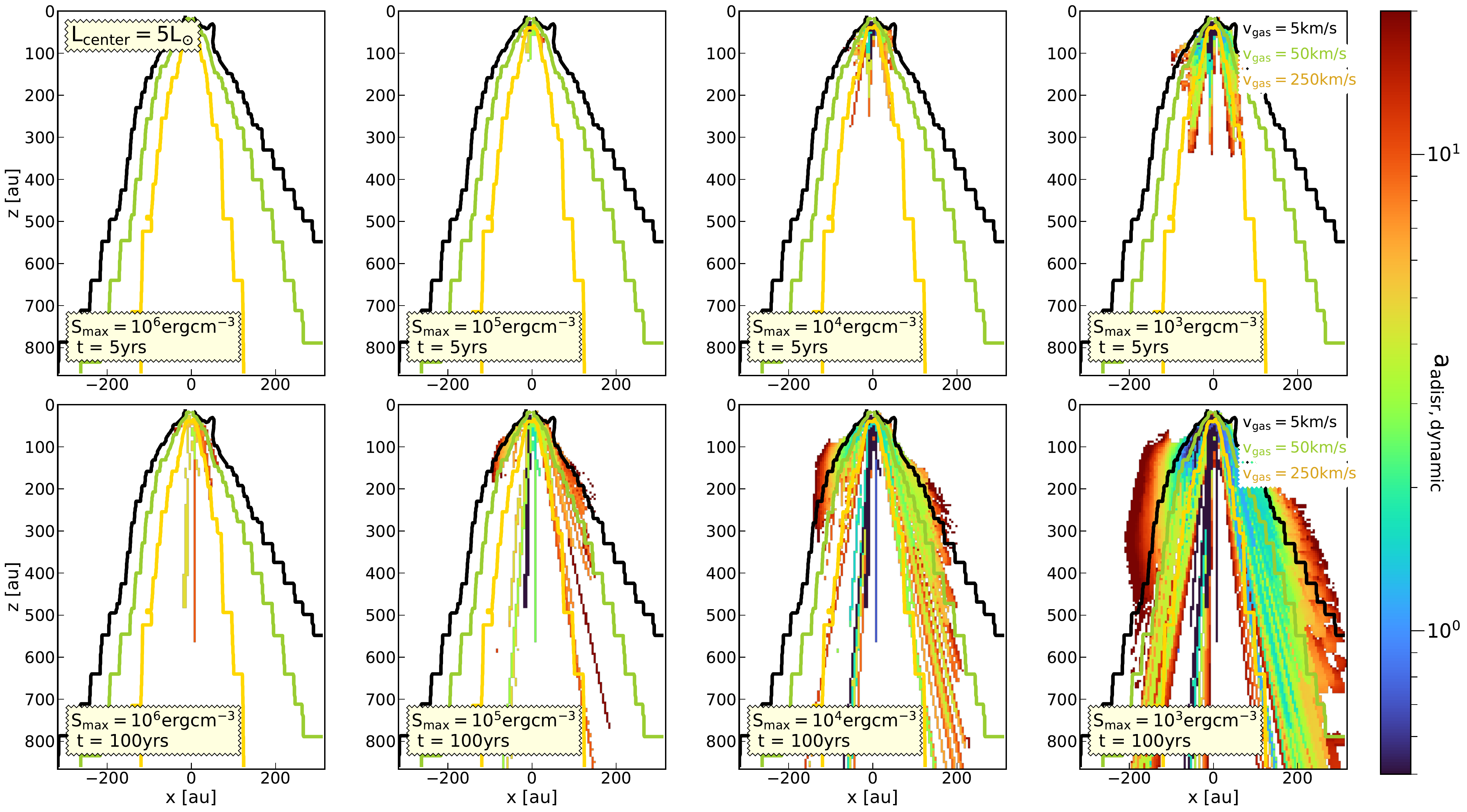}
    \caption{Similar results to Figure \ref{fig:RATD_5Lsun} but for the bolometric luminosity of $5L_{\odot}$. RATD is not operated for dust grains having $S_{\rm max} \geq 10^{4}\erg\cm^{-3}$. It can only suppress the migration of large grains having $S_{\rm max} = 10^{3}\erg\cm^{-3}$ beyond $\geq 3\mum$ from the outflow cavity.} 
     \label{fig:RATD_5Lsun}
\end{figure*}

\begin{figure*}
\centering
    \includegraphics[width=\textwidth,height=\textheight,keepaspectratio]{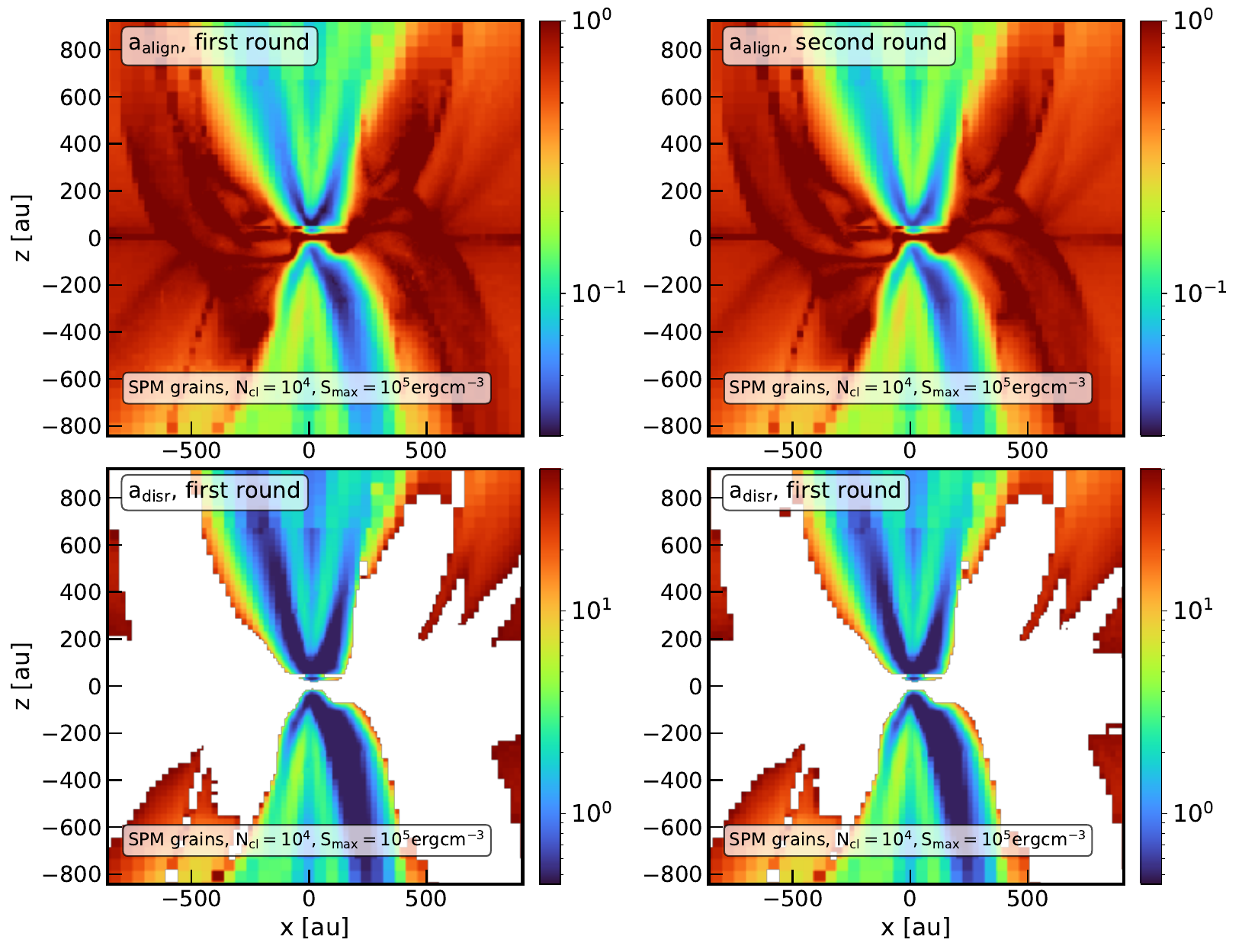}
    \caption{Comparison of the alignment and disruption state before (left column) and after (right column) taking into account the change in grain size distribution by RATD when performing 3D MCRT. The upper row shows results for the minimum alignment size $a_{\rm align}$, and the lower row relates to the minimum disruption size $a_{\rm disr,POLARIS}$, considering SPM grains with $N_{\rm cl} = 10^{4}$ and $S_{\rm max} = 10^{5}\erg\cm^{-3}$. In the second MCRT simulation, a lower fraction of sub-micron grains inside the outflow cavity can be aligned with $\B$ and a lower fraction of micron-sized grains can be rotationally disrupted owing to the increased UV-optical-NIR dust extinction by enhanced amount of small grains by RATD. However, the difference between the two cases is insignificant due to the small change in the size distribution of small grains shown in Figure \ref{fig:eta_midplane}.} 
     \label{fig:2nd_time}
\end{figure*}
 
As discussed in Section \ref{sec:polaris_eta}, the partial conversion of large grains (a fraction of $f_{\rm disr}$) to smaller sizes by RATD may change the size distribution of grains below $a_{\rm disr,POLARIS}$. We show the spatial distribution of the new power law $\alpha$ for such enhanced small grains in Figure \ref{fig:eta_midplane}. For grains having $S_{\rm max} = 10^{3}\erg\cm^{-3}$ (first panel), sub-micron grains below $a_{\rm disr,POLARIS} \sim 0.5\mum$ inside the jet and outflow cavity will follow the size distribution with slope $\alpha \sim -3.64$, micron-sized grains below $10\mum$ inside the outflow cavity, inner envelope, and envelope will follow the distribution with slope $\alpha \sim -3.54$. The size distribution will be shallower by increasing the maximum tensile strength that narrows the disruption range. For grains having $S_{\rm max} = 10^{7}\erg\cm^{-3}$, only sub-micron and micron-sized grains below $a_{\rm disr,POLARIS} \sim 2-3\mum$ inside the jet and outflow cavity follow the modified size distribution with $\alpha \sim -3.6$. Dust grains beyond the outflow cavity will keep their standard MRN distribution with $\alpha = -3.5$ considered in our model.
  
\section{Effect of bolometric luminosity on the destruction of dust grains by RATD}\label{sec:appen_Lstar}
To understand the condition for activating RATD inside intermediate-mass protostar, we show in this section the minimum disruption size $a_{\rm disr,dynamic}$ obtained inside the jet and outflow in case $L_{\rm center} = 20L_{\odot}$ (Figure \ref{fig:RATD_20Lsun}) and $L_{\rm center} = 5L_{\odot}$ (Figure \ref{fig:RATD_5Lsun}). We consider the dynamic approach (Section \ref{sec:vgrain_omegat}) and apply it to model RATD activities for grains having $S_{\rm max} = 10^{6}, 10^{5}, 10^{4}, 10^{3}\erg\cm^{-3}$ after 5 and 100 yrs. The map focuses on $600\times 800$ au below the protostar. 

With lower bolometric luminosity, RATD remove the appearance of large micron-sized grains inside the jet and outflow weaker, i.e., larger $a_{\rm disr,dynamic}$, owing to the inefficient RAT acting on dust grains. For $L_{\rm center} = 20L_{\odot}$ (Figure \ref{fig:RATD_20Lsun}), RATD can make micron-sized grains having $S_{\rm max} \sim 10^{3}-10^{4}\erg\cm^{-3}$ and size below $a_{\rm disr,dynamic}\sim 1 -3\mum$ to dominant the outflow cavity after $\sim 100$ yr (third and fourth columns). But it cannot totally remove the existence of large micron-sized grains having $S_{\rm max} \geq 10^{5}\erg\cm^{-3}$ from the outflow cavity (first and second columns). For grains propagating with a few hundred km/s along the jet, except grains having $S_{\rm max} = 10^{3}\erg\cm^{-3}$, large grains can freely migrate outward regardless of the operation of RATD inside the core. In case of $L_{\rm center} = 5L_{\odot}$ (Figure \ref{fig:RATD_5Lsun}), RATD can only let micron-sized grains having $S_{\rm max} = 10^{4}\erg\cm^{-3}$ (fourth column) below $a_{\rm disr,dynamic} \sim 3\mum$ to dominant the dust population inside the outflow cavity. So except for grains having $S_{\rm max} \leq 10^{4}\erg\cm^{-3}$, RATD will be terminated when the bolometric luminosity reduces to below $20L_{\odot}$, similar to the finding in \cite{Valentin_2023b}.

\section{Alignment and Disruption state under the change of dust opacity by RATD}\label{sec:appen_RATD_more_step}
As discussed in Section \ref{sec:polaris_stokes}, modifying grain size distribution under RATD activities will change the spectral energy density distribution acting on protostellar dust grains. More IR photons (mid-infrared (MIR) to far-infrared (FIR)) can escape due to the removal of VLGs, but ultraviolet (UV) to near-IR (NIR) photons will be extinct stronger owing to the enhancement of small grains. Consequently, it can modify the grain alignment and disruption size and change the behavior of polarized dust emission. To take into account this issue, we recalculate the dust extinction and scattering cross-section over size distribution at wavelength $\lambda$ $\langle C_{\rm ext} \rangle$ and $\langle C_{\rm sca} \rangle$ used to determine the photon-dust interactions during the MCRT simulation (see \citealt{Lucy_1999}, \citealt{Reissl_2014}, \citealt{Reissl_2016} for detailed MCRT calculation).

Since MCRT simulation does not care in detail about the alignment state of dust grains (\citealt{Lucy_1999}), the change of the average extinction and scattering cross-sections under RATD activities can be described as follows:

\bea 
\langle C_{\rm ext} \rangle \approx \int_{\rm a_{\rm min}}^{\rm a_{\rm disr,POLARIS}} C_{\rm ext} a^{\alpha} da  
+ \int_{\rm adisr,POLARIS}^{\rm a_{\rm max}} (1 - f_{\rm disr}(a)) C_{\rm ext} a^{-3.5} da,
\ena

\bea 
\langle C_{\rm sca} \rangle \approx \int_{\rm a_{\rm min}}^{\rm a_{\rm disr,POLARIS}} C_{\rm sca} a^{\alpha} da + \int_{\rm adisr,POLARIS}^{\rm a_{\rm max}} (1 - f_{\rm disr}(a)) C_{\rm sca} a^{-3.5} da,
\ena
where $C_{\rm ext} = (2C_{\rm ext,\parallel} + C_{\rm ext,\perp})/3$ and $C_{\rm sca} = (2C_{\rm sca,\parallel} + C_{\rm sca,\perp})/3$ are the extinction and scattering cross-section of random orient grain of size $a$.

Knowing the new average extinction and scattering cross-section, we perform the MCRT simulation again and use the new radiation field distribution to determine the new dust temperature, grain alignment, and grain disruption sizes.
 
The change of $a_{\rm align}$ (upper row) and $a_{\rm disr,POLARIS}$ (lower row) after the first round (left column) and second round (right column) of MCRT simulation is illustrated in Figure \ref{fig:2nd_time}. Here we show results for SPM grains with high $N_{\rm cl} = 10^{4}$ and low $S_{\rm max} = 10^{5}$, in which all micron-sized grains above $1\mum$ are removed by RATD in the outflow cavity (Figure \ref{fig:adisr_midplane}, first panel and Figure \ref{fig:f_disr}, fourth column). We focus on the inner 800 au around the protostar, where the effect of RATD is strongest on dust grains. One can see that inside the outflow cavity, less sub-micron grains can be aligned with $\B$, i.e., larger $a_{\rm align}$, and less micron-sized grains are destroyed by RATD, i.e., larger $a_{\rm disr,POLARIS}$, after taking into account the change in size distribution. This issue can be explained by the enhanced UV-NIR extinction resulting from increased small grains below $2\mum$ by RATD activities. The lower RAT efficiency acting on sub-micron and micron-sized grains leads them to slower rotation and weaker alignment/disruption than results obtained before RATD happens. In contrast, the disruption size inside the envelope decreases compared with results in the first round of MCRT simulation, i.e., lower $a_{\rm disr,POLARIS}$, due to the reduced FIR extinction as VLGs are removed. For SPM grains with lower $N_{\rm cl} < 10^{4}$, the above issue may be fainter because of weaker enrichment of small grains by RATD, i.e., smaller $f_{\rm disr}$. But for composite grains whose minimum disruption size is about $\sim 10\mum$ (Figure \ref{fig:adisr_midplane}, fourth panel), the removal of VLGs can reduce FIR extinction, allowing more smaller sizes to be disrupted by RATD (similar to results of VLGs in the envelope shown in Figure \ref{fig:2nd_time}).

However, the difference between the alignment and disruption size after the first and second rounds of MCRT simulation is nearly negligible. Thus, using results from the first MCRT simulation for interpreting dust polarization properties in Section \ref{sec:ratd_polarization}. is valid for our intermediate-mass protostar with $L_{\rm center} = 100L_{\odot}$. This treatment can be applied to both intermediate-mass and low-mass stars with lower central luminosity due to the inefficient RATD effect acting on dust grains. For massive stars that can produce very high stellar luminosity of $L_{\rm center} \sim 10^{3} - 10^{5}L_{\odot}$, the significant change in size distribution by RATD can cause remarkable changes in the alignment and disruption sizes between different MCRT simulations, especially for grains at thousand au scale which mostly be affected by FIR wavelengths.

\end{document}